\documentclass[journal]{IEEEtran}
\usepackage{lineno}
\usepackage{cite}
\usepackage{amsmath,amssymb,amsfonts}
\usepackage{graphicx}
\usepackage{textcomp}
\usepackage{xcolor}
\usepackage{balance}
\usepackage{cite}
\usepackage{hyperref}
\usepackage{amsmath,amssymb,amsfonts}
\usepackage{amsmath}
\usepackage{amsthm}
\DeclareMathOperator*{\argmax}{argmax}

\usepackage{graphicx}
\usepackage{textcomp}
\usepackage{xcolor}
\usepackage{graphicx}
\usepackage{float}
\usepackage{subfigure}
\usepackage{amsmath}
\usepackage{amsfonts,amssymb}
\usepackage{mathrsfs}
\usepackage{mathtools}
\usepackage{bm}
\usepackage{multirow}
\usepackage{array}
\usepackage{amssymb}
\usepackage{amsmath}
\usepackage{cite}
\usepackage{url}
\usepackage{xcolor}
\usepackage{cite,graphicx,amsmath,amssymb}
\usepackage{subfigure}
\usepackage{fancyhdr}
\usepackage{mdwmath}
\usepackage{mdwtab}
\usepackage{caption}
\usepackage{amsthm}
\usepackage{setspace}
\usepackage{bm}
\usepackage{mathtools}
\usepackage{dsfont}
\usepackage{bbm}
\usepackage{algorithm}
\usepackage{algorithmic}
\usepackage{cases}
\newtheorem{remark}{Remark}
\newtheorem{theorem}{Theorem}

\newtheorem{lemma}{Lemma}

\newtheorem{corollary}{Corollary}

\makeatletter
\newcommand{\biggg}{\bBigg@{3}}
\newcommand{\Biggg}{\bBigg@{3.5}}
\makeatother
\begin{document}
\title{Performance Analysis of Holographic MIMO Based Integrated Sensing and Communications}

\author{Boqun~Zhao,~\IEEEmembership{Graduate Student Member,~IEEE,}~Chongjun~Ouyang,~\IEEEmembership{Member,~IEEE,}\\Xingqi~Zhang,~\IEEEmembership{Senior Member,~IEEE,}
	and Yuanwei~Liu,~\IEEEmembership{Fellow,~IEEE}\vspace{-10pt}
\thanks{B. Zhao and X. Zhang are with the Department of Electrical and Computer Engineering, University of Alberta, Edmonton AB, T6G 2R3, Canada (email: \{boqun1, xingqi.zhang\}@ualberta.ca).}
\thanks{C. Ouyang was with the School of Electrical and Electronic Engineering, University College Dublin, Dublin, D04 V1W8, Ireland, and is now with the School of Electronic Engineering and Computer Science, Queen Mary University of London, London, E1 4NS, U.K. (e-mail: c.ouyang@qmul.ac.uk).}
\thanks{Y. Liu is with the Department of Electrical and Electronic Engineering, The University of Hong Kong, Hong Kong (email: yuanwei@hku.hk).}
}
\maketitle

\begin{abstract}
A holographic multiple-input multiple-output (MIMO)-based integrated sensing and communications (ISAC) framework is proposed for both downlink and uplink scenarios. The spatial correlation is incorporated into the communication channel modeling, while a spherical wave-based model is used to characterize the sensing link. By considering both instantaneous and statistical channel state information, closed-form expressions are derived for sensing rates (SRs), communication rates (CRs), and outage probabilities under various ISAC designs. This enables an investigation into the theoretical performance limits of the proposed holographic MIMO-based ISAC (HISAC) framework. Further insights are gained by examining the high signal-to-noise ratio (SNR) slopes and diversity orders. Specifically: \romannumeral1) for the downlink case, a sensing-centric (S-C) design and a communications-centric (C-C) design are investigated using different beamforming strategies, and a Pareto optimal design is proposed to characterize the attainable SR-CR region; \romannumeral2) for the uplink case, the S-C design and the C-C design differ in the interference cancellation order between the communication and sensing signals, with the rate region obtained through a time-sharing strategy. Numerical results are provided to demonstrate that HISAC systems outperform both conventional MIMO-based ISAC systems and holographic MIMO-based frequency-division sensing and communications systems, underscoring the superior performance of the HISAC framework.
\end{abstract}
\begin{IEEEkeywords}
Holographic multiple-input multiple-output (HMIMO), integrated sensing and communications (ISAC), performance analysis.	
\end{IEEEkeywords}
\section{Introduction}
The concept of integrated sensing and communications (ISAC) has garnered significant attention from both academia and industry, due to its potential contributions to the development of next-generation networks \cite{Zhang2021_JSTSP}. A key feature that distinguishes ISAC is its ability to share the same hardware, power, frequency, and time resources for both communications and sensing. This contrasts with the conventional frequency-division sensing and communications (FDSAC) approach, which requires separate infrastructures and frequency bands for each function. As a result, ISAC is expected to outperform FDSAC in terms of spectrum efficiency, energy consumption, and hardware demands \cite{LiuAn,meng_1}.

The emergence of holographic multiple-input multiple-output (MIMO), inspired by the promising beamforming gains of massive MIMO \cite{chen2021structured} with its large aperture \cite{wangzhe}, has gained significant prominence. Holographic MIMO (HMIMO) is characterized by larger and denser arrays, with antennas spaced at intervals smaller than half the wavelength \cite{holographic, holographic2, var_cal}. This technology is expected to greatly enhance wireless transmission capabilities, particularly with its potential integration into next-generation networks \cite{holographic}. Leveraging these advantages, recent studies \cite{holoISAC2,holoISAC} have applied HMIMO to ISAC, which we refer to as holographic ISAC (HISAC). These studies demonstrate its potential to improve both sensing and communication (S\&C) performance through optimized beamforming strategies.

Research on HISAC is still in its early stages. In \cite{holoISAC2}, the authors provided a comprehensive overview of the hardware structure and working principles of HISAC, which identified challenges and outlined future research opportunities for implementing HISAC networks. A holographic beamforming scheme was proposed in \cite{holoISAC}, which demonstrated superior performance compared to conventional MIMO systems where the antenna spacing is greater than or equal to half a wavelength. In \cite{holo_ISAC3}, an analog and digital beamforming optimization algorithm for multi-band HISAC systems was designed to overcome the physical limitations of ultra-wideband systems. The study in \cite{HISAC} explored a HISAC system utilizing continuous-aperture arrays for both the transmitter and receiver, addressing a joint transmit-receive beamforming optimization problem aimed at balancing multi-target sensing and multi-user communications. Additionally, \cite{Apurba_1} developed an artificial intelligence (AI)-based framework integrating HISAC into a cell-free network to ensure efficient power allocation for beamforming in the desired direction. In \cite{Apurba_2}, also based on an AI framework, the authors optimized dense-location-based power allocation for holographic beamforming by incorporating HMIMO-assisted integrated sensing, localization, and communications. While these studies primarily focus on hardware design and beamforming strategies for HISAC, the performance limits of HISAC remain largely unexplored.

From an information-theoretic perspective, the performance limits of S\&C can be evaluated by the sensing rate (SR) and communication rate (CR), respectively \cite{LiuAn,ouyang2022integrated}. SR measures the system's ability to estimate environmental information through sensing processes, while CR gauges the system's data transmission capacity via communication processes \cite{ouyang2022integrated}. A comprehensive analysis of these two metrics offers valuable insights into the overall performance and effectiveness of ISAC in seamlessly merging S\&C functions.

Some performance analyses have been conducted for single-antenna and conventional MIMO-based ISAC systems. In \cite{ISAC_performance1}, the authors analyzed the performance of ISAC supported by a single-antenna micro base station (BS) under non-orthogonal downlink transmission. Another study \cite{ISAC_performance3} derived achievable performance bounds for a receiver that observes both communications and radar echoes using the same frequency allocation. In \cite{con_MIMO} and \cite{Con_mimo_uplink}, distinct dual-functional S\&C (DFSAC) scenarios were considered, which characterized the downlink and uplink S\&C performance of conventional MIMO-based ISAC, under the assumption of independent and identically distributed (i.i.d.) communication channels. Additionally, in \cite{meng_2}, the authors proposed a novel cooperative scheme for MIMO-ISAC networks, deriving the Cramér–Rao bound and communication rate expressions using stochastic geometry.

However, the distinctive feature of HMIMO lies in the deployment of an exceptionally large number of sub-wavelength spaced antennas, which induces spatial correlation among the antenna elements across the entire array\footnote{In our work, we adopt the definition of HMIMO from \cite{channel_model}, which characterizes HMIMO systems as large-aperture MIMO systems with ultra-dense antenna deployments. Specifically, the antenna spacing in HMIMO systems is set to sub-wavelength levels to form an approximately continuous electromagnetic aperture. In contrast, conventional MIMO systems are defined as those where the antenna spacing is equal to or greater than half a wavelength to minimize spatial correlation. It is evident that, for a given array aperture size and antenna element dimensions, HMIMO systems will have a higher number of antennas than conventional MIMO systems. This results in a larger array gain but also introduces the effects of spatial correlation, which must be accounted for in system design.}. Consequently, the channel fading in HMIMO exhibits inherent spatial correlation, making the use of simplistic i.i.d. fading models unsuitable. Due to this unique characteristic, analyzing the S\&C performance of HMIMO proves to be a challenging task and has not yet been thoroughly explored.

Motivated by the aforementioned research gaps, this article undertakes a comprehensive performance analysis of a HISAC system from the information-theoretic perspective. To address the challenges posed by HMIMO, we derive closed-form expressions for the SR and CR, thereby offering insights into the S\&C performance of HISAC. The main contributions of this article can be summarized as follows:
\begin{itemize}
\item We propose a HISAC framework for both downlink and uplink scenarios, accounting for spatial correlation within the communication channel by leveraging a channel model based on the approximated Fourier plane-wave series expansion \cite{channel_model}. For the sensing link, we employ a spherical wave-based free-space deterministic model.
\item Considering different types of channel state information (CSI) available to the BS, namely instantaneous CSI (I-CSI) and statistical CSI (S-CSI), we derive closed-form expressions for SRs, CRs, and outage probabilities (OPs) based on the proposed channel models and various ISAC designs. For the downlink scenario, we explore sensing-centric (S-C) and communications-centric (C-C) designs, aiming to optimize beamforming strategies that maximize SR and CR, respectively. The SR-CR region is further characterized through a Pareto optimal beamforming design. In the uplink case, we examine S-C and C-C designs with distinct interference cancellation orders for S\&C signals at the BS, and the rate region is derived using a time-sharing strategy between the two designs.    
\item We present numerical results to demonstrate that: \romannumeral1) HISAC achieves higher SRs and CRs than conventional MIMO-based ISAC in both downlink and uplink scenarios; \romannumeral2) HISAC provides more degrees of freedom (DoFs) than FDSAC, and the achievable SR-CR regions of FDSAC are fully encompassed within those of HISAC. 
\end{itemize}

The remainder of this article is organized as follows. Section \ref{system} introduces the conceptual framework of HISAC, including both the communication and sensing models. Sections \ref{downlink} and \ref{uplink} analyze the downlink and uplink S\&C performance, respectively, with a focus on the outcomes for both I-CSI and S-CSI. Section \ref{numerical} presents numerical results to validate the accuracy of the theoretical insights. Finally, Section \ref{conclusion} concludes the article.
\subsubsection*{Notation}
Throughout this paper, scalars, vectors, and matrices are denoted by non-bold, bold lower-case, and bold upper-case letters, respectively. For the matrix $\mathbf{A}$, $[\mathbf{A}]_{i,j}$, ${\mathbf{A}}^{\mathsf{T}}$, ${\mathbf{A}}^{*}$, ${\mathbf{A}}^{\mathsf{H}}$, and $\mathrm{rank}\left( \mathbf{A} \right)$ denote the $(i,j)$th entry, transpose, conjugate, transpose conjugate, and rank of $\mathbf{A}$, respectively. For the square matrix $\mathbf{B}$, ${\mathbf{B}}^{\frac{1}{2}}$, ${\mathbf{B}}^{-1}$, ${\mathsf{tr}}(\mathbf{B})$, and $\det(\mathbf{B})$ denote the principal square root, inverse, trace, and determinant of $\mathbf{B}$, respectively. The notation $[\mathbf{a}]_i$ denotes the $i$th entry of vector $\mathbf{a}$, and ${\mathsf{diag}}\{\mathbf{a}\}$ returns a diagonal matrix whose diagonal elements are entries of $\mathbf{a}$. The notations $\lvert a\rvert$ and $\lVert \mathbf{a} \rVert$ denote the magnitude and norm of scalar $a$ and vector $\mathbf{a}$, respectively. The identity matrix with dimensions $N\times N$ is represented by ${\mathbf{I}}_N$, and the zero matrix is denoted by $\mathbf{0}$. The matrix inequalities ${\mathbf{A}}\succeq{\mathbf{B}}$ and ${\mathbf{A}}\succ{\mathbf{B}}$ imply that $\mathbf{A}-{\mathbf{B}}$ is positive semi-definite and positive definite, respectively. The sets $\mathbbmss{Z}$, $\mathbbmss{R}$, and $\mathbbmss{C}$ stand for the integer, real, and complex spaces, respectively, and notation $\mathbbmss{E}\{\cdot\}$ represents mathematical expectation. The mutual information between random variables $X$ and $Y$ conditioned on $Z$ is shown by $I\left(X;Y|Z\right)$, and $\otimes$ denotes the Kronecker product. The modulus operator and the floor function are represented by $\mathsf{mod}(\cdot,\cdot)$ and $\lfloor \cdot \rfloor$, respectively. Finally, ${\mathcal{CN}}({\bm\mu},\mathbf{X})$ is used to denote the circularly-symmetric complex Gaussian distribution with mean $\bm\mu$ and covariance matrix $\mathbf{X}$.

\begin{figure}[!t]
    \centering
    \subfigbottomskip=0pt
	\subfigcapskip=0pt
    \subfigure[An ISAC system.]
    {
        \includegraphics[height=1.4in]{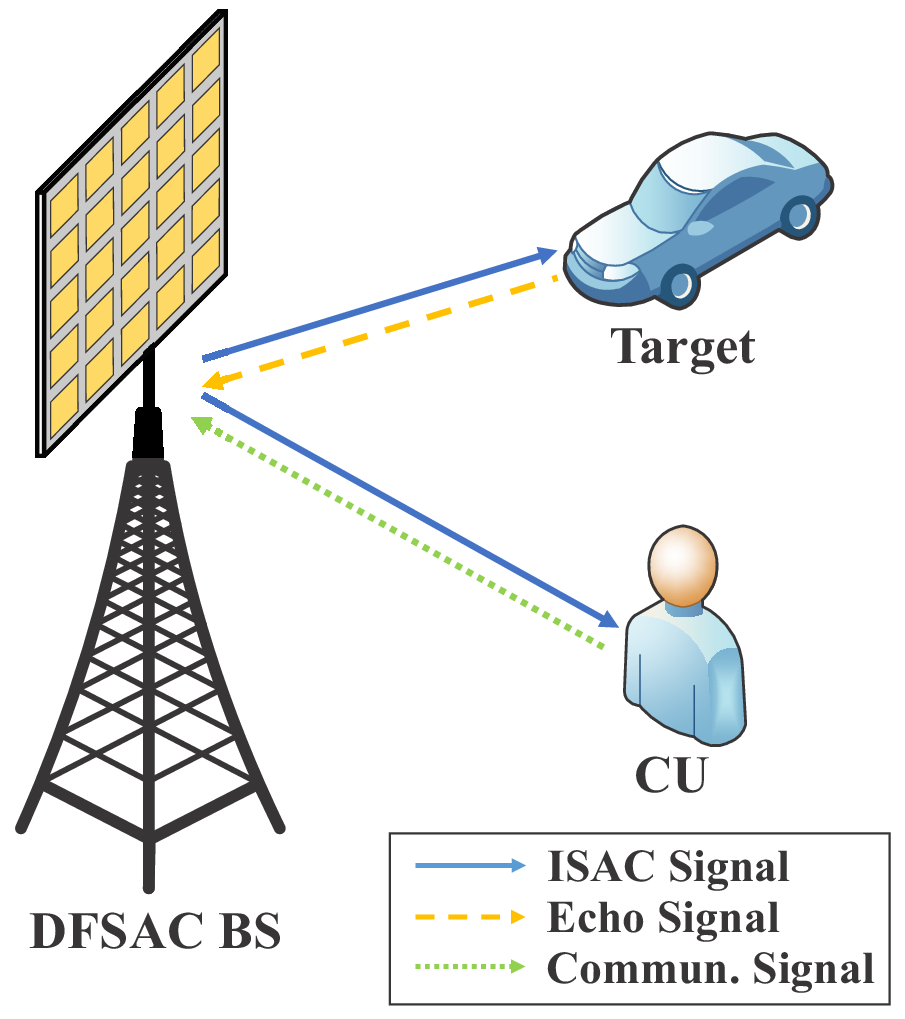}
	   \label{model_a}	
    }
    \quad
   \subfigure[Geometry of HMIMO.]
    {
        \includegraphics[height=1.8in]{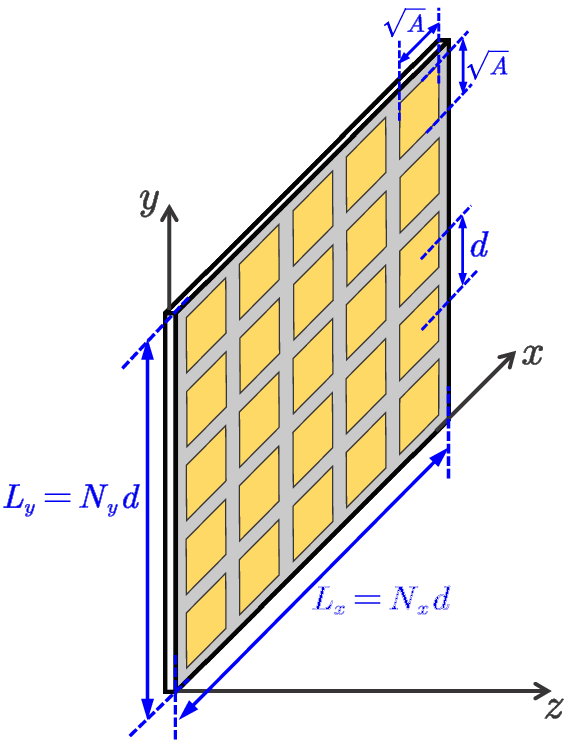}
	   \label{model_b}	
    }
\caption{Illustration of HISAC.}
\vspace{-10pt}
    \label{model}
\end{figure}

\section{System Model}\label{system}
Consider a downlink/uplink ISAC system as depicted in {\figurename} \ref{model_a}, where a DFSAC BS is equipped with a HMIMO containing $N$ antennas. The BS aims to serve a single-antenna communication user (CU) while simultaneously sensing a single target. As shown in {\figurename} \ref{model_b}, we assume that the HMIMO is deployed on the $x$-$y$ plane with $N=N_xN_y$, where $N_x$ and $N_y$ denote the number of antennas along the $x$-axis and $y$-axis, respectively. The aperture size of each antenna is denoted as $\sqrt{A}\times \sqrt{A}$, and the inter-element distance is denoted as $d$ satisfying $\sqrt{A}\leq d<\lambda/2$, where $\lambda$ denotes the wavelength\footnote{The size of an antenna is typically very small, generally less than half a wavelength. For instance, the aperture size of an isotropic antenna is $\frac{\lambda^2}{4\pi}$.}. Hence, the HMIMO has the size of $L_x\times L_y$ with $L_x=N_xd$ and $L_y=N_yd$, and centered at $\mathbf{c}=\left[ L_x/2,L_y/2,0 \right] ^{\mathsf{T}}$. By indexing the antennas row-by-row, the location of the ${n}$th antenna for $n=1,\ldots,N$ can be characterized as follows:
{
\begin{align}
\mathbf{p}_{n}&=\left[ p_{{n},x},p_{{n},y},p_{{n},z} \right] ^{\mathsf{T}}\notag\\
&=\left[ \mathrm{mod}\left( {n}-1,N_x \right) d,\left\lfloor \frac{{{n}-1}}{N_y} \right\rfloor d,0 \right]^{\mathsf{T}}.
\end{align}
}

\vspace{-5pt}
\subsection{Downlink Signal Model}
In downlink HISAC, both communications and sensing are achieved using a single DFSAC signal. This approach leverages the same signal to perform both functions simultaneously. The signal is broadcast into the environment to deliver data to the communication user while also being reflected by the target for sensing purposes, as illustrated in {\figurename} \ref{model_a}. By utilizing the DFSAC signal, communication and sensing operations effectively share the same space-time-frequency-power resource block.

Let $\mathbf{X}=\left[{\mathbf x}_1 \ldots {\mathbf x}_L\right]\in{\mathbbmss{C}}^{N\times L}$ be a DFSAC signal sent from the BS, where $L$ denotes the length of the communication frame/sensing pulse. From a sensing perspective, ${\mathbf x}_\ell\in{\mathbbmss{C} }^{{N}\times1}$ denotes the sensing snapshot transmitted during the $\ell$th time slot for $\ell=1,\ldots,L$. From a communication perspective, ${\mathbf x}_\ell$ corresponds to the $\ell$th data symbol vector. Under the proposed HISAC framework, the downlink ISAC signal $\mathbf{X}$ is given by
{
\begin{align}\label{dual_function_signal_matrix}
\mathbf{X}=\sqrt{p}\mathbf{w} \mathbf{s}^{\mathsf{H}} ,
\end{align}
}where $\mathbf{w}\in{\mathbbmss{C}}^{{N}\times 1}$ represents the normalized beamforming vector with $\lVert \mathbf{w} \rVert ^2=1$, $p$ is the power budget, and $\mathbf{s}=\left[{s}_1 \ldots {s}_L\right]^\mathsf{H}\in{\mathbbmss{C} }^{L\times 1}$ denotes the unit-power data stream intended for the CU with $L^{-1}\lVert \mathbf{s} \rVert^2=1$.
\subsubsection{Sensing Model}
We assume that the target is located at $\mathbf{r}_{\mathrm{s}}=\left[ r_{\mathrm{s},x},r_{\mathrm{s},y},r_{\mathrm{s},z} \right] ^{\mathsf{T}}$. To accurately capture the characteristics of the line-of-sight sensing links, we model the sensing links based on spherical-wave propagation, which accounts for the  path loss variations across each antenna element. Let $\mathbf{h}_{\mathrm{s}} \in \mathbbmss{C} ^{N\times 1}$ denote the sensing link between the BS and the target. The channel coefficient between the ${n}$th antenna and the target is modeled as \cite{usw1}
\begin{equation} \label{sensing_model}
\left[ \mathbf{h}_{\mathrm{s}} \right] _n =\sqrt{\frac{A\mu_{{\mathsf{a}}}}{4\pi r_{\rm{s}}^2}}{{{\rm{e}}}}^{-{{\rm{j}}}k_0\left\| \mathbf{r}_{\mathrm{s}}-\mathbf{p}_{n}\right\|},
\end{equation}
for $n=1,\ldots,N$, where $r_{\rm{s}}=\lVert \mathbf{r}_{\mathrm{s}}-\mathbf{c} \rVert$ is the propagation distance, $k_0=\frac{2\pi}{\lambda}$ is the wavenumber, and $\mu_{\mathsf{a}}$ measures the channel power contributed by a unit antenna aperture at the reference distance of $1$ m \cite{bjornson2020rayleigh}. For simplicity, we define $\alpha_0\triangleq A\mu_{{\mathsf{a}}}$ as the channel power at the reference distance of $1$ m.

When transmitting the DFSAC signal matrix $\mathbf{X}$ for target sensing, the received reflected echo signal at the BS can be written as follows:
\begin{equation}\label{reflected_echo_signal_matrix}
{\mathbf{Y}}_{\mathrm{s}}={\mathbf{G}}{\mathbf{X}}+{\mathbf{N}}_{\mathrm{s}},
\end{equation}
where $\mathbf{G}\in{\mathbbmss{C} }^{N\times N}$ denotes the target response matrix, and ${\mathbf{N}}_{\mathrm{s}}\in{\mathbbmss{C} }^{N\times L}$ denotes the additive white Gaussian noise (AWGN) matrix with each entry having mean zero and variance $\sigma_{\rm{s}}^2$. Furthermore, the target response matrix can be modeled by the round-trip channel as follows: \cite{Tang2019_TSP} 
{
\begin{equation}\label{G_model}
\mathbf{G}=\beta \mathbf{h}_{\mathrm{s}}\mathbf{h}_{\mathrm{s}}^{\mathsf{T}},
\end{equation}
}where $\beta\sim{\mathcal{CN}}\left(0,\alpha_{\rm{s}}\right)$ denotes the complex amplitude of the target with the average strength of $\alpha_{\rm{s}}$. Substituting \eqref{dual_function_signal_matrix} and \eqref{G_model} into \eqref{reflected_echo_signal_matrix} gives
\begin{equation}\label{echo}
\mathbf{Y}_{\mathrm{s}}=\sqrt{p}\beta \mathbf{h}_{\mathrm{s}}\mathbf{h}_{\mathrm{s}}^{\mathsf{T}}\mathbf{ws}^{\mathsf{H}}+\mathbf{N}_{\mathrm{s}}.
\end{equation}

We assume that the location of the target is perfectly tracked and focus on the estimation of the reflection coefficient $\beta$. This sensing task aims to extract environmental information contained within $\beta$ from the echo signal ${\mathbf{Y}}_{\mathrm{s}}$, given the foreknowledge of the DFSAC signal $\mathbf{X}$. The information-theoretic limit for this sensing task can be quantified by the sensing mutual information (MI) that characterizes the MI between ${\mathbf{Y}}_{\mathrm{s}}$ and $\beta$ conditioned on $\mathbf{X}$ \cite{LiuAn,ouyang2022integrated}. On this basis, we employ the SR as the metric for sensing performance evaluation\footnote{Under our considered sensing model, maximizing the SR is equivalent to minimizing the mean-square error in the estimation of the target response $\mathbf{G}$ \cite{ouyang2022integrated,boqun_twc}. More details can be found in Appendix \ref{Appendix:0}.}, which is defined as the sensing MI per unit time \cite{Tang2019_TSP,ouyang2022integrated}.  Assuming that each DFSAC symbol lasts 1 unit time, the SR is written as follows:
{
\begin{equation}\label{SR_define}
\mathcal{R}_{\rm{d},\mathrm{s}}=L^{-1}I\left( \mathbf{Y}_{\mathrm{s}};\beta |\mathbf{X} \right).   
\end{equation}
}In particular, $\mathcal{R}_{\rm{d},\mathrm{s}}$ can be calculated in the following form.
\vspace{-5pt}
\begin{lemma}\label{SR_lemma}
Given $\mathbf{w}$, the SR can be expressed as follows: 
{
\begin{align}\label{SR_Downlink_Basic}
\mathcal{R}_{\rm{d},\mathrm{s}}=L^{-1}\log _2( 1+{p}{\sigma_{\rm{s}}^{-2}}L\alpha _{\mathrm{s}}\lVert \mathbf{h}_{\mathrm{s}} \rVert ^2\lvert \mathbf{h}_{\mathrm{s}}^{\mathsf{T}}\mathbf{w} \rvert^2 ).
\end{align}
}\end{lemma}
\vspace{-5pt}
\begin{IEEEproof}
Please refer to Appendix \ref{Appendix:A} for more details.   
\end{IEEEproof}
\subsubsection{Communication Model}
We assume that the CU is situated at a position of $\mathbf{r}_{\mathrm{c}}=\left[ r_{\mathrm{c},x},r_{\mathrm{c},y},r_{\mathrm{c},z} \right] ^{\mathsf{T}}$. Due to the sub-half-wavelength antenna spacing in the holographic array and the presence of scatterers, spatial correlation for the small-scale fading between antennas becomes significant. Therefore, we model the communication channel response $\mathbf{h}_{\mathrm{c}}\in \mathbbmss{C}^{N\times 1}$ as a correlated Rayleigh-distributed vector, i.e., $\mathbf{h}_{\mathrm{c}}\sim \mathcal{C} \mathcal{N} \left( \mathbf{0},\mathbf{R} \right)$, where $\mathbf{R}={\mathbbmss{E}}\{{\mathbf{h}}_{\rm{c}}{\mathbf{h}}_{\rm{c}}^{\mathsf{H}}\}\succeq{\mathbf{0}}$ is the correlation matrix. 

To further characterize the spatial correlation, we employ a correlation model based on the Fourier plane-wave series expansion of an electromagnetic random channel \cite{channel_model}, which effectively captures the spatial correlation effects under scattering environment. Following the methodology in \cite{channel_model}, the spatial impulse response $[\mathbf{h}_{\mathrm{c}}]_n$ can be derived based on a four-dimensional (4D) Fourier plane-wave representation: 
{
\begin{align}\label{channel_origin}
[\mathbf{h}_{\mathrm{c}}]_n=\frac{1}{4\pi^2}&\iiiint_{\mathcal{D} \times \mathcal{D}}{a_{\rm{c}}\left( k_x,k_y,\mathbf{r}_{\rm{c}} \right) H_a\left( k_x,k_y,\kappa _x,\kappa _y \right) }\notag\\
& \times a_{\rm{b}}\left( \kappa_x,\kappa_y,\mathbf{p}_{n} \right) dk_xdk_yd\kappa_xd\kappa_y,
\end{align}
}where $\bm{\kappa}=[\kappa_x,\kappa_y,\kappa_z]^{\mathsf{T}}$ with $\kappa_z=({k_0^2-\kappa_{x}^{2}-\kappa_{y}^{2}})^{\frac{1}{2}}$, $\mathbf{k}=[k_x,k_y,k_z]^{\mathsf{T}}$ with $k_z=(k_0^2-k_{x}^{2}-k_{y}^{2})^{\frac{1}{2}}$, 
{
\begin{align}
&a_{\rm{b}}\left( \kappa _x,\kappa_y,\mathbf{p}_{n} \right)
={\rm{e}}^{-{\rm{j}}k_0( \kappa_xp_{n,x}+\kappa_yp_{n,y}+\kappa_zp_{n,z} )},\\
&a_{\rm{c}}\left( k_x,k_y,\mathbf{r}_{\rm{c}} \right)={\rm{e}}^{{\rm{j}}k_0\left( k_xr_{\mathrm{c},x} +k_yr_{\mathrm{c},y} +k_zr_{\mathrm{c},z} \right)}
\end{align}
}denote the transmit and receive responses at ${\mathbf{p}}_n$ and $\mathbf{r}_{\rm{c}}$, respectively, and the integration region $\mathcal{D}$ is given by $\mathcal{D} =\{(x,y)\in\mathbbmss{R}^2|x^{2}+y^{2}\leq k_0^2\}$. Besides, $H_a(k_x,k_y,\kappa_x,\kappa_y)$ in \eqref{channel_origin} is refereed to as the angular or wavenumber response between the transmit propagation direction $\frac{\bm{\kappa }}{\left\| \bm{\kappa } \right\|}$ of the BS and the receive propagation direction $\frac{\mathbf{k }}{\left\| \mathbf{k } \right\|}$ of the CU, which is complex Gaussian distributed with mean zero and variance determined by the scattering environment and the array geometry \cite{channel_model}.
\vspace{-5pt}
\begin{remark}
By comparing \eqref{sensing_model} with \eqref{channel_origin}, we observe that the sub-wavelength antenna spacing achieved through HMIMO deployment has a greater impact on the communication model than on the sensing model. Specifically, this sub-wavelength spacing does not significantly alter the sensing model in terms of its physical or mathematical representation. However, as will be demonstrated later, this spacing contributes to a larger array gain, thereby improving sensing performance. Furthermore, in the context of ISAC, where communication and sensing mutually influence each other, the impact of sub-wavelength spacing on the communication model ultimately affects sensing performance as well.
\end{remark}
\vspace{-5pt}
According to \cite{var_cal,channel_model}, the channel description in \eqref{channel_origin} can be discretized via the Karhunen-Loève expansion as follows:
\begin{align} \label{hc_i}
\mathbf{h}_{\mathrm{c}}=\sqrt{N}\sum_{(m_x,m_y)\in\mathcal{E}}H(m_x,m_y)\mathbf{a}(m_x,m_y), 
\end{align}
where $[\mathbf{a}(m_x,m_y)]_n=\frac{1}{\sqrt{N}}a_{\rm{b}}(\frac{2\pi m_x}{L_x},\frac{2\pi m_y}{L_y},\mathbf{p}_n){\rm{e}}^{{\rm{j}}k_0r_{\mathrm{c},z}}$ for $n=1,\ldots,N$, and $\mathcal{E}=\{(m_x,m_y)\in\mathbbmss{Z}^2:(\frac{m_x\lambda}{L_x})^2+(\frac{m_y\lambda}{L_y})^2\leq 1\}$. The set $\{H(m_x,m_y)\sim\mathcal{CN}(0,\sigma^2(m_x,m_y))\}_{(m_x,m_y)\in\mathcal{E}}$ consists of $\lvert\mathcal{E}\rvert\triangleq {\mathsf{n}}$ statistically independent complex Gaussian variables, each with variance $\sigma^2(m_x,m_y)=A\mu_{\mathsf{i}}\sigma_{\mathsf{i}}^2(m_x,m_y)$ \cite{bjornson2020rayleigh}. Here, $A$ represents the area of an antenna, $\mu_{\mathsf{i}}$ is the average intensity attenuation, and $\{\sigma_{\mathsf{i}}^2(m_x,m_y)\}$ is determined by \cite[Eq. (70)]{var_cal}. As a result, $\mathbf{h}_{\mathrm{c}}$ can be reexpressed as follows: \cite{channel_model}
\begin{equation}\label{commun_model}
\mathbf{h}_{\mathrm{c}}=\mathbf{U}\bm{\Sigma }^{\frac{1}{2}}\overline{\mathbf{h}} ,
\end{equation}
where $\overline{\mathbf{h}}\sim \mathcal{CN}(\mathbf{0},\mathbf{I}_{\mathsf{n}})$ contains ${\mathsf{n}}$ independently distributed Gaussian variables, $\bm{\Sigma }\in \mathbbmss{C} ^{{\mathsf{n}}\times {\mathsf{n}}}$ is a diagonal matrix whose diagonal elements are $\{{N}\sigma^2(m_x,m_y)\}_{(m_x,m_y)\in\mathcal{E}}$, and $\mathbf{U}\in \mathbbmss{C}^{N \times {\mathsf{n}}}$ is a deterministic semi-unitary matrix collecting the ${\mathsf{n}}$ vectors $\{\mathbf{a}(m_x,m_y)\}_{(m_x,m_y)\in\mathcal{E}}$ with $\mathbf{U}^{\mathsf{H}}\mathbf{U}=\mathbf{I}_N$. Based on \eqref{commun_model}, the correlation matrix $\mathbf{R}$ can be written as follows\footnote{We use a stochastic multipath fading model \eqref{commun_model} to describe the communication link and a line-of-sight model \eqref{sensing_model} for the sensing link. The multipath fading model is employed because it provides key insights into communication performance, such as diversity order and high-SNR slope, which will be discussed in detail later. In contrast, the sensing link typically involves a round-trip BS-target-BS path. Including additional scatterers in this transmission would result in multiple path losses, leading to significantly reduced power at the BS. Therefore, we omit the links involving other scatterers and rely on \eqref{sensing_model} and \eqref{G_model} to describe the sensing link. This approach is also widely adopted in the current literature; see \cite{holoISAC,con_MIMO,Con_mimo_uplink,meng_2}.}:
\begin{align}\label{correlation_matrix}
\mathbf{R}=\mathbbmss{E} \left\{ \mathbf{h}_{\mathrm{c}}\mathbf{h}_{\mathrm{c}}^{\mathsf{H}} \right\} =\mathbf{U}\bm{\Sigma } \mathbf{U}^{\mathsf{H}}.
\end{align}
Note that \eqref{correlation_matrix} can be regarded as the eigendecomposition (ED) of $\mathbf{R}$ with $\mathrm{rank}\left( \mathbf{R} \right)=\mathsf{n}$ and its $\mathsf{n}$ positive eigenvalues stored in $\{{N}\sigma^2(m_x,m_y)\}_{(m_x,m_y)\in\mathcal{E}}\triangleq\{\lambda_1\geq\ldots\geq\lambda_{\mathsf{n}}>0\}$. 

The received downlink signal at the CU is given by
{
\begin{align}
\mathbf{y}_{\mathrm{c}}^{\mathsf{T}}=\mathbf{h}_{\mathrm{c}}^{\mathsf{T}}\mathbf{X}+\mathbf{n}_{\mathrm{c}}^{\mathsf{T}}=\sqrt{p}\mathbf{h}_{\mathrm{c}}^{\mathsf{T}}\mathbf{ws}^{\mathsf{H}}+\mathbf{n}_{\mathrm{c}}^{\mathsf{T}},
\end{align}
}where $\mathbf{n}_{\mathrm{c}}\in \mathbb{C} ^{L\times 1}$ denotes the AWGN vector with each entry having mean zero and variance $\sigma_{\rm{c}}^2$. Consequently, the downlink CR can be written as follows:
\begin{align}\label{CR_Downlink_Basic}
\overline{\mathcal{R}}_{\mathrm{d},\mathrm{c}}=\log _2( 1+p/\sigma_{\rm{c}}^2\lvert\mathbf{h}_{\mathrm{c}}^{\mathsf{T}}\mathbf{w}\rvert^2).   
\end{align}
\subsubsection*{Downlink Beamforming Design}
Referring to \eqref{SR_Downlink_Basic} and \eqref{CR_Downlink_Basic}, we find that both the SR and CR are influenced by the beamformer $\mathbf{w}$. However, variations in communication and sensing channels, along with different quality of service (QoS) requirements, present challenges in designing beamforming methods that effectively integrate these two functions into a single signal. To address this, we propose three beamforming designs in Section {\ref{downlink}}. The first design, called the \emph{C-C (communications-centric) design}, aims to maximize the CR, which prioritizes communication performance at the expense of sensing capabilities. The second design, termed the \emph{S-C (sensing-centric) design}, focuses on maximizing the SR, which achieves the highest sensing performance but sacrifices communication efficiency. Finally, we propose a \emph{Pareto Optimal design} to characterize the trade-off boundary of the SR-CR region, which provides a balanced optimization between communication and sensing performance. These beamforming designs are essential for enabling efficient integration of communication and sensing functionalities within a single space-time-frequency-power resource block.
\subsection{Uplink Signal Model}
For uplink HISAC, we assume that the communication and sensing components share the same space-time-frequency resource block. In this setup, the DFSAC BS aims to extract environmental information from the reflected sensing echoes while simultaneously detecting the data symbols sent by the CU. The sensing and communication signals are assumed to be perfectly synchronized at the BS by using properly designed synchronization sequences. As a result, the BS observes the following superposed S\&C signal:
\begin{align} \label{BS_receive}
\mathbf{Y}_{\rm{u}}=\sqrt{p_{\mathrm{c}}}\mathbf{h}_{\mathrm{c}} \mathbf{s}_{\mathrm{c}}^\mathsf{H}+\sqrt{p_{\rm{s}}}\mathbf{G}{\mathbf{w}}{\mathbf{s}}_{\mathrm{s}}^\mathsf{H}+\mathbf{N}_{\rm{u}},
\end{align}
where $p_{\mathrm{c}}$ is the communication power budget, $\mathbf{s}_{\mathrm{c}}=\left[ s_{\mathrm{c},1}\ldots s_{\mathrm{c},L} \right]^\mathsf{H}\in{\mathbbmss{C} }^{ L\times 1}$ denotes the message sent by the CU subject to ${\mathbbmss{E}}\{\mathbf{s}_{\mathrm{c}}\mathbf{s}_{\mathrm{c}}^{\mathsf{H}}\}={\mathbf{I}_L}$, $p_{\mathrm{s}}$ is the sensing power budget, $\mathbf{s}_{\mathrm{s}}=\left[ s_{\mathrm{s},1}\ldots s_{\mathrm{s},L} \right]^\mathsf{H}\in{\mathbbmss{C} }^{ L\times 1}$ denotes the sensing pulse subject to $L^{-1}\lVert \mathbf{s}_{\mathrm{s}} \rVert^2=1$, and $\mathbf{N}_{\rm{u}}\in{\mathbbmss{C} }^{N \times L}$ is the AWGN matrix with each entry having mean zero and variance $\sigma_{\rm{u}}^2$.

\subsubsection*{Uplink Interference Cancellation Design}
To address the inter-functionality interference (IFI) between the S\&C signals, the method of successive interference cancellation (SIC) with two different interference cancellation orders can be employed \cite{ouyang2022integrated}. The first is named the \emph{S-C SIC}, where the BS first detects the communication signal $\mathbf{s}_{\mathrm{c}}$ by treating the sensing signal as interference, and then $\sqrt{p_{\mathrm{c}}}\mathbf{h}_{\mathrm{c}} \mathbf{s}_{\mathrm{c}}^\mathsf{H}$ is subtracted from the superposed signal $\mathbf{Y}_{\rm{u}}$, with the remaining part utilized for sensing the target response. The second one is termed the \emph{C-C SIC}, where the BS first senses the target response $\mathbf{G}$ by treating the communication signal as interference, and then $\sqrt{p_{\rm{s}}}\mathbf{G}{\mathbf{w}}{\mathbf{s}}_{\mathrm{s}}^\mathsf{H}$ is subtracted from $\mathbf{Y}_{\rm{u}}$, with the rest part used for detecting the communication signal. Clearly, \emph{the first SIC order enhances communication performance, while the second one improves sensing performance}, which will be further discussed in Section \ref{uplink}. It is worth noting that this approach allows the communication and sensing components to share the same time-frequency-spatial resource block.

In the context of the given HISAC framework, our aim is to assess its S\&C performance in both downlink and uplink scenarios. In the following pages, we will integrate two distinct CSI assumptions into our analyses: \emph{I-CSI (instantaneous CSI)} and \emph{S-CSI (statistical CSI)}. In the I-CSI case, the BS is assumed to possess perfect knowledge of $\mathbf{h}_{\mathrm{c}}$, while in the S-CSI case, the BS only has information about the distribution of $\mathbf{h}_{\mathrm{c}}$ and its correlation matrix $\mathbf{R}$.
\section{Downlink Performance Analysis} \label{downlink}
In this section, we analyze the S\&C performance of downlink HISAC. Specifically, the communication performance is evaluated using the ergodic CR (ECR) and OP, whereas the sensing performance is evaluated through the SR.
\subsection{Instantaneous Channel State Information}
\subsubsection{Sensing-Centric Design}
Under the S-C design, the beamforming vector $\mathbf{w}$ is set to maximize the downlink SR, and the optimal beamforming vector satisfies
{
\begin{align}
\mathbf{w}&=\argmax\nolimits_{\lVert\mathbf{w}\rVert^2=1}\mathcal{R}_{\rm{d},\mathrm{s}}\notag\\
&=\argmax\nolimits_{\lVert\mathbf{w}\rVert^2=1}\lvert \mathbf{h}_{\mathrm{s}}^{\mathsf{T}}\mathbf{w} \rvert^2
={\lVert \mathbf{h}_{\mathrm{s}}\rVert^{-1}}\mathbf{h}_{\mathrm{s}}^{*}.
\end{align}
}By substituting $\mathbf{w}=\frac{\mathbf{h}_{\mathrm{s}}^{*}}{\lVert \mathbf{h}_{\mathrm{s}}\rVert}$ into \eqref{SR_Downlink_Basic}, we write the SR as follows:
{
\begin{align}
\mathcal{R}_{\mathrm{d},\mathrm{s}}^{\mathrm{s}}=L^{-1}\log _2( 1+{p}{\sigma_{\rm{s}}^{-2}}L\alpha _{\mathrm{s}}\lVert \mathbf{h}_{\mathrm{s}} \rVert ^4).
\end{align}
}The following theorem provides a closed-form expression for $\mathcal{R}_{\mathrm{d},\mathrm{s}}^{\mathrm{s}}$ as well as its high-SNR approximation.
\vspace{-5pt}
\begin{theorem}\label{SC_SR_theorem}
In the S-C design, the downlink SR is given by
\begin{align}\label{SC_SR}
\mathcal{R} _{\mathrm{d},\mathrm{s}}^{\mathrm{s}}=\frac{1}{L}\log _2\left( 1+\frac{pL\alpha _{\mathrm{s}}\alpha_0^2N^2}{16\pi^2 \sigma_{\rm{s}}^{2}r_{\mathrm{s}}^{4}} \right) .
\end{align}
When $p\rightarrow \infty $, the SR satisfies
{
\begin{align}\label{SC_SR_p}
\mathcal{R} _{\mathrm{d},\mathrm{s}}^{\mathrm{s}}\approx \frac{1}{L}\log _2{p}+\frac{1}{L}\log _2\left( \frac{L\alpha _{\mathrm{s}}\alpha_0^2N^2}{16\pi^2\sigma_{\rm{s}}^{2} r_{\mathrm{s}}^{4}} \right).
\end{align}
}\end{theorem}
\vspace{-5pt}
\begin{IEEEproof}
Based on \eqref{sensing_model}, we have $\lVert \mathbf{h}_{\mathrm{s}} \rVert ^4=\frac{\alpha_0^2N^2}{16\pi^2r_{\mathrm{s}}^{4}}$, which yields \eqref{SC_SR}. Let $p\rightarrow \infty $ and apply the fact of $\lim_{x\rightarrow\infty}\frac{\log_2(1+x)}{\log_2{x}}=1$ to \eqref{SC_SR}, which yields \eqref{SC_SR_p}. 
\end{IEEEproof} 
To glean more insights into the sensing performance, we examine the high-SNR slope and power offset for the SR, which are, respectively, defined as follows: \cite{mimo} 
	\begin{subequations}
	\begin{align}
		\mathcal{S} _{\mathrm{d},\mathrm{s}}^{\mathrm{s}}&=\lim_{p\rightarrow \infty}\frac{\mathcal{R} _{\mathrm{d},\mathrm{s}}^{\mathrm{s}}}{\log _2p},\\
		\mathcal{O} _{\mathrm{d},\mathrm{s}}^{\mathrm{s}}&=\lim_{p\rightarrow \infty}\left(\log_2p-\frac{\mathcal{R} _{\mathrm{d},\mathrm{s}}^{\mathrm{s}}}{\mathcal{S} _{\mathrm{d},\mathrm{s}}^{\mathrm{s}}}\right) .   
	\end{align}
	\end{subequations}	
High-SNR slope, also referred to as the maximum multiplexing gain, is a commonly used performance metric that describes the rate as a function of the transmit power at high SNR, represented on a log scale. In the context of multi-antenna communications, it measures the spatial DoFs, reflecting the number of available spatial dimensions. Further, high-SNR power offset, which is the zero-order term independent of power, illustrates the power shift with respect to the curve of $\log_2(p)$, enabling different rates to be distinguished even if they have the same high-SNR slope. Consequently, the rate can be enhanced by increasing the high-SNR slope or reducing the high-SNR power offset.
\vspace{-5pt}
\begin{remark}
The results in \textbf{Theorem \ref{SC_SR_theorem}} suggest that the high-SNR slope and power offset of the SR achieved by the S-C design are given by $\frac{1}{L}$ and $\log _2\left( \frac{16\pi^2\sigma_{\rm{s}}^{2} r_{\mathrm{s}}^{4}}{L\alpha _{\mathrm{s}}\alpha_0^2N^2} \right)$, respectively.
\end{remark}
\vspace{-5pt}
As discussed in \eqref{sensing_model}, we have $\alpha_0=A\mu_{{\mathsf{a}}}$, where $\mu_{{\mathsf{a}}}$ represents the distance-dependent channel power gain, and $A$ denotes the area of a single element on the HMIMO array. Substituting $\alpha_0=A\mu_{{\mathsf{a}}}$ into \eqref{SC_SR} gives the SR as follows:
\begin{align}\label{SC_SR_Array_Occupation_Ratio}
\mathcal{R} _{\mathrm{d},\mathrm{s}}^{\mathrm{s}}=\frac{1}{L}\log _2\left( 1+\frac{pL\alpha _{\mathrm{s}}\mu_{{\mathsf{a}}}^2(AN)^2}{16\pi^2 \sigma_{\rm{s}}^{2}r_{\mathrm{s}}^{4}} \right) .
\end{align}
By defining the array occupation ratio (AOR) as $\eta_{\mathsf{aor}}=\frac{A}{d^2}\in(0,1]$, we have
\begin{align}
\frac{AN}{L_xL_y}=\frac{AN}{N_xN_y d^2}=\eta_{\mathsf{aor}}\Leftrightarrow AN = L_xL_y \eta_{\mathsf{aor}}.
\end{align}
Consequently, \eqref{SC_SR_Array_Occupation_Ratio} can be rewritten as follows:
\begin{equation}\label{SR_Array_Aperture_Relation}
\mathcal{R} _{\mathrm{d},\mathrm{s}}^{\mathrm{s}}=\frac{1}{L}\log _2\left( 1+\frac{pL\alpha _{\mathrm{s}}\mu_{{\mathsf{a}}}^2(L_xL_y \eta_{\mathsf{aor}})^2}{16\pi^2 \sigma_{\rm{s}}^{2}r_{\mathrm{s}}^{4}} \right) .
\end{equation}
\vspace{-5pt}
\begin{remark}
The results in \eqref{SR_Array_Aperture_Relation} suggest that, under our considered model, the maximum SR, i.e., the S-C SR, is proportional to the product of the HMIMO array's aperture size $L_xL_y$ and its AOR $\eta_{\mathsf{aor}}$, which represents the effective radiation aperture of the entire array.
\end{remark}
\vspace{-5pt}
Given $L_xL_y$ and $A$, reducing the antenna spacing will add more antennas and increase the AOR, thereby enhancing the SR. However, when the antenna spacing approaches $\sqrt{A}$, i.e., when the antennas are deployed edge-to-edge, the AOR reaches its maximum value of $\eta_{\mathsf{aor}}=1$. Under this condition, the SR will approach its upper limit as follows:
\begin{align}\label{Sensing_Rate_Bound}
\lim_{\eta_{\mathsf{aor}}\rightarrow1}\mathcal{R} _{\mathrm{d},\mathrm{s}}^{\mathrm{s}}=\frac{1}{L}\log _2\left( 1+\frac{pL\alpha _{\mathrm{s}}\mu_{{\mathsf{a}}}^2(L_xL_y)^2}{16\pi^2 \sigma_{\rm{s}}^{2}r_{\mathrm{s}}^{4}} \right) .
\end{align}
These arguments imply that increasing the density of antenna deployment does not result in an unbounded increase in the SR. Instead, the SR is inherently limited by the aperture size of the HMIMO array. Moreover, since $\sqrt{A}<\frac{\lambda}{2}$ is typically satisfied, the bound in \eqref{Sensing_Rate_Bound} cannot be achieved by a conventional MIMO system that uses an array with antenna spacing equal to or greater than $\frac{\lambda}{2}$ to minimize spatial correlation.

We next analyze the communication performance under the S-C design. By inserting $\mathbf{w}={\lVert \mathbf{h}_{\mathrm{s}}\rVert^{-1}}\mathbf{h}_{\mathrm{s}}^{*}$ into \eqref{CR_Downlink_Basic}, we obtain the CR as follows:
\begin{align} \label{SC_CR_def}
\overline{\mathcal{R}}_{\mathrm{d},\mathrm{c}}^{\mathrm{s}}=\log_2( 1+p/\sigma_{\rm{c}}^2\lVert\mathbf{h}_{\mathrm{s}}\rVert^{-2}\lvert\mathbf{h}_{\mathrm{s}}^{\mathsf{H}}\mathbf{h}_{\mathrm{c}}\rvert^2).
\end{align}
The following theorem provides a closed-form expression for the ECR ${\mathcal{R}}_{\mathrm{d},\mathrm{c}}^{\mathrm{s}}=\mathbbmss{E}\{\overline{\mathcal{R}}_{\mathrm{d},\mathrm{c}}^{\mathrm{s}}\}$ and its high-SNR approximation.
\vspace{-5pt}
\begin{theorem}\label{SC_CR_theorem}
The ECR achieved by the S-C design is
\begin{align}\label{SC_CR}
\mathcal{R} _{\mathrm{d},\mathrm{c}}^{\mathrm{s}}=-\frac{1}{\ln 2}{\rm{e}}^{{\sigma_{\rm{c}}^2\varOmega}/{p}}\mathrm{Ei}( -{\sigma_{\rm{c}}^2\varOmega}/{p}  ) ,
\end{align}
where $\varOmega=\frac{N}{\mathbf{b}^{\mathsf{H}}\mathbf{Rb}}$ with $\left[ \mathbf{b} \right] _n=\mathrm{e}^{-\mathrm{j}k_0\left\| \mathbf{r}_{\mathrm{s}}-\mathbf{p}_n \right\|}$ for $n=1,\ldots,N$, and $\mathrm{Ei}\left( x \right) =-\int_{-x}^{\infty}{\frac{{\rm{e}}^{-t}}{t}dt}$ is the exponential integral function. When $p\rightarrow \infty $, the ECR satisfies
{
\begin{align}\label{SC_CR_p}
\mathcal{R} _{\mathrm{d},\mathrm{c}}^{\mathrm{s}}\approx \log_2{p}-\log_2{{\sigma_{\rm{c}}^2}}-\log _2\varOmega -{\mathcal{C}}/{\ln 2},     
\end{align}
}where $\mathcal{C}$ is the Euler constant.
\end{theorem}
\vspace{-5pt}
\begin{IEEEproof}
Please refer to Appendix \ref{Appendix:B} for more details.
\end{IEEEproof}
\vspace{-5pt}
\begin{remark}
The results in \textbf{Theorem \ref{SC_CR_theorem}} suggest that the high-SNR slope and power offset of the ECR achieved by the S-C design are given by $1$ and $\log_2({{\sigma_{\rm{c}}^2}}\varOmega)+\frac{\mathcal{C}}{\ln 2}$, respectively.
\end{remark}
\vspace{-5pt}
We then consider a special case where the antennas are deployed edge-by-edge, i.e., $\eta_{\mathsf{aor}}=1$. It follows that $N=\frac{L_xL_y}{A}$, and thus $\lim_{\eta_{\mathsf{aor}}\rightarrow1}\mathcal{R} _{\mathrm{d},\mathrm{c}}^{\mathrm{s}}=-\frac{1}{\ln 2}{\rm{e}}^{{\frac{L_xL_y\sigma_{\rm{c}}^2}{A{p}\mathbf{b}^{\mathsf{H}}\mathbf{Rb}}}}\mathrm{Ei}( -{\frac{\sigma_{\rm{c}}^2L_zL_y}{Ap\mathbf{b}^{\mathsf{H}}\mathbf{Rb}}}  )$, which is an upper bound for the CR.

Furthermore, the OP of the CR is defined as follows:
{
\begin{align}\label{OP_def}
\mathcal{P} _{\mathrm{d}}^{\mathrm{s}}=\mathrm{Pr}(\overline{\mathcal{R}}_{\mathrm{d},\mathrm{c}}^{\mathrm{s}}<\mathcal{R}_0), 
\end{align}
}where $\mathcal{R}_0>0$ is the target rate. The following theorem provides a closed-form expression for the OP as well as its high-SNR approximation.
\vspace{-5pt}
\begin{theorem}\label{SC_OP_theorem}
The OP achieved by the S-C design is given by
\begin{align}\label{SC_OP}
\mathcal{P} _{\mathrm{d}}^{\mathrm{s}}=1-{\rm{e}}^{-{\sigma_{\rm{c}}^2\varOmega}(2^{\mathcal{R}_0}-1)/p}.
\end{align}
When $p\rightarrow \infty $, the OP satisfies
\begin{align}\label{SC_OP_p}
\mathcal{P} _{\mathrm{d}}^{\mathrm{s}}\approx {\sigma_{\rm{c}}^2\varOmega}(2^{\mathcal{R}_0}-1)/p.   
\end{align}
\end{theorem}
\vspace{-5pt}
\begin{IEEEproof}
Please refer to Appendix \ref{Appendix:B} for more details.
\end{IEEEproof}
We next examine the diversity order and array gains based the asymptotic OP. The diversity order and array gain under the S-C design are, respectively, defined as follows: \cite{mimo}
	\begin{align}
		\mathcal{D} _{\mathrm{d}}^{\mathrm{s}}=-\lim_{p\rightarrow \infty}\frac{\log _2 \mathcal{P} _{\mathrm{d}}^{\mathrm{s}} }{\log _2p}, \
		\mathcal{A} _{\mathrm{d}}^{\mathrm{s}}=\lim_{p\rightarrow \infty}\frac{\left(\mathcal{P} _{\mathrm{d}}^{\mathrm{s}}\right)^{-{1}/{\mathcal{D} _{\mathrm{d}}^{\mathrm{s}}}}}{p}.
	\end{align}
Diversity order, also referred to as the maximum diversity gain, characterizes the slope of the OP as a function of the transmit power at high SNR, represented on a log-log scale, which describes how fast the OP decay as the transmit power grows. The array gain specifies the power gain of the OP with respect to a baseline OP of $p^{-\mathcal{D} _{\mathrm{d}}^{\mathrm{s}}}$. We note that increasing the diversity order or array gain can improves the OP, benefiting the ISAC performance. 
\vspace{-5pt}
\begin{remark}
The results in \textbf{Theorem \ref{SC_OP_theorem}} suggest that the diversity order and array gain of the OP achieved by the S-C design are given by $1$ and ${\sigma_{\rm{c}}^{-2}\varOmega^{-1}}(2^{\mathcal{R}_0}-1)^{-1}$, respectively.
\end{remark}
\vspace{-5pt}
\subsubsection{Communications-Centric Design}
Next we focus on the C-C design. With I-CSI, the beamforming vector that maximizes the downlink CR is given by
{
\begin{align}
\mathbf{w}&=\argmax\nolimits_{\lVert\mathbf{w}\rVert^2=1}\overline{\mathcal{R}}_{\mathrm{d},\mathrm{c}}\notag\\
&=\argmax\nolimits_{\lVert\mathbf{w}\rVert^2=1}\lvert \mathbf{h}_{\mathrm{c}}^{\mathsf{T}}\mathbf{w} \rvert^2
={\lVert \mathbf{h}_{\mathrm{c}}\rVert^{-1}}{\mathbf{h}_{\mathrm{c}}^{*}}.
\end{align}
}Upon substituting $\mathbf{w}={\lVert \mathbf{h}_{\mathrm{c}}\rVert^{-1}}{\mathbf{h}_{\mathrm{c}}^{*}}$ into \eqref{CR_Downlink_Basic}, we express the CR as follows:
{
\begin{align}\label{eq_Rcc}
\overline{\mathcal{R}}_{\mathrm{d},\mathrm{c}}^{\mathrm{c}}=\log_2( 1+p/{\sigma_{\rm{c}}^2}\lVert \mathbf{h}_{\mathrm{c}} \rVert ^2 ),
\end{align}
}A closed-form expression for the OP $\mathcal{P} _{\mathrm{d}}^{\mathrm{c}}=\mathrm{Pr}(\overline{\mathcal{R}}_{\mathrm{d},\mathrm{c}}^{\mathrm{c}}<\mathcal{R}_0)$ is given as follows.
\vspace{-5pt}
\begin{theorem}\label{CC_OP_theorem}
The OP achieved by the C-C design is given by
{
\begin{align}\label{CC_OP}
\mathcal{P} _{\mathrm{d}}^{\mathrm{c}}=\frac{\lambda _{\mathsf{n}}^{\mathsf{n}}}{\prod_{n=1}^{\mathsf{n}}{\lambda_n}}\sum_{k=0}^{\infty}{\frac{\delta _k\Upsilon( \mathsf{n}+k,\sigma_{\rm{c}}^2/p\lambda _{\mathsf{n}}^{-1}( 2^{\mathcal{R}_0}-1 )  )}{\left( \mathsf{n}+k-1 \right) !}},
\end{align}
}where $\Upsilon\left( s,x \right) =\int_0^x{t^{s-1}{\rm{e}}^{-t}dt}$ denotes the lower incomplete gamma function, $\delta _{0}=1$, and $\delta_k$ ($k>0$) can be calculated recursively as $\delta _{k}=\frac{1}{k}\sum_{i=1}^{k}{[\sum_{n=1}^{\mathsf{n}}{(1-{\lambda_{\mathsf{n}}}/{\lambda_{n}})^i}]}\delta_{k-i}$. When $p\rightarrow \infty $, the OP satisfies
{
\begin{align}\label{CC_OP_p}
\mathcal{P} _{\mathrm{d}}^{\mathrm{c}}\approx\frac{(2^{\mathcal{R}_0}-1)^{\mathsf{n}}\sigma_{\rm{c}}^{2\mathsf{n}}}{p^{\mathsf{n}}{\mathsf{n}}!\prod_{n=1}^{\mathsf{n}}{\lambda_n}}.     
\end{align}
}\end{theorem}
\vspace{-5pt}
\begin{IEEEproof}
Please refer to Appendix \ref{Appendix:C} for more details.
\end{IEEEproof}
\vspace{-5pt}
\begin{remark}
The results in \textbf{Theorem \ref{CC_OP_theorem}} suggest that the diversity order and array gain of the OP achieved by the C-C design are given by $\mathsf{n}$ and $({\mathsf{n}}!\prod_{n=1}^{\mathsf{n}}{\lambda_n})^{\frac{1}{\mathsf{n}}}{\sigma_{\rm{c}}^{-2}}(2^{\mathcal{R}_0}-1)^{-1}$, respectively.
\end{remark}
\vspace{-5pt}
Next, we turn our attention to the ECR, ${\mathcal{R}}_{\mathrm{d},\mathrm{c}}^{\mathrm{c}}=\mathbbmss{E}\{\overline{\mathcal{R}}_{\mathrm{d},\mathrm{c}}^{\mathrm{c}}\}$. For clarity, we define the following function: 
\begin{align}
\zeta \left( {\mathbf{A}},a \right) &\triangleq\frac{{\lambda}_{{\mathbf{A}},\mathsf{r}}^{\mathsf{r}}\log _2\mathrm{e}}{\prod_{r=1}^{\mathsf{r}}{{\lambda}_{{\mathbf{A}},r}}}\sum\nolimits_{k=0}^{\infty}{\sum\nolimits_{\mu =0}^{\mathsf{r}+k-1}{\frac{\delta _k}{(\mathsf{r}+k-1-\mu )!}}}\notag\\
&\times \left[ \frac{(-1)^{\mathsf{r}+k-\mu}\mathrm{e}^{1/a{\lambda}_{{\mathbf{A}},\mathsf{r}}^{-1}}}{(a{\lambda}_{{\mathbf{A}},\mathsf{r}})^{\mathsf{r}+k-1-\mu}}\mathrm{Ei}\left( -\frac{1}{a{\lambda}_{{\mathbf{A}},\mathsf{r}}} \right) \right. \notag\\
&\left. +\sum_{u=1}^{\mathsf{r}+k-1-\mu}{( u-1) !\left( -\frac{1}{a{\lambda}_{{\mathbf{A}},\mathsf{r}}} \right) ^{\mathsf{r}+k-1-\mu -u}} \right],  \label{ECR_Basic_Function}      
\end{align}
where ${\lambda}_{{\mathbf{A}},1}\geq\ldots\geq{\lambda}_{{\mathbf{A}},\mathsf{r}}$ denote the positive eigenvalues of a semi-definite matrix $\mathbf{A}\succeq{\mathbf{0}}$ with $\mathsf{r}=\mathrm{rank}\left( \mathbf{A} \right) $. It is important to note the semi-definiteness of $\mathbf{A}$ because the function $\zeta \left( \mathbf{A},a \right)$ is used to express the ergodic rate of a fading channel, where the received SNR is a weighted sum of several gamma-distributed variables with positive weights $\{{\lambda}_{{\mathbf{A}},r}\}_{r=1}^{\mathsf{r}}$, as detailed in Appendix \ref{Appendix:C}.  
\vspace{-5pt}
\begin{theorem}\label{CC_CR_theorem}
The ECR achieved by the C-C design is 
{
\begin{align}\label{CC_CR}
{\mathcal{R}}_{\mathrm{d},\mathrm{c}}^{\mathrm{c}}=\zeta \left( \mathbf{R},p/\sigma _{\mathrm{c}}^{2} \right) . 
\end{align}
}When $p\rightarrow \infty $, the ECR satisfies
{
\begin{equation}\label{CC_CR_p}
\begin{split}
{\mathcal{R}}_{\mathrm{d},\mathrm{c}}^{\mathrm{c}}\approx\log_2{p}-\log_2{{\sigma_{\rm{c}}^2}}+\upsilon_{\mathbf{R}},
\end{split}
\end{equation}
}where $v_{\mathbf{R}}=\frac{\lambda_{\mathsf{n}}^{\mathsf{n}}\log_2{\rm{e}}}{\prod_{n=1}^{\mathsf{n}}{\lambda_n}}
\sum_{k=0}^{\infty}{\delta_k\left(\psi({\mathsf{n}}+k)+\ln(\lambda_{{\mathsf{n}}})\right)}$.
\end{theorem}
\vspace{-5pt}
\begin{IEEEproof}
Please refer to Appendix \ref{Appendix:C} for more details.
\end{IEEEproof}
\vspace{-5pt}
\begin{remark}
The results in \textbf{Theorem \ref{CC_CR_theorem}} suggest that the high-SNR slope and power offset of the ECR achieved by the C-C design are given by $1$ and $\log_2{{\sigma_{\rm{c}}^2}}-v_{\mathbf{R}}$, respectively.
\end{remark}
\vspace{-5pt}
In the previous section, we discussed the SR and CR when the antennas are deployed in an edge-to-edge manner. However, for the C-C design, the variation in antenna spacing $d$ affects the correlation matrix, making a quantitative analysis of the impact of $d$ on the SR and CR a challenging task. As a compromise, we will use simulation results in Section \ref{numerical} to illustrate the effect of $d$ on the HISAC performance.

By substituting $\mathbf{w}={\lVert \mathbf{h}_{\mathrm{c}}\rVert^{-1}}{\mathbf{h}_{\mathrm{c}}^{*}}$ into \eqref{SR_Downlink_Basic}, we obtain the SR as follows:
{
\begin{align}\label{CC_SR_def}
\overline{\mathcal{R}}_{\rm{d},\mathrm{s}}^{\rm{c}}=\frac{1}{L}\log_2\left( 1+\frac{p}{\sigma_{\rm{s}}^2}L\alpha _{\mathrm{s}}\lVert \mathbf{h}_{\mathrm{s}} \rVert^2\frac{\lvert \mathbf{h}_{\mathrm{c}}^{\mathsf{H}}\mathbf{h}_{\mathrm{s}}\rvert^2}{\lVert\mathbf{h}_{\mathrm{c}}\rVert^2}\right).     
\end{align}
}To account for the statistics of $\mathbf{h}_{\mathrm{c}}$, we define the average SR as ${\mathcal{R}}_{\rm{d},\mathrm{s}}^{\rm{c}}\triangleq\mathbbmss{E}\{ \overline{\mathcal{R}}_{\rm{d},\mathrm{s}}^{\rm{c}}\} $ to assess the sensing performance.
\vspace{-5pt}
\begin{theorem}\label{cc_SR_theorem}
The average SR achieved by the C-C design is
{
\begin{align}\label{CC_SR}
{\mathcal{R}}_{\rm{d},\mathrm{s}}^{\rm{c}}=L^{-1}(\upsilon_{\bm{\Delta}}-\upsilon_{\mathbf{R}}),
\end{align}
}where $\upsilon_{\bm{\Delta}}$ is obtained by replacing $\{\lambda_n\}_{n=1}^{\mathsf{n}}$ in $\upsilon_{\mathbf{R}}$ with the positive eigenvalues of $\mathbf{\Delta }=\mathbf{R}^{\frac{1}{2}}( p/\sigma_{\rm{s}}^2L\alpha _{\mathrm{s}}\lVert\mathbf{h}_{\mathrm{s}}\rVert^2\mathbf{h}_{\mathrm{s}}\mathbf{h}_{\mathrm{s}}^{\mathsf{H}}+\mathbf{I}_N)\mathbf{R}^{\frac{1}{2}}$. When $p\rightarrow \infty $, the average SR satisfies
{
\begin{align}\label{CC_SR_p}
{\mathcal{R}}_{\rm{d},\mathrm{s}}^{\rm{c}}\approx L^{-1}( \log_2{p}-\log _2\varXi -{\mathcal{C}}/{\ln 2}- \upsilon_\mathbf{R}),     
\end{align}
}where $\varXi =\frac{16\pi^2\sigma_{\rm{s}}^{2} r_{\mathrm{s}}^{4}}{L\alpha _{\mathrm{s}}\alpha_0^2N^2}\varOmega$.
\end{theorem}
\vspace{-5pt}
\begin{IEEEproof}
Please refer to Appendix~\ref{Appendix:D} for more details.
\end{IEEEproof}
\vspace{-5pt}
\begin{remark}
The results in \textbf{Theorem \ref{cc_SR_theorem}} suggest that the high-SNR slope and power offset of the SR achieved by the S-C design are given by $\frac{1}{L}$ and $\log _2\varXi +\frac{\mathcal{C}}{\ln 2}+ \upsilon_\mathbf{R}$, respectively.
\end{remark}
\vspace{-5pt}
\vspace{-5pt}
\begin{remark} \label{do_ins_compare}
The above arguments suggest that for downlink HISAC with I-CSI, the beamforming design \emph{does not influence the high-SNR slope}. However, it does \emph{impact the diversity order, array gain, and high-SNR power offset}. Since ${\mathsf{n}}\geq 1$, the C-C design can achieve a higher diversity order than the S-C design.
\end{remark}
\vspace{-5pt}
\subsubsection{Pareto Optimal Design}
In practical scenarios, the beamforming vector $\mathbf{w}$ can be customized to meet diverse quality-of-service requirements, introducing a trade-off between S\&C performance. To evaluate this trade-off, we investigate the Pareto boundary of the SR-CR region. Specifically, the Pareto boundary comprises SR-CR pairs where it is impossible to enhance one of the rates without simultaneously reducing the other \cite{pareto}. Any rate pair situated on this boundary can be identified by solving the following problem: \cite{pareto}
{
\begin{subequations}\label{Problem_CR_SR_Tradeoff}
\begin{align}
\max_{\mathbf{w},\mathcal{R}} \quad &\mathcal{R} 
\\
\mathrm{s}.\mathrm{t}.\quad &{L}^{-1}\log_2(1+p/\sigma_{\rm{s}}^2L\alpha_{\mathrm{s}}\lVert\mathbf{h}_{\mathrm{s}}\rVert^2\lvert\mathbf{h}_{\mathrm{s}}^{\mathsf{T}}\mathbf{w}\rvert^2)\geq\tau \mathcal{R},\label{ins_b}
\\
&\log_2(1+p/\sigma_{\rm{c}}^2\lvert \mathbf{h}_{\mathrm{c}}^{\mathsf{T}}\mathbf{w} \rvert^2) \geq (1-\tau)\mathcal{R} ,\label{ins_c}\\
&\lVert \mathbf{w} \rVert ^2=1,\label{ins_d}
\end{align}    
\end{subequations}
}where $\tau\in\left[0,1\right]$ denotes a particular rate-profile parameter. The complete Pareto boundary is obtained by solving the problem \eqref{Problem_CR_SR_Tradeoff} with $\tau$ ranging from $0$ to $1$. The solution to problem \eqref{Problem_CR_SR_Tradeoff} is given as follows.
\vspace{-5pt}
\begin{theorem}\label{pareto_theo}
For a given $\tau$, the optimal beamforming vector is given by
\begin{align}\label{optimal_w}
\mathbf{w}_{\tau}=\begin{cases}
	{\lVert \mathbf{h}_{\mathrm{c}}\rVert^{-1}}{\mathbf{h}_{\mathrm{c}}^{*}},  &\tau \in \left[ 0,  \frac{\overline{\mathcal{R}}_{\rm{d},\mathrm{s}}^{\rm{c}}}{{\overline{\mathcal{R}}_{\mathrm{d},\mathrm{c}}^{\mathrm{c}}+\overline{\mathcal{R}}_{\rm{d},\mathrm{s}}^{\rm{c}}}}\right]\\
	a\mathbf{h}_1^{*}+b\mathbf{h}_2^{*}{\rm{e}}^{-{\rm{j}}\angle \rho}, &\tau \in \left(\frac{\overline{\mathcal{R}}_{\rm{d},\mathrm{s}}^{\rm{c}}}{{\overline{\mathcal{R}}_{\mathrm{d},\mathrm{c}}^{\mathrm{c}}+\overline{\mathcal{R}}_{\rm{d},\mathrm{s}}^{\rm{c}}}},\frac{\mathcal{R}_{\mathrm{d},\mathrm{s}}^{\mathrm{s}}}{\overline{\mathcal{R}}_{\mathrm{d},\mathrm{c}}^{\mathrm{s}}+\mathcal{R}_{\mathrm{d},\mathrm{s}}^{\mathrm{s}}}\right)\\
	{\lVert \mathbf{h}_{\mathrm{s}}\rVert^{-1}}{\mathbf{h}_{\mathrm{s}}^{*}},  &\tau \in \left[\frac{\mathcal{R}_{\mathrm{d},\mathrm{s}}^{\mathrm{s}}}{\overline{\mathcal{R}}_{\mathrm{d},\mathrm{c}}^{\mathrm{s}}+\mathcal{R}_{\mathrm{d},\mathrm{s}}^{\mathrm{s}}}, 1\right]
\end{cases}.
\end{align}
where $\mathbf{h}_1=\sqrt{p/\sigma_{\rm{c}}^2}\mathbf{h}_{\mathrm{c}} $, $\mathbf{h}_2=\sqrt{\frac{p\alpha_0L\alpha _sN}{4\pi \sigma_{\rm{s}}^2r_{\mathrm{s}}^{2}}}\mathbf{h}_{\mathrm{s}}$, $\rho =\mathbf{h}_{1}^{\mathsf{H}}\mathbf{h}_2$, $a=\frac{\mu _1\sqrt{2^{( 1-\tau ) \mathcal{R} ^{\star}}-1}}{\varsigma }$ and $b=\frac{\mu _2\sqrt{2^{\tau L\mathcal{R} ^{\star}}-1}}{\varsigma}$ with $\varsigma$ for normalization, $\mu _1=\frac{\xi _2}{\chi}$ and $\mu _2=\frac{\xi _1}{\chi}$ with $\xi _1=\lVert \mathbf{h}_1 \rVert ^2-\sqrt{\frac{2^{\left( 1-\tau \right) \mathcal{R}^\star}-1}{2^{\tau L\mathcal{R}^\star}-1}}\lvert \rho \rvert$, $\xi _2=\lVert \mathbf{h}_2 \rVert ^2-\sqrt{\frac{2^{\tau L\mathcal{R}^\star}-1}{2^{( 1-\tau ) \mathcal{R}^\star}-1}}\lvert \rho \rvert$, and $\chi =\xi _12^{\tau L\mathcal{R}^\star}\tau L\ln 2+\xi _22^{( 1-\tau ) \mathcal{R}^\star}\left( 1-\tau \right) \ln 2$. $\mathcal{R}^\star$ denotes the solution to the equation $\xi _1( 2^{\tau L\mathcal{R}}-1 ) +\xi _2( 2^{( 1-\tau ) \mathcal{R}}-1 )= \lVert \mathbf{h}_1 \rVert ^2\lVert \mathbf{h}_2 \rVert ^2-\lvert\rho\rvert^2$.
\end{theorem}
\vspace{-5pt}
\begin{IEEEproof}
Please refer to Appendix \ref{Appendix:E} for more details.
\end{IEEEproof}
\vspace{-5pt}
\begin{remark}
The results in \textbf{Theorem \ref{pareto_theo}} suggest that the Pareto optimal beamforming vector lies in the plane spanned by $\mathbf{h}_{\mathrm{c}}^{*}$ and $\mathbf{h}_{\mathrm{s}}^{*} {\rm{e}}^{-{\rm{j}}\angle \rho}$.
\end{remark}
\vspace{-5pt} 
Given $\tau$, let $\mathcal{R}_{\mathrm{d},\mathrm{s}}^{\tau}$ and $\mathcal{R}_{\mathrm{d},\mathrm{c}}^{\tau}$ denote the SR and the CR achieved by Pareto optimal beamforming vector $\mathbf{w}_{\tau}$, respectively. Accordingly, we have $\mathcal{R}_{\mathrm{d},\mathrm{s}}^{\tau}\in [ \overline{\mathcal{R}}_{\rm{d},\mathrm{s}}^{\rm{c}},\mathcal{R} _{\mathrm{d},\mathrm{s}}^{\mathrm{s}} ] $ and $\mathcal{R}_{\mathrm{d},\mathrm{c}}^{\tau} \in [ \overline{\mathcal{R}}_{\mathrm{d},\mathrm{c}}^{\mathrm{s}},  \overline{\mathcal{R}}_{\mathrm{d},\mathrm{c}}^{\mathrm{c}}]$ with $\mathcal{R}_{\mathrm{d},\mathrm{s}}^{0}=\overline{\mathcal{R}}_{\rm{d},\mathrm{s}}^{\rm{c}}$, $\mathcal{R}_{\mathrm{d},\mathrm{s}}^{1}=\mathcal{R} _{\mathrm{d},\mathrm{s}}^{\mathrm{s}}$, $\mathcal{R}_{\mathrm{d},\mathrm{c}}^{0}=\overline{\mathcal{R}}_{\mathrm{d},\mathrm{c}}^{\mathrm{c}}$, and $\mathcal{R}_{\mathrm{d},\mathrm{c}}^{1}=\overline{\mathcal{R}}_{\mathrm{d},\mathrm{c}}^{\mathrm{s}}$. By exploiting the Sandwich theorem, the high-SNR slopes and diversity orders of any SR-CR pair on the Pareto boundary can be obtained. Finally, letting $({\mathcal{R}}_{\mathrm{s}},{\mathcal{R}}_{\mathrm{c}})$ denote the achievable SR-CR pair, the SR-CR region achieved by HISAC with I-CSI can be written as follows:
{
\begin{align}\label{Rate_Regio_ISAC}
\mathcal{C}_{\mathrm{d}}=\left\{\left({\mathcal{R}}_{\mathrm{s}},{\mathcal{R}}_{\mathrm{c}}\right)\left|
\begin{aligned}
{\mathcal{R}}_{\mathrm{s}}\in[0,{\mathcal{R}_{\mathrm{d},\mathrm{s}}^{\tau}}],
{\mathcal{R}}_{\mathrm{c}}\in[0,{\mathcal{R}_{\mathrm{d},\mathrm{c}}^{\tau}}],\tau\in\left[0,1\right]
\end{aligned}
\right.\right\}.
\end{align}
}\subsection{Statistical Channel State Information}
Having investigated the I-CSI case, we now move to the S-CSI case where the BS has no knowledge of $\mathbf{h}_{\mathrm{c}}$ and only knows $\mathbf{R}$.
\subsubsection{Sensing-Centric Design} 
Since the optimal beamforming vector under the S-C design, i.e., ${\lVert \mathbf{h}_{\mathrm{s}}\rVert^{-1}}{\mathbf{h}_{\mathrm{s}}^{*}}$, is independent of $\mathbf{h}_{\mathrm{c}}$, the S\&C performance of the S-C design in the S-CSI case is the same as that in the I-CSI case.
\subsubsection{Communications-Centric Design}
Since the instantaneous information of $\mathbf{h}_{\mathrm{c}}$ is unknown to the BS, the C-C beamforming vector is set to maximize the ECR instead of the instantaneous CR. As a result, the optimal beamforming vector satisfies
{
\begin{align}
\mathbf{w}&=\argmax\nolimits_{\lVert\mathbf{w}\rVert^2=1}\mathbbmss{E}\{\overline{\mathcal{R}}_{\mathrm{d},\mathrm{c}}\}\\
&=\argmax\nolimits_{\lVert\mathbf{w}\rVert^2=1}\mathbbmss{E}\{\log _2( 1+p/\sigma_{\rm{c}}^2\lvert\mathbf{h}_{\mathrm{c}}^{\mathsf{T}}\mathbf{w}\rvert^2)\}.\label{Average_CR_C_C_Beam}
\end{align}
}The solution to the above problem is summarized as follows.
\vspace{-5pt}
\begin{lemma}\label{Lemma_S_CSI_Downlink}
The optimal beamforming vector under the C-C design with S-CSI is given by the normalized principal eigenvector of $\mathbf{R}$, i.e., $\mathbf{a}(m_x^{\star},m_y^{\star})\triangleq{\mathbf{a}}_{\star}$, where
{
\begin{align}
(m_x^{\star},m_y^{\star})=\argmax\nolimits_{(m_x,m_y)\in\mathcal{E}}\sigma^2(m_x,m_y).
\end{align}
}\end{lemma}
\vspace{-5pt}
\begin{IEEEproof}
Please refer to Appendix \ref{Appendix:F} for more details.
\end{IEEEproof}
By setting $\mathbf{w}={\mathbf{a}}_{\star}$, we obtain 
{
\begin{align}
{\mathbf{a}}_{\star}^{\mathsf{H}}{\mathbf{R}}{\mathbf{a}}_{\star}=\max\nolimits_{(m_x,m_y)\in\mathcal{E}}\sigma^2(m_x,m_y)=\lambda_{1}.
\end{align}
}We then present the analytical results for the ECR and OP in the following theorems.
\vspace{-5pt}
\begin{theorem}\label{CC_st_CR_the}
With S-CSI, the ECR achieved by the C-C design can be written as follows:
{
\begin{align}
\mathcal{R} _{\mathrm{d},\mathrm{c}}^{\mathrm{c}}=-{\rm{e}}^{{\sigma_{\rm{c}}^2}\lambda_1^{-1}/{p} }\mathrm{Ei}( -{\sigma_{\rm{c}}^2}\lambda_1^{-1}/{p}  )\log_2{\rm{e}} .    
\end{align}
}When $p\rightarrow \infty $, the ECR satisfies
{
\begin{align}
\mathcal{R} _{\mathrm{d},\mathrm{c}}^{\mathrm{c}}\approx \log_2{p}-\log_2{{\sigma_{\rm{c}}^2}}+\log _2\lambda_1 -{\mathcal{C}}/{\ln 2}.     
\end{align}
}\end{theorem}
\vspace{-5pt}
\begin{IEEEproof}
Similar to the proof of \textbf{Theorem~\ref{SC_CR_theorem}}.
\end{IEEEproof}
\vspace{-5pt}
\begin{theorem}\label{CC_st_OP_the}
With S-CSI, the OP achieved by the C-C design can be written as follows:
{
\begin{align}
\mathcal{P} _{\mathrm{d}}^{\mathrm{c}}=1-{\rm{e}}^{-{\sigma_{\rm{c}}^2}\lambda_1^{-1}(2^{\mathcal{R}_0}-1)/p}.    
\end{align}
}When $p\rightarrow \infty $, the OP satisfies
{
\begin{align}
\mathcal{P} _{\mathrm{d}}^{\mathrm{c}}\approx{\sigma_{\rm{c}}^2}\lambda_1^{-1}(2^{\mathcal{R}_0}-1)/p.   
\end{align}
}\end{theorem}
\vspace{-5pt}
\begin{IEEEproof}
Similar to the proof of \textbf{Theorem~\ref{SC_OP_theorem}}.
\end{IEEEproof}
\vspace{-5pt}
\begin{remark}
The results in \textbf{Theorem \ref{CC_st_CR_the}} and \textbf{\ref{CC_st_OP_the}} suggest that the high-SNR slope and diversity order of the CR achieved by the C-C design with S-CSI are both equal to $1$.
\end{remark}
\vspace{-5pt}
In the sequel, we analyze the SR achieved by the C-C design under the S-CSI case.
\vspace{-5pt}
\begin{theorem}\label{cc_SR_theorem_S_CSI}
Let us define $\varGamma\triangleq\frac{L\alpha _{\mathrm{s}}\alpha_0^2N}{16\pi ^2r_{\mathrm{s}}^{4}}\lvert \sum_{n=1}^N{a_{\rm{b}}^{*}(\frac{2\pi m_x}{L_x},\frac{2\pi m_y}{L_y},\mathbf{p}_n)}{\rm{e}}^{-{\rm{j}}k_0r_{\mathrm{c},z}-{\rm{j}}k_0\lVert \mathbf{r}_{\mathrm{s}}-\mathbf{p}_n \rVert} \rvert^2$. The SR achieved by the C-C design with S-CSI can be written as follows:
{
\begin{align}
\mathcal{R} _{\mathrm{d},\mathrm{s}}^{\mathrm{c}}=L^{-1}\log _2\left( 1+p/\sigma_{\rm{s}}^2\varGamma \right) ,    
\end{align}
}When $p\rightarrow \infty $, the SR satisfies
{
\begin{align}
\mathcal{R} _{\mathrm{d},\mathrm{s}}^{\mathrm{c}}\approx L^{-1}\log _2p+L^{-1}\log _2(\varGamma/\sigma_{\rm{s}}^2).  
\end{align}
}\end{theorem}
\vspace{-5pt}
\begin{IEEEproof}
Similar to the proof of \textbf{Theorem~\ref{SC_SR_theorem}}.
\end{IEEEproof}
\vspace{-5pt}
\begin{remark}
The results in \textbf{Theorem \ref{cc_SR_theorem_S_CSI}} suggest that the high-SNR slope and power offset of the SR achieved by the C-C design with S-CSI are given by $1/L$ and $\log _2(\sigma_{\rm{s}}^2/\varGamma)$, respectively.
\end{remark}
\vspace{-5pt}
In addition to this, the following conclusion can be found.
\vspace{-5pt}
\begin{remark} \label{do_st_compare}
The above arguments imply that for downlink HISAC with S-CSI, the beamforming design \emph{does not affect the high-SNR slope and diversity order}, but it does \emph{impact the array gain and high-SNR power offset}. Specifically, under the C-C design, the diversity order achieved with I-CSI is higher than that achieved with S-CSI.
\end{remark}

\subsubsection{Pareto Optimal Design}
We next characterize the downlink rate region in the case of S-CSI. Based on the proof in Appendix \ref{Appendix:G}, the ECR $\mathbbmss{E}\{\overline{\mathcal{R}}_{\mathrm{d},\mathrm{c}}\}$ is monotonically increasing with ${\mathbf{w}^{\mathsf{T}}\mathbf{Rw}^{*}}$. As a result, the rate region can be characterized by first solving the following problem:
{
\begin{subequations}\label{problem_sta}
\begin{align}
\max_{\mathbf{w},x} \quad &x \\
\mathrm{s}.\mathrm{t}.\quad &{\mathbf{w}^{\mathsf{T}}\mathbf{Rw}^{*}} \ge ( 1-\tau ) x ,~\lvert \mathbf{h}_{\mathrm{s}}^{\mathsf{T}}\mathbf{w} \rvert^2\ge \tau x,\\
&\lVert\mathbf{w}\rVert^2=1,
\end{align}
\end{subequations}
}for $\tau\in[0,1]$, and then calculating the corresponding SR-ECR pair. By defining $\mathbf{W}\triangleq \mathbf{w}^{*}\mathbf{w}^{\mathsf{T}}$ and $\mathbf{H}\triangleq \mathbf{h}_{\mathrm{s}}\mathbf{h}_{\mathrm{s}}^{\mathsf{H}}$, problem \eqref{problem_sta} can be rewritten as follows:
{
\begin{subequations}\label{sdr}
\begin{align}
\max_{\mathbf{W},x} \quad &x \\
\mathrm{s}.\mathrm{t}.\quad &\mathrm{tr}( \mathbf{RW} ) \geq ( 1-\tau ) x  , \ \mathrm{tr}\left( \mathbf{HW} \right) \geq \tau x , \\
&\mathrm{tr}( \mathbf{W} ) =1, \ \mathbf{W}\succeq \mathbf{0}, \ \mathrm{rank}\left( \mathbf{W} \right) =1,
\end{align}
\end{subequations}
}which is a non-convex problem whose optimal solution generally requires a brute-force search. Let $(\mathcal{R}_{\mathrm{d},\mathrm{s}}^{\tau},\mathcal{R}_{\mathrm{d},\mathrm{c}}^{\tau})$ denote the optimal SR-CR pair corresponding to $\tau$. Then the high-SNR slopes and diversity orders of $(\mathcal{R}_{\mathrm{d},\mathrm{s}}^{\tau},\mathcal{R}_{\mathrm{d},\mathrm{c}}^{\tau})$ can be derived by using the Sandwich theorem. Besides, the rate region can be characterized as follows:
{
\begin{align}\label{Rate_Regio_ISAC1}
\mathcal{C}_{\mathrm{d}}=\left\{\left({\mathcal{R}}_{\mathrm{s}},{\mathcal{R}}_{\mathrm{c}}\right)\left|
\begin{aligned}
{\mathcal{R}}_{\mathrm{s}}\in[0,{\mathcal{R}_{\mathrm{d},\mathrm{s}}^{\tau}}],
{\mathcal{R}}_{\mathrm{c}}\in[0,{\mathcal{R}_{\mathrm{d},\mathrm{c}}^{\tau}}],\tau\in\left[0,1\right]
\end{aligned}
\right.\right\}.
\end{align}}

Unfortunately, obtaining the entire rate region is computationally inefficient. As a compromise, we next provide an inner bound of the rate region. The non-convexity of \eqref{sdr} lies in the rank-one constraint $\mathrm{rank}( \mathbf{W} ) =1$. For this constraint, the conventional approach is applying semidefinite relaxation (SDR), where we first solve the problem by ignoring the rank-one constraint, and then construct a rank-one solution with Gaussian randomization method if the solution obtained from the relaxed problem is not rank-one \cite{SDR}, as outlined in \textbf{Algorithm \ref{algorithm1}}. Let $\overline{\mathcal{C}}_{\mathrm{d}}$ denote the SR-CR region achieved by the SDR-based method. Then we have $\overline{\mathcal{C}}_{\mathrm{d}}\subseteq{\mathcal{C}}_{\mathrm{d}}$. The boundary of $\overline{\mathcal{C}}_{\mathrm{d}}$ thus serves an inner bound for $\overline{\mathcal{C}}_{\mathrm{d}}$, whose tightness will be verified by the numerical results presented in Section \ref{numerical}. 

\begin{algorithm}[!t]
	\algsetup{linenosize=\tiny} \scriptsize
	\caption{SDR-based Method for Solving Problem \eqref{problem_sta}}
	\label{algorithm1}
	\begin{algorithmic}[1]
		\STATE Given an SDR solution $\mathbf{W}^{\star}$ for \eqref{sdr} and the number of randomizations $M$;
		\STATE Perform the Cholesky decomposition $\mathbf{W}^{\star}={\mathbf{P}}^{*}{\mathbf{P}}^{\mathsf{T}}$.
		\FORALL{$m=1:M$} 
		\STATE Generate $\mathbf{d}_{m}\sim \mathcal{C} \mathcal{N} \left( \mathbf{0},\mathbf{I}_N \right) $ and compute $\tilde{\mathbf{w}}_m=\mathbf{Pd}_m\sim \mathcal{C} \mathcal{N} \left( \mathbf{0},\mathbf{W}^{\star} \right) $;
		\STATE Construct a feasible point $\mathbf{w}_m=\frac{{\tilde{\mathbf{w}}}_m}{\lVert {\tilde{\mathbf{w}}}_m \rVert}$;     
		\ENDFOR
		\STATE Update $m^{\star}=\argmax _{m=1,\ldots,M}\min ( \frac{1}{\tau}{{\mathbf{w}_m^{\mathsf{T}}}{\mathbf{H}}{\mathbf{w}}_m^{*}},\frac{1}{1-\tau}{{\mathbf{w}}_m^{\mathsf{T}}{\mathbf{R}}{\mathbf{w}}_m^{*}}) $;
		\STATE Output ${\mathbf{w}}_{m^{\star}}$ as the solution.
	\end{algorithmic}
\end{algorithm}

\subsection{HMIMO Based FDSAC}
We consider HMIMO based FDSAC, referred to as FDSAC for simplicity, as a baseline scenario. Different from downlink ISAC where S\&C share the total bandwidth and power simultaneously, FDSAC involves splitting the total bandwidth into two sub-bands: one dedicated solely to communications and the other to sensing. Additionally, the overall power is also divided into two separate allocations for S\&C, respectively. Specifically, we assume that $\kappa\in \left[ 0,1 \right]$ fraction of the total bandwidth and $\iota\in \left[ 0,1 \right]$ fraction of the total power is used for communications, and the other is used for sensing. Consequently, the SR achieved by FDSAC is given by $\mathcal{R} _{\mathrm{d},\mathrm{s}}^{\mathrm{f}}=\frac{1-\kappa}{L}\log _2( 1+\frac{1-\iota}{1-\kappa}p/\sigma_{\rm{s}}^2L\alpha _{\mathrm{s}}\lVert \mathbf{h}_{\mathrm{s}} \rVert ^4 )$. As for communications, the CRs in the cases of I-CSI and S-CSI can be written as $\mathcal{R} _{\mathrm{d},\mathrm{c}}^{\mathrm{f}}=\kappa\log_2( 1+\frac{\iota}{\kappa}p/{\sigma_{\rm{c}}^2}\lVert \mathbf{h}_{\mathrm{c}} \rVert ^2 )$ and $\mathcal{R} _{\mathrm{d},\mathrm{c}}^{\mathrm{f}}=\kappa\log _2( 1+\frac{\iota}{\kappa}p/\sigma_{\rm{c}}^2\lvert\mathbf{h}_{\mathrm{c}}^{\mathsf{T}}\mathbf{a}_{\star}\rvert^2)$, respectively. Note that the CR and SR achieved by FDSAC can be analyzed in a similar way we analyze those achieved by HISAC. For the sake of reference, we summarize the results related to diversity order and high-SNR slope in Table \ref{table1}.
\begin{table}[!t]
\center
\scalebox{0.8}{
\begin{tabular}{|c|c|c|c|c|c|c|}\hline
\multicolumn{1}{|c|}{\multirow{3}{*}{Design}} & \multicolumn{2}{c|}{Sensing} & \multicolumn{4}{c|}{Communications} \\ \cline{2-7}
&I-CSI & S-CSI & \multicolumn{2}{c|}{I-CSI} & \multicolumn{2}{c|}{S-CSI}\\ \cline{2-7}
 & $\mathcal{S}$  & $\mathcal{S}$  & $\mathcal{S}$ & $\mathcal{D}$   & $\mathcal{S}$ & $\mathcal{D}$\\ \hline
 S-C & $1/L$ & $1/L$  & $1$ & $1$ & $1$ & $1$ \\ \hline
  C-C & $1/L$ & $1/L$  & $1$ & $\mathsf{n}$ & $1$ & $1$ \\ \hline
 Pareto Optimal & $1/L$ & $1/L$  & $1$ & $[1,\mathsf{n}]$ & $1$ & $1$ \\ \hline
 FDSAC & $\left(1\!-\!\kappa\right)/L$ & $\left(1\!-\!\kappa\right)/L$  & $\kappa$ & $\mathsf{n}$ & $\kappa$ & $1$\\ \hline
\end{tabular}
}
\caption{Downlink High-SNR Slope ($\mathcal{S}$) and Diversity Order ($\mathcal{D}$) with I-CSI and S-CSI.}
\vspace{-7pt}
\label{table1}
\end{table}
\vspace{-5pt} 
\begin{remark} \label{do_fdsac}
The results in Table \ref{table1} indicate that HISAC can achieve higher high-SNR slopes for both downlink SR and CR than FDSAC, providing more DoFs, which demonstrates the effective combination of S\&C in ISAC. 
\end{remark}
\vspace{-4pt}
The SR-CR region of downlink FDSAC satisfies 
{
\begin{align}\label{Rate_Regio_FDSAC}
\mathcal{C}_{\mathrm{d}}^{\mathrm{f}}=\left\{\left({\mathcal{R}}_{\mathrm{s}},{\mathcal{R}}_{\mathrm{c}}\right)\left|
\begin{aligned}
&{\mathcal{R}}_{\mathrm{s}}\in[0,{\mathcal{R}_{\mathrm{d},\mathrm{s}}^{\rm{f}}}],
{\mathcal{R}}_{\mathrm{c}}\in[0,{\mathcal{R}_{\mathrm{d},\mathrm{c}}^{\rm{f}}}],\\
&\kappa\in \left[ 0,1 \right],\iota\in \left[ 0,1 \right]
\end{aligned}
\right.\right\}.
\end{align}
}
\section{Uplink Performance Analysis} \label{uplink}
Having analyzed the downlink performance of HISAC, we now shift our fucus on the uplink performance.
\subsection{Instantaneous Channel State Information}
\subsubsection{Communications-Centric SIC}
Let us first study the sensing performance achieved by the C-C SIC, where the target response signal is firstly estimated by regarding the communication signal as interference. From a worst-case design perspective, we can treat the aggregate interference-plus-noise $\mathbf{Z}_{\mathrm{c}}=\sqrt{p_{\mathrm{c}}}\mathbf{h}_{\mathrm{c}} \mathbf{s}_{\mathrm{c}}^\mathsf{H}+\mathbf{N}_{\mathrm{u}}$ as the Gaussian noise \cite{GaussianNoise}.  Under this consideration, we conclude the following lemma.
\vspace{-5pt}
\begin{lemma}\label{up_CC_SR_lem}
The SR achieved by the C-C SIC is given by
{
\begin{align}
\overline{\mathcal{R}} _{\mathrm{u},\mathrm{s}}^{\mathrm{c}}\!=\!\frac{1}{L}\log _2\!\left[ 1\!+\!\frac{p_{\mathrm{s}}L\alpha _{\mathrm{s}}\!\left| \mathbf{h}_{\mathrm{s}}^{\mathsf{T}}\mathbf{w} \right|^2}{\sigma _{\mathrm{u}}^{2}}\!\left( \!\left\| \mathbf{h}_{\mathrm{s}} \right\| ^2\!-\!\frac{p_{\mathrm{c}}\left| \mathbf{h}_{\mathrm{s}}^{\mathsf{H}}\mathbf{h}_{\mathrm{c}} \right|^2}{p_{\mathrm{c}}\left\| \mathbf{h}_{\mathrm{c}} \right\| ^2\!+\!\sigma _{\mathrm{u}}^{2}} \right) \!\right]  .
\end{align}    
}\end{lemma}
\vspace{-5pt}
\begin{IEEEproof}
Please refer to Appendix \ref{Appendix:G} for more details.
\end{IEEEproof}
The results in \textbf{Lemma \ref{up_CC_SR_lem}} suggest that the SR is maximized when ${\mathbf{w}}=\lVert{\mathbf{h}}_{\rm{s}}\rVert^{-1}{\mathbf{h}}_{\rm{s}}^{*}$. The following theorem provide a closed-form expression for the average SR ${\mathcal{R}}_{\mathrm{u},\mathrm{s}}^{\mathrm{c}}={\mathbbmss{E}}\{\overline{\mathcal{R}}_{\mathrm{u},\mathrm{s}}^{\mathrm{c}}\}$ and its high-SNR approximation.
\vspace{-5pt}
\begin{theorem}\label{up_CC_SR_the}
The average SR achieved by the C-C SIC is 
{
\begin{align}\label{up_CC_SR}
\mathcal{R} _{\mathrm{u},\mathrm{s}}^{\mathrm{c}}=\frac{1}{L}\left[\zeta \left( \bm{\Theta} ,\frac{p_{\mathrm{c}}}{{\sigma_{\rm{u}}^2}}\varPsi^{-1} \right) -\zeta \left( \mathbf{R},\frac{p_{\rm{c}}}{\sigma_{\rm{u}}^2} \right) +\log _2\varPsi\right],    
\end{align}
}where $\zeta(\cdot,\cdot)$ is defined in \eqref{ECR_Basic_Function}, $\bm{\Theta }=\mathbf{R}+\frac{p_{\rm{s}}}{\sigma_{\rm{u}}^2}L\alpha _{\mathrm{s}}\left\| \mathbf{h}_{\mathrm{s}} \right\| ^2\mathbf{R}^{\frac{1}{2}}( \left\| \mathbf{h}_{\mathrm{s}} \right\| ^2\mathbf{I}_N-\mathbf{h}_{\mathrm{s}}\mathbf{h}_{\mathrm{s}}^{\mathsf{H}}) \mathbf{R}^{\frac{1}{2}}\succeq{\mathbf{0}}$, and $\varPsi =\frac{p_{\rm{s}}}{\sigma_{\rm{u}}^2}L\alpha _{\mathrm{s}}\lVert\mathbf{h}_{\mathrm{s}} \rVert^4+1$. When ${p}_{\rm{s}}\rightarrow \infty $, its high-SNR approximation satisfies
{
\begin{align}\label{up_CC_SR_p}
\!\!\mathcal{R} _{\mathrm{u},\mathrm{s}}^{\mathrm{c}}\!\approx\!\frac{1}{L}\!\left[\log _2p_{\rm{s}}\!+\!\zeta\! \left( \bm{\tilde{\Theta}} ,\frac{p_{\mathrm{c}}}{{\sigma_{\rm{u}}^2}\tilde{\varPsi}}\! \right)\! \!-\!\zeta\! \left( \mathbf{R},\frac{p_{\mathrm{c}}}{{\sigma_{\rm{u}}^2}} \right) \! +\!\log _2\tilde{\varPsi}\right]\!,   
\end{align}
}where $\tilde{\bm{\Theta}}=L\alpha _{\mathrm{s}}\left\| \mathbf{h}_{\mathrm{s}} \right\| ^2\mathbf{R}^{\frac{1}{2}}( \left\| \mathbf{h}_{\mathrm{s}} \right\| ^2\mathbf{I}_N-\mathbf{h}_{\mathrm{s}}\mathbf{h}_{\mathrm{s}}^{\mathsf{H}} ) \,\,\mathbf{R}^{\frac{1}{2}}\succeq{\mathbf{0}} $, and $\tilde{\varPsi}=L\alpha _{\mathrm{s}}\left\| \mathbf{h}_{\mathrm{s}} \right\| ^4/{\sigma_{\rm{u}}^2}$.
\end{theorem}
\vspace{-5pt}
\begin{IEEEproof}
Please refer to Appendix~\ref{Appendix:H} for more details.
\end{IEEEproof}
\vspace{-5pt}
\begin{remark}
\textbf{Theorem \ref{up_CC_SR_the}} suggests that the high-SNR slope of the SR achieved by the C-C SIC is given by $1/L$.
\end{remark}
\vspace{-5pt}
After estimating the target response, the echo signal $\sqrt{p_{\rm{s}}}\mathbf{G}{\mathbf{w}}{\mathbf{s}}_{\mathrm{s}}^\mathsf{H}$ can be removed from the received superposed S\&C signal. The remaining communication signal is then decoded using the optimal detection vector ${\lVert \mathbf{h}_{\mathrm{c}} \rVert}^{-1}{\mathbf{h}_{\mathrm{c}}}$. Consequently, the CR is given by $\overline{\mathcal{R}}_{\mathrm{u},\mathrm{c}}^{\mathrm{c}}=\log_2(1+\frac{p_{\rm{c}}}{\sigma_{\rm{u}}^2}\lVert{\mathbf{h}}_{\rm{c}}\rVert^2)$. In this scenario, the ECR ${\mathcal{R}}_{\mathrm{u},\mathrm{c}}^{\mathrm{c}}={\mathbbmss{E}}\{\overline{\mathcal{R}}_{\mathrm{u},\mathrm{c}}^{\mathrm{c}}\}$ and the OP $\mathcal{P} _{\mathrm{u}}^{\mathrm{c}}=\Pr(\overline{\mathcal{R}}_{\mathrm{u},\mathrm{s}}^{\mathrm{c}}<{\mathcal{R}}_0)$ can be analyzed in a similar way we analyze those achieved by the downlink C-C design with I-CSI. Further details are omitted for brevity.
\subsubsection{Sensing-Centric SIC}
In the context of S-C SIC, the BS first detects the communication signal from the CU, considering the echo signal $\sqrt{p_{\rm{s}}}\mathbf{G}{\mathbf{w}}{\mathbf{s}}_{\mathrm{s}}^\mathsf{H}$ as interference. From a worst-case design perspective, the aggregate interference-plus-noise  $\mathbf{Z}_{\mathrm{s}}=\sqrt{p_{\rm{s}}}\mathbf{G}{\mathbf{w}}{\mathbf{s}}_{\mathrm{s}}^\mathsf{H}+\mathbf{N}_{\mathrm{u}}$ is treated as the Gaussian noise \cite{GaussianNoise}. Under this circumstance, the uplink ECR and OP are given in the following theorems, respectively.
\vspace{-5pt}
\begin{theorem}\label{up_SC_CR_the}
The uplink ECR of the S-C design is given by
{
\begin{align}\label{up_SC_CR}
\mathcal{R} _{\mathrm{u},\mathrm{c}}^{\mathrm{s}}=\zeta ( \mathbf{\Phi },p_{\rm{c}}/\sigma_{\rm{u}}^2 ) ,  
\end{align}
}where $\bm{\Phi }=\mathbf{R}^{\frac{1}{2}}( p_{\rm{s}}/\sigma_{\rm{u}}^2\alpha _{\mathrm{s}}\left\| \mathbf{h}_{\mathrm{s}} \right\| ^2\mathbf{h}_{\mathrm{s}}\mathbf{h}_{\mathrm{s}}^{\mathsf{H}}+\mathbf{I}_N ) ^{-1}\mathbf{R}^{\frac{1}{2}}\succeq{\mathbf{0}}$. When ${p}_{\mathrm{c}}\rightarrow \infty $, the ECR satisfies
{
\begin{align}\label{up_SC_CR_p}
\mathcal{R} _{\mathrm{u},\mathrm{c}}^{\mathrm{s}}\approx\log _2{p_{\rm{c}}}-\log_2{\sigma_{\rm{u}}^2}+\upsilon_{\bm{\Phi }} ,
\end{align}
}where $\upsilon_{\bm{\Phi }}$ is obtained by replacing $\{\lambda_n\}_{n=1}^{\mathsf{n}}$ in $\upsilon_{\mathbf{R}}$ with the positive eigenvalues of $\bm{\Phi }$.
\end{theorem}
\vspace{-5pt}
\begin{IEEEproof}
Please refer to Appendix~\ref{Appendix:I} for more details.
\end{IEEEproof}
\vspace{-5pt}
\begin{theorem}\label{up_SC_OP_the}
The OP achieved by the S-C SIC is written as
{
\begin{align}\label{up_SC_OP}
\mathcal{P} _{\mathrm{u}}^{\mathrm{s}}=\frac{\lambda _{\Phi,\mathsf{m}}^{\mathsf{m}}}{\prod_{m=1}^{\mathsf{m}}{\lambda_{\Phi,m}}}\sum_{k=0}^{\infty}{\frac{\delta'_{k}\Upsilon \!\left( {\mathsf{m}}\!+\!k,\sigma_{\rm{u}}^2/p_{\rm{c}}\lambda_{\Phi,\mathsf{m}}^{-1}( 2^{\mathcal{R}_0}\!-\!1) \right)  }{( {\mathsf{m}}+k-1 ) !}},
\end{align}
}where $\lambda_{\Phi,1}\geq\ldots\geq\lambda_{\Phi,\mathsf{m}}$ denote the positive eigenvalues of the matrix $\bm{\Phi}$ with $\mathsf{m}=\mathrm{rank}\left( \mathbf{\Phi} \right)$, $\delta' _{0}=1$, and $\delta'_{k}$ ($k>0$) can be obtained recursively by $\delta'_{k}=\frac{1}{k}\sum_{i=1}^{k}{[\sum_{m=1}^{\mathsf{m}}{( 1-{\lambda_{\Phi,\mathsf{m}}}/{\lambda_{\Phi,m}} ) ^i}]}\delta' _{k-i}$. When $p_{\mathrm{c}}\rightarrow \infty $, the OP satisfies
{
\begin{align}\label{up_SC_OP_p}
\mathcal{P} _{\mathrm{u}}^{\mathrm{s}}
\approx\frac{(2^{\mathcal{R}_0}-1)^{\mathsf{m}}\sigma_{\rm{u}}^{2\mathsf{m}}}{p_{\rm{c}}^{\mathsf{m}}{\mathsf{m}}!\prod_{m=1}^{\mathsf{m}}{\lambda_{\Phi,m}}}.     
\end{align}
}\end{theorem}
\vspace{-3pt}
\begin{IEEEproof}
Similar to the proof of \textbf{Theorem \ref{CC_OP_theorem}}.
\end{IEEEproof}
\vspace{-5pt}
\begin{remark}
The results in \textbf{Theorem \ref{up_SC_OP_the}} suggest that the diversity order and array gain of the OP achieved by the S-C sic are given by $\mathsf{m}$ and $({\mathsf{m}}!\prod_{m=1}^{\mathsf{m}}{\lambda_{\Phi,m}})^{\frac{1}{\mathsf{m}}}{\sigma_{\rm{u}}^{-2}}(2^{\mathcal{R}_0}-1)^{-1}$, respectively.
\end{remark}
\vspace{-5pt}
After detecting the communication signal, the BS subtracts it from the received signal, utilizing the remaining part to extract information from the target response matrix $\mathbf{G}$, which yields a same SR as in \textbf{Theorem~\ref{SC_SR_theorem}}, i.e., $\mathcal{R} _{\mathrm{u},\mathrm{s}}^{\mathrm{s}}=\mathcal{R} _{\mathrm{d},\mathrm{s}}^{\mathrm{s}}$.
\vspace{-5pt}
\begin{remark} \label{up_ins_compare}
For the uplink scenario with I-CSI, the SIC order does not affect the high-SNR slopes of both the CR and SR, whereas it does affect the diversity order and high-SNR slope. Specifically, since $\mathsf{m}={\rm{rank}}(\bm\Phi)\leq\rm{rank}(\mathbf{R})=\mathsf{n}$, the diversity order achieved by the C-C SIC is no smaller than that achieved by the S-C SIC.
\end{remark}
\vspace{-5pt}
\subsection{Statistical Channel State Information}
Having investigated the S\&C performance in the case of I-CSI, we shift our focus to the case of S-CSI. In this scenario, the BS lacks complete knowledge of the communication channel $\mathbf{h}_{\rm{c}}$. Consequently, the BS can only utilize the S-CSI to design the detection vector for the communication signal, which is denoted as $\mathbf{v}$. It is assumed that the BS possesses knowledge of the effective channel $\mathbf{v}^{\mathsf{H}}\mathbf{h}_{\rm{c}}\in{\mathbbmss{C}}$, the acquisition of which incurs significantly lower signaling overhead than obtaining $\mathbf{h}_{\rm{c}}$.
\subsubsection{Communications-Centric SIC}
We commence by analyzing the SR achieved by the C-C SIC.
\vspace{-5pt}
\begin{theorem}\label{up_st_CC_SR_the}
By treating $\mathbf{Z}_{\mathrm{c}}$ as Gaussian noise, the SR is given by  
{
\begin{equation}\label{up_st_CC_SR}
\mathcal{R} _{\mathrm{u},\mathrm{s}}^{\mathrm{c}}=L^{-1}\log _2\!\left( 1+p_{\rm{s}}L\alpha _{\mathrm{s}}\lVert \mathbf{h}_{\mathrm{s}} \rVert ^2\mathbf{h}_{\mathrm{s}}^{\mathsf{H}}\left( p_{\rm{c}}\mathbf{R}+\sigma_{\rm{u}}^2\mathbf{I}_N\right) ^{\!-1}\mathbf{h}_{\mathrm{s}} \right) .    
\end{equation}
}When ${p}_{\rm{s}}\rightarrow \infty $, the SR satisfies
{
\begin{equation}\label{up_st_CC_SR_p}
\begin{split}
\mathcal{R} _{\mathrm{u},\mathrm{s}}^{\mathrm{c}}&\approx L^{-1}\log _2p_{\rm{s}}\\
&+L^{-1}\log _2( 
L\alpha _{\mathrm{s}}\lVert \mathbf{h}_{\mathrm{s}} \rVert ^2\mathbf{h}_{\mathrm{s}}^{\mathsf{H}}( p_{\rm{c}}\mathbf{R}\!+\!\sigma_{\rm{u}}^2\mathbf{I}_N) ^{-1}\mathbf{h}_{\mathrm{s}}
).
\end{split}
\end{equation}
}\end{theorem}
\vspace{-3pt}
\begin{IEEEproof}
Similar to the proof of \textbf{Theorem~\ref{up_CC_SR_the}}.
\end{IEEEproof}
\vspace{-5pt}
\begin{remark}
\textbf{Theorem \ref{up_st_CC_SR_the}} suggests that the high-SNR slope of the SR achieved by the C-C SIC is given by $L^{-1}$.
\end{remark}
\vspace{-5pt}
After estimating the target response, the echo signal can be removed from ${\mathbf{Y}}_{\rm{u}}$ with the remaining communication signal being without interference. The resulting ECR $\mathcal{R} _{\mathrm{u},\mathrm{c}}^{\mathrm{c}}$ and OP can thus be analyzed by following similar steps shown in \textbf{Theorem \ref{CC_st_CR_the}} and \textbf{\ref{CC_st_OP_the}}, respectively.

\subsubsection{Sensing-Centric SIC}
Then we move to the S-C design where the communication signal is first detected by considering the echo signal as interference. Under the S-C design, the uplink CR is written as follows:
{
\begin{align} \label{up_SC_st_CR}
\overline{\mathcal{R}}_{\mathrm{u},\mathrm{c}}^{\mathrm{s}}=\log _2\!\left( 1+\frac{{p}_{\mathrm{c}}\lvert \mathbf{v}^{\mathsf{H}}\mathbf{h}_{\mathrm{c}} \rvert^2}{p_{\rm{s}}\alpha _{\mathrm{s}}\lVert \mathbf{h}_{\mathrm{s}} \rVert ^2\lvert \mathbf{v}^{\mathsf{H}}\mathbf{h}_{\mathrm{s}} \rvert^2+\sigma_{\rm{u}}^2}\! \right) ,    
\end{align}
}where $\mathbf{v}$ denotes the communication detection vector with $\lVert \mathbf{v} \rVert ^2=1$. Recalling that the BS only has the statistical information of $\mathbf{h}_{\rm{c}}$, $\mathbf{v}$ should be designed to maximize the ECR $\mathcal{R} _{\mathrm{u},\mathrm{c}}^{\mathrm{s}}=\mathbbmss{E}\{\overline{\mathcal{R}}_{\mathrm{u},\mathrm{c}}^{\mathrm{s}}\}$. The following lemma gives the optimal detection vector that maximizes the ECR.
\vspace{-5pt}
\begin{lemma}\label{Lemma_S_CSI_Uplink}
The optimal detection vector under the S-C design with S-CSI is given by
{
\begin{align}
\mathbf{v}_{\star}=\frac{( p_{\rm{s}}/\sigma_{\rm{u}}^2\alpha _{\mathrm{s}}\left\| \mathbf{h}_{\mathrm{s}} \right\| ^2\mathbf{h}_{\mathrm{s}}\mathbf{h}_{\mathrm{s}}^{\mathsf{H}}+\mathbf{I}_N )^{-\frac{1}{2}}\mathbf{u}_{\star}}{\mathbf{u}_{\star}^{\mathsf{H}}( p_{\rm{s}}/\sigma_{\rm{u}}^2\alpha _{\mathrm{s}}\left\| \mathbf{h}_{\mathrm{s}} \right\| ^2\mathbf{h}_{\mathrm{s}}\mathbf{h}_{\mathrm{s}}^{\mathsf{H}}+\mathbf{I}_N )^{-1}\mathbf{u}_{\star}},
\end{align}
}where $\mathbf{u}_{\star}$ is the principal eigenvector of $( p_{\rm{s}}/\sigma_{\rm{u}}^2\alpha _{\mathrm{s}}\left\| \mathbf{h}_{\mathrm{s}} \right\| ^2\mathbf{h}_{\mathrm{s}}\mathbf{h}_{\mathrm{s}}^{\mathsf{H}}+\mathbf{I}_N )^{-\frac{1}{2}}{\mathbf{R}}( p_{\rm{s}}/\sigma_{\rm{u}}^2\alpha _{\mathrm{s}}\left\| \mathbf{h}_{\mathrm{s}} \right\| ^2\mathbf{h}_{\mathrm{s}}\mathbf{h}_{\mathrm{s}}^{\mathsf{H}}+\mathbf{I}_N )^{-\frac{1}{2}}$. 
\end{lemma}
\vspace{-5pt}
\begin{IEEEproof}
Please refer to Appendix \ref{Appendix:J} for more details.
\end{IEEEproof}
By setting $\mathbf{v}=\mathbf{v}_{\star}$, the communication performance is studied in the following theorems.
\vspace{-5pt}
\begin{theorem}\label{up_CC_st_CR_the}
The uplink ECR achieved by the S-C design is derived as
{
\begin{equation}
\mathcal{R} _{\mathrm{u},\mathrm{c}}^{\mathrm{s}}=-\frac{1}{\ln 2}{\rm{e}}^{\frac{\sigma_{\rm{u}}^2}{p_{\rm{c}}\varkappa}}\mathrm{Ei}( -\frac{\sigma_{\rm{u}}^2}{p_{\rm{c}}\varkappa}  ).    
\end{equation}
}When $p_{\rm{c}}\rightarrow\infty$, the ECR satisfies
{
\begin{align}
\mathcal{R} _{\mathrm{u},\mathrm{c}}^{\mathrm{s}}\approx\log _2p_{\rm{c}}-\log_2\sigma_{\rm{u}}^2+\log _2\varkappa -{\mathcal{C}}\log_2{\rm{e}}.    
\end{align}
}\end{theorem}
\vspace{-5pt}
\begin{IEEEproof}
Similar to the proof of \textbf{Theorem~\ref{SC_CR_theorem}}.
\end{IEEEproof}
\vspace{-5pt}
\begin{theorem}\label{up_CC_st_OP_the}
The OP of the S-C design is given by
{
\begin{align}
\mathcal{P} _{\mathrm{u}}^{\mathrm{s}}=1-{\rm{e}}^{-\frac{\sigma_{\rm{u}}^2(2^{\mathcal{R}_0}-1)}{p_{\rm{c}}\varkappa}} .    
\end{align}
}When $p_{\rm{c}}\rightarrow\infty$, the OP satisfies
{
\begin{align}
\tilde{\mathcal{P}}_{\mathrm{u}}^{\mathrm{s}}\approx\frac{\sigma_{\rm{u}}^2(2^{\mathcal{R}_0}-1)}{p_{\rm{c}}\varkappa}.
\end{align}
}\end{theorem}
\vspace{-5pt}
\begin{IEEEproof}
Similar to the proof of \textbf{Theorem~\ref{SC_OP_theorem}}.
\end{IEEEproof}
\vspace{-5pt}
\begin{remark}
The results in \textbf{Theorem \ref{up_CC_st_CR_the}} and \textbf{\ref{up_CC_st_OP_the}} suggest that the high-SNR slope and diversity order of the CR achieved by the S-C SIC with S-CSI are both equal to $1$.
\end{remark}
\vspace{-5pt}
After decoding the communication signal, the sensing information can be estimated without interference, which yields a same average SR as in \textbf{Theorem~\ref{SC_SR_theorem}}, i.e., $\mathcal{R} _{\mathrm{u},\mathrm{s}}^{\mathrm{s}}=\mathcal{R} _{\mathrm{d},\mathrm{s}}^{\mathrm{s}}$.

Based on the analysis above for the uplink HISAC, we can make the following conclusion.
\vspace{-5pt}
\begin{remark} \label{up_st_compare}
For the uplink scenario with S-CSI, the SIC order does not affect either the high-SNR slopes or diversity orders. Besides, I-CSI yields higher diversity orders than S-CSI.
\end{remark}

\subsection{Rate Region Characterization}
To characterize the uplink SR-CR region, we utilize the time-sharing strategy \cite{mimo}, where the S-C SIC is applied with probability $\epsilon  $ and the C-C SIC is applied with probability $1-\epsilon  $. For a given $\epsilon$, letting $( \mathcal{R} _{\mathrm{u},\mathrm{s}}^{\epsilon},{\mathcal{R}} _{\mathrm{u},\mathrm{c}}^{\epsilon } ) $ represent the achievable SR-CR pair, we have $\mathcal{R} _{\mathrm{u},\mathrm{s}}^{\epsilon }=\epsilon  {\mathcal{R}} _{\mathrm{u},\mathrm{s}}^{\mathrm{s}}+( 1-\epsilon  ) {\mathcal{R}} _{\mathrm{u},\mathrm{s}}^{\mathrm{c}}$ and $\mathcal{R} _{\mathrm{u},\mathrm{c}}^{\epsilon }=\epsilon  {\mathcal{R}} _{\mathrm{u},\mathrm{c}}^{\mathrm{s}}+( 1-\epsilon  ) {\mathcal{R}} _{\mathrm{u},\mathrm{c}}^{\mathrm{c}}$. By exploiting the sandwich theorem, we can obtain the high-SNR slopes and diversity orders of any uplink SR-CR pair achieved through the time-sharing strategy, which are summarized in Table \ref{table2}.

\begin{table}[!t]
\center
\scalebox{0.8}{
\begin{tabular}{|c|c|c|c|c|c|c|}\hline
\multicolumn{1}{|c|}{\multirow{3}{*}{Design}} & \multicolumn{2}{c|}{Sensing} & \multicolumn{4}{c|}{Communications} \\ \cline{2-7}
&I-CSI & S-CSI & \multicolumn{2}{c|}{I-CSI} & \multicolumn{2}{c|}{S-CSI}\\ \cline{2-7}
 & $\mathcal{S}$ & $\mathcal{S}$  & $\mathcal{S}$ & $\mathcal{D}$   & $\mathcal{S}$ & $\mathcal{D}$\\ \hline
 S-C & $1/L$ & $1/L$ & $1$ & $\mathsf{m}$ & $1$ & $1$ \\ \hline
 C-C & $1/L$ & $1/L$ & $1$ & $\mathsf{n}$ & $1$ & $1$ \\ \hline
 Time-Sharing & $1/L$ & $1/L$  & $1$ & $[\mathsf{m},\mathsf{n}]$ & $1$ & $1$ \\ \hline
 FDSAC & $\left(1\!-\!\kappa\right)/L$ & $\left(1\!-\!\kappa\right)/L$  & $\kappa$ & $\mathsf{n}$ & $\kappa$ & $1$\\ \hline
\end{tabular}
}
\caption{Uplink High-SNR Slope ($\mathcal{S}$) and Diversity Order ($\mathcal{D}$) with I-CSI and S-CSI.}
\vspace{-8pt}
\label{table2}
\end{table}

Denoting the achievable SR-CR pair as $(\mathcal{R} _{\mathrm{s}},\mathcal{R} _{\mathrm{c}})$, the uplink rate regions achieved by HISAC reads
{
\begin{align}
\!\mathcal{C} _{\mathrm{u}}\!=\!\left\{ \!\left( \mathcal{R} _{\mathrm{s}},\mathcal{R} _{\mathrm{c}} \right) \left| \!\!\begin{array}{c}
	\mathcal{R} _{\mathrm{u},\mathrm{s}}\!\in\! \left[ 0,{\mathcal{R}} _{\mathrm{u},\mathrm{s}}^{\epsilon} \right] \!,\mathcal{R} _{\mathrm{u},\mathrm{c}}\!\in \!\left[ 0,{\mathcal{R}} _{\mathrm{u},\mathrm{c}}^{\epsilon} \right] \!,\epsilon \in\! \left[ 0,1 \right]\\
\end{array} \right.\!\!\!\! \right\} .
\end{align}
}

\begin{figure*}[!t]
	\centering
	\subfigbottomskip=0pt
	\subfigcapskip=0pt
	\setlength{\abovecaptionskip}{3pt}
	\subfigure[CR.]
	{
		\includegraphics[height=2.1in]{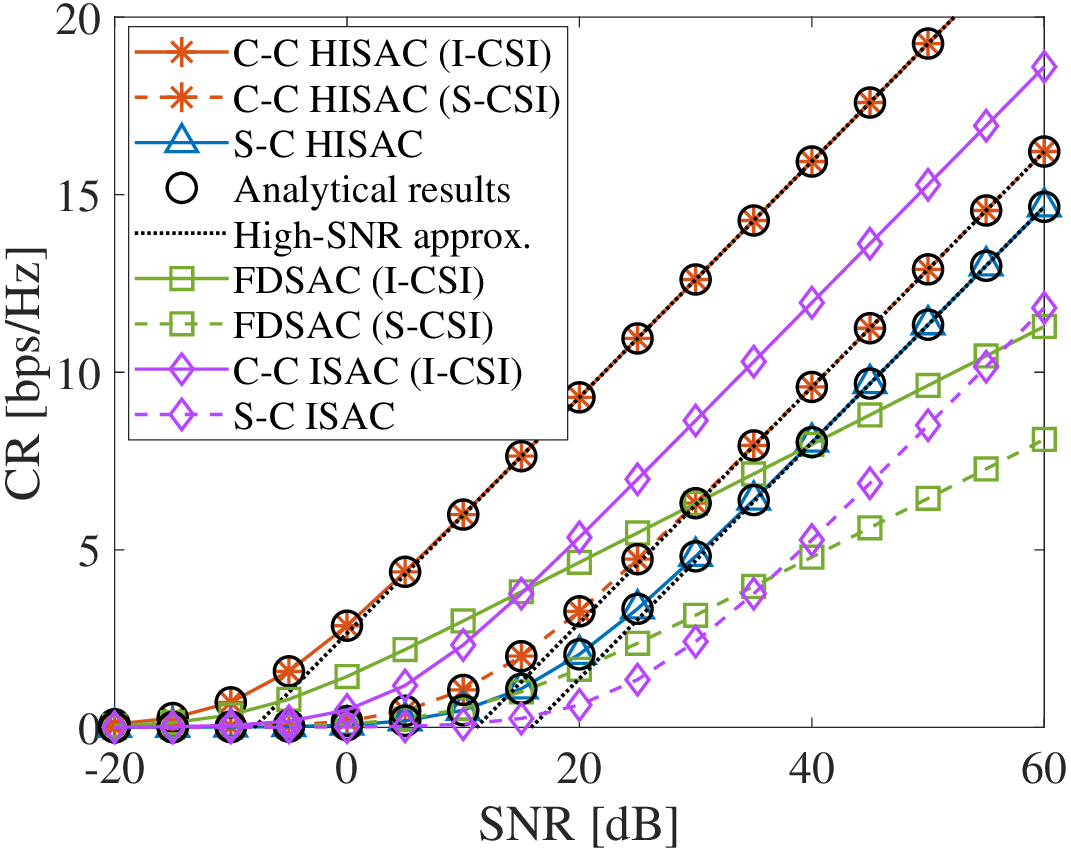}
		\label{do_CR}	
	}
	\subfigure[OP.]
	{
		\includegraphics[height=2.1in]{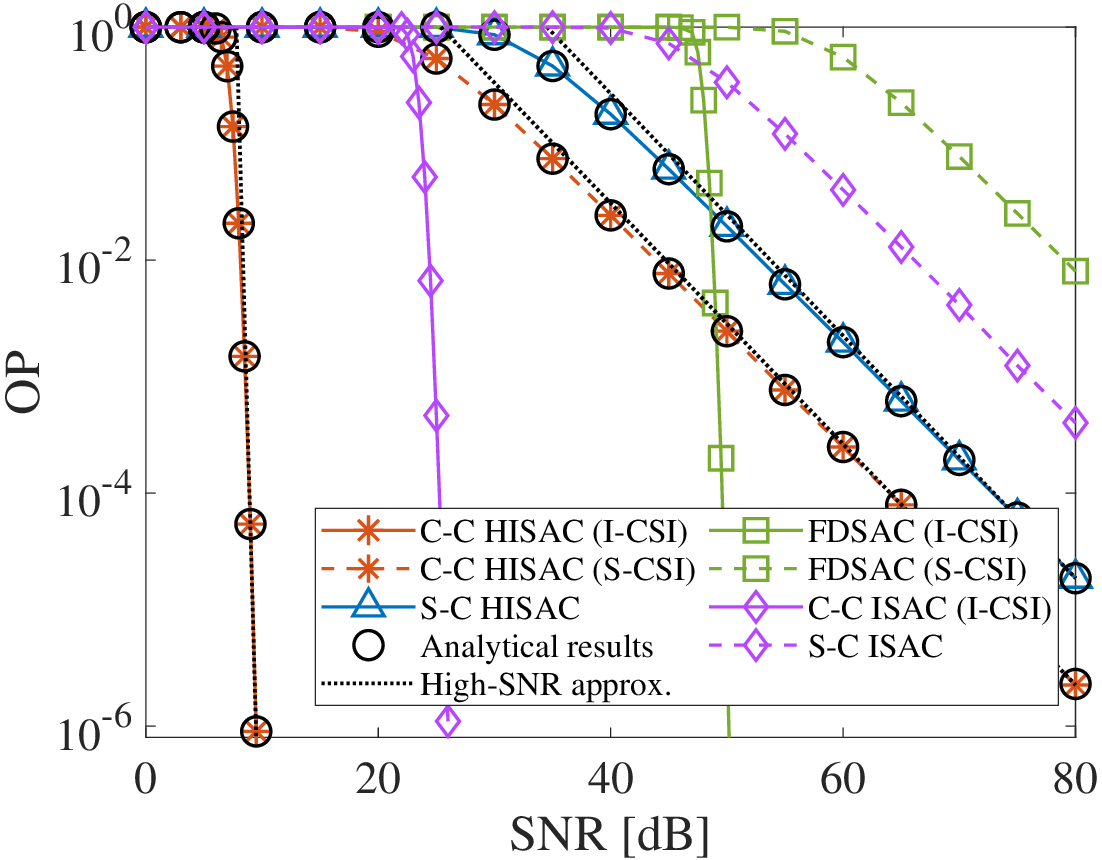}
		\label{do_OP}	
	}
	\caption{Downlink communication performance.}
	\vspace{-10pt}
	\label{do_Commun}
\end{figure*}

\subsection{HMIMO based FDSAC}
In uplink scenario, we also consider FDSAC as the baseline scheme, wherein $\kappa\in \left[ 0,1 \right]$ fraction of the total bandwidth is allocated to communications, and the other is used for sensing. Consequently, the SR achieved by FDSAC is given by $\mathcal{R} _{\mathrm{u},\mathrm{s}}^{\mathrm{f}}=\frac{1-\kappa}{L}\log _2( 1+\frac{1}{1-\kappa}p_{\rm{s}}/\sigma_{\rm{u}}^2L\alpha _{\mathrm{s}}\lVert \mathbf{h}_{\mathrm{s}} \rVert ^4 )$. As for communications, the CRs in the cases of I-CSI and S-CSI can be written as $\mathcal{R} _{\mathrm{u},\mathrm{c}}^{\mathrm{f}}=\kappa\log_2( 1+\frac{1}{\kappa}p_{\rm{c}}/{\sigma_{\rm{u}}^2}\lVert \mathbf{h}_{\mathrm{c}} \rVert ^2 )$ and $\mathcal{R} _{\mathrm{u},\mathrm{c}}^{\mathrm{f}}=\kappa\log _2( 1+\frac{1}{\kappa}p_{\rm{c}}/\sigma_{\rm{u}}^2\lvert\mathbf{v}_{\star}^{\mathsf{H}}\mathbf{h}_{\mathrm{c}}\rvert^2)$, respectively. Note that the CR and SR achieved by FDSAC can be analyzed in a similar way we analyze those achieved by HISAC. For the sake of reference, we summarize the results related to diversity order and high-SNR slope in Table \ref{table2}.
\vspace{-5pt}
\begin{remark} \label{up_fdsac}
The results in Table \ref{table2} indicate that HISAC can achieve higher high-SNR slopes for both uplink SR and CR than FDSAC, providing more DoFs in terms of both S\&C.    
\end{remark}
Finally, the SR-CR region of uplink FDSAC satisfies  
{
\begin{align}\label{Rate_Regio_FDSAC2}
\mathcal{C}_{\mathrm{u}}^{\mathrm{f}}=\left\{\left({\mathcal{R}}_{\mathrm{s}},{\mathcal{R}}_{\mathrm{c}}\right)\left|
\begin{aligned}
{\mathcal{R}}_{\mathrm{s}}\in[0,{\mathcal{R}_{\mathrm{u},\mathrm{s}}^{\rm{f}}}],
{\mathcal{R}}_{\mathrm{c}}\in[0,{\mathcal{R}_{\mathrm{u},\mathrm{c}}^{\rm{f}}}],\kappa\in \left[ 0,1 \right]
\end{aligned}
\right.\right\}.
\end{align}
}

\section{Numerical Results} \label{numerical}
In this section, numerical results are provided to evaluate the S\&C performance of the proposed systems and verify the derived analytical results. Without otherwise specification, the simulation parameter settings are defined as follows: $p/\sigma_{\rm{s}}^2=p/\sigma_{\rm{c}}^2=30$ dB, $p_{\rm{c}}/\sigma_{\rm{u}}^2=p_{\rm{s}}/\sigma_{\rm{u}}^2=30$ dB,  $\lambda =0.125$ m, $L_x=L_y=5\lambda$, $d={\lambda}/{4}$, $A={\lambda^2}/{64}$, $\alpha_0=A\mu_{{\mathsf{a}}}=A\mu_{\mathsf{i}}=0$ dB, $\mathcal{R}_0=12$ bps/Hz, $L=4$, and $\kappa=\iota=0.5$. The CU and the target are located at $\mathbf{r}_{\mathrm{c}}=\left[ 1\ \mathrm{m},1\ \mathrm{m},5\ \mathrm{m} \right] $ and $\mathbf{r}_{\mathrm{s}}=\left[ 0\ \mathrm{m},0\ \mathrm{m},3\ \mathrm{m} \right] $, respectively. For comparison, we also presents the simulation results for conventional MIMO based ISAC, referred to as ``ISAC'' in the plots, where the antenna spacing is set as $d=\frac{5}{8}\lambda$ and the communication channels for each antenna are assumed as i.i.d. Rayleigh channels, while the other simulation settings are same with those of the HISAC given above.

\subsection{Downlink Performance}
{\figurename} {\ref{do_CR}} and {\figurename} {\ref{do_OP}} plot the downlink CR and OP versus the communication SNR $p/\sigma_{\rm{c}}^2$, respectively. It can be seen that the C-C HISAC with I-CSI attains the best CR and OP, whereas FDSAC achieves the poorest communication performance. The derived analytical results match the simulation results well, and the high-SNR approximations precisely follow the simulation results in the high-SNR regime. Moreover, it is worth noting that regardless of the CSI assumption, the CRs of both C-C HISAC and S-C HISAC exhibit the same high-SNR slope, which is higher than that of FDSAC. However, under the I-CSI case, the downlink diversity order achieved by the C-C HISAC is significantly higher than the S-C HISAC, while the diversity orders achieved by different designs with S-CSI are identical. These observations validate the conclusions drawn in \textbf{Remark \ref{do_ins_compare}} and \textbf{\ref{do_st_compare}}.  

\begin{figure} [!t]
	\centering
	\includegraphics[height=2.1in]{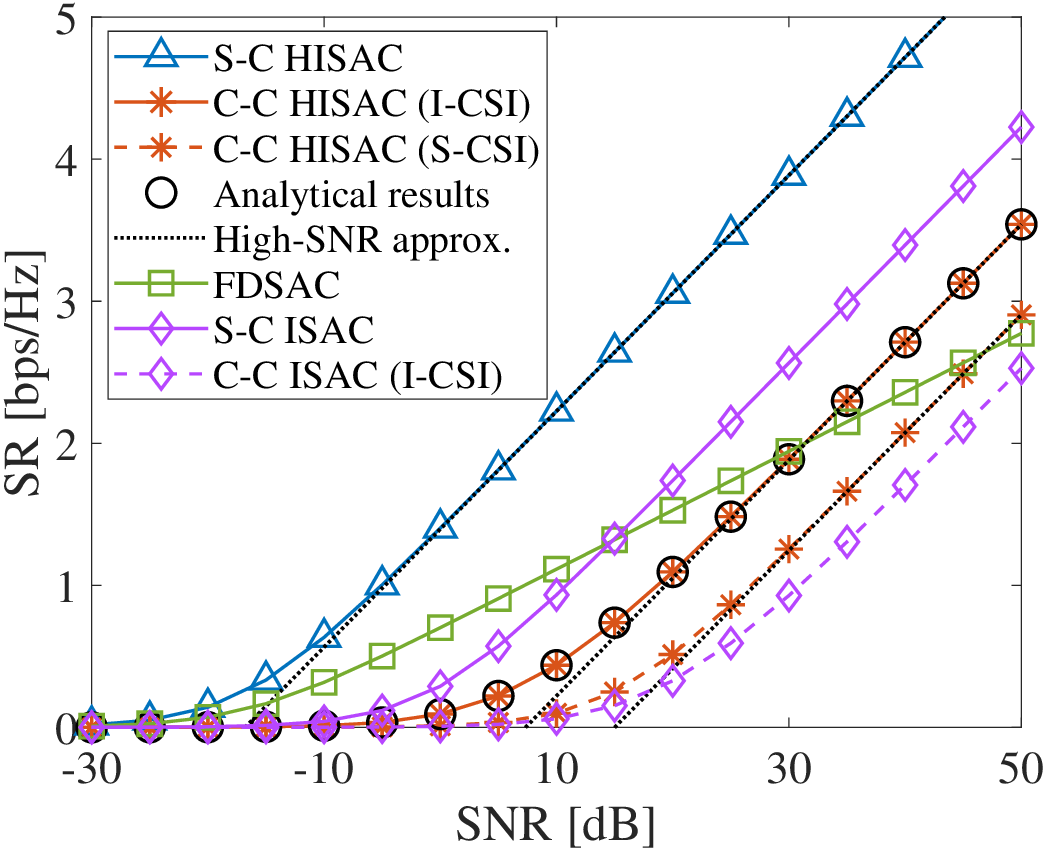}
	\caption{Downlink SR.}
	\vspace{-7pt}
	\label{do_SR}
\end{figure}

Fig.~\ref{do_SR} shows the downlink SR versus the sensing SNR $p/\sigma_{\rm{s}}^2$, validating the accuracy of the analytical and approximated results. It is evident that S-C HISAC exhibits the most superior SR performance, while the SRs achieved by the S-C HISAC and the C-C HISAC yield the same high-SNR slope in both CSI cases. Through a joint examination of Fig.~\ref{do_CR} and Fig.~\ref{do_SR}, we observe that both the CRs and SRs of HISAC demonstrate higher high-SNR slopes than FDSAC, aligning with the statement in \textbf{Remark \ref{do_fdsac}}. Additionally, while the CSI has no impact on the HISAC performance under the S-C design, it does influence the HISAC performance under the C-C design. Specifically, both CR and SR of the C-C design in the I-CSI case are larger than those in the S-CSI case, with the compensation of higher signaling overhead. It is also noticeable that the performance of downlink HISAC surpass conventional ISAC in terms of both S\&C capabilities. In particular, with I-CSI, HISAC achieves higher CR and SR than ISAC, while both systems exhibit identical high-SNR slopes. On the other hand, for outage performance, due to the larger antenna spacing within the conventional MIMO resulting in fewer antennas ($M=L_x/d\times L_y/d=64$), the downlink diversity order of ISAC, which equals to the number of antennas \cite{con_MIMO}, is lower than that of HISAC equivalent to the rank of $\mathbf{R}$ ($\mathsf{n}=88$ under this setting). 

\begin{figure} [!t]
	\centering
	\includegraphics[height=2.05in]{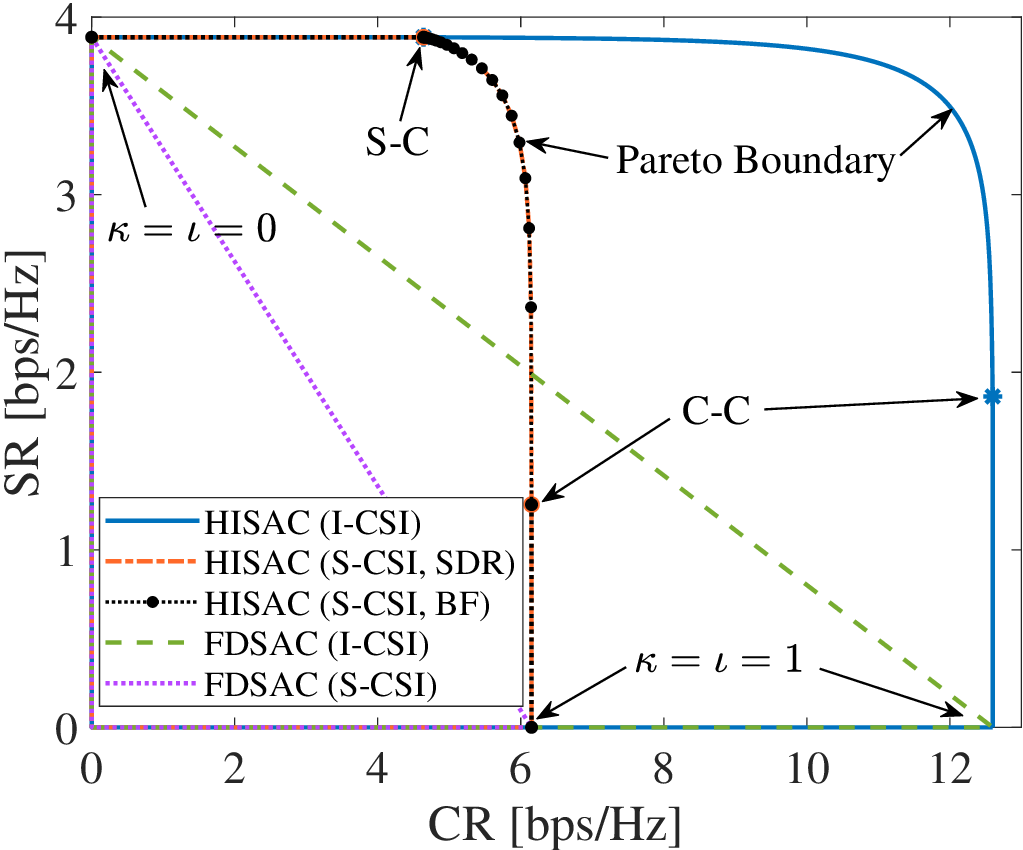}
	\caption{Downlink rate regions.}
	\vspace{-7pt}
	\label{do_region}
\end{figure}

\begin{figure} [!t]
	\centering
	\includegraphics[height=2.05in]{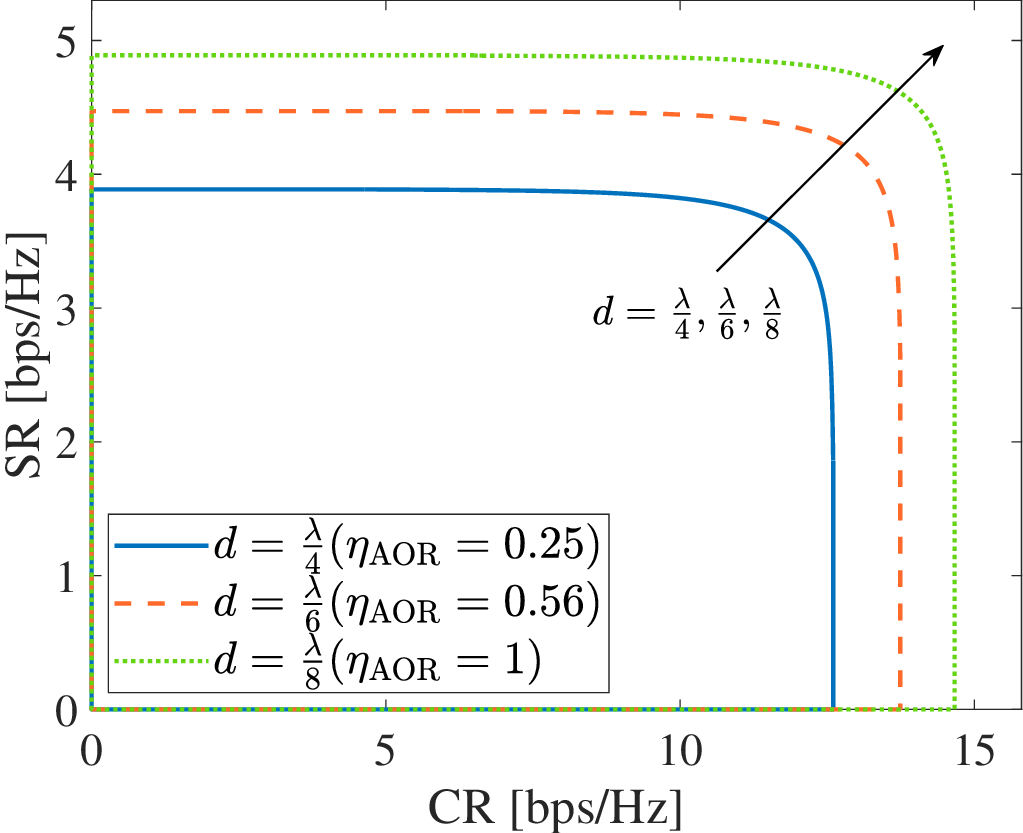}
	\caption{Downlink rate regions with I-CSI for different antenna spacing.}
	\vspace{-10pt}
	\label{region_d}
\end{figure}

In Fig.~\ref{do_region}, the downlink SR-CR regions achieved by the HISAC system and the baseline FDSAC system with different CSI are presented. For HISAC with either I-CSI or S-CSI, the two marked points on the graph represent the S-C and C-C designs, respectively. The curve section connecting the two points illustrates the Pareto boundary of the downlink HISAC's rate region, determined by solving either the problem \eqref{Problem_CR_SR_Tradeoff} or \eqref{problem_sta} for values of $\tau$ ranging from $0$ to $1$. As anticipated, it is evident that with I-CSI, HISAC can achieve a more extensive rate region than in the S-CSI case. Additionally, it is crucial to emphasize that in both CSI scenarios, the donwlink SR-CR region achieved by FDSAC is entirely encompassed by the rate region of HISAC. This observation highlights the superior S\&C performance of HISAC compared to FDSAC. Last but not least, it can be seen from the above graph that for the S-CSI case, the rate region achieved by the SDR-based method is nearly coincident with that achieved by the brute-force (BF) search-based Pareto optimal design. This means that the rank-one relaxation in problem \eqref{sdr} is tight in our considered system.

Moreover, Fig.~\ref{region_d} shows the rate regions of HISAC using I-CSI with different values of $d$. We note that reducing the antenna spacing, which increases the AOR and allows for more antennas to be deployed, expands the attainable SR-CR region. In particular, when $d=\sqrt{A}=\lambda/8$ ($\eta_{\mathsf{aor}}=1$), meaning the antennas are placed edge-to-edge, the region reaches its maximum boundary.

\begin{figure}[!t]
	\centering
	\subfigbottomskip=0pt
	\subfigcapskip=-2pt
	\setlength{\abovecaptionskip}{2pt}
	\subfigure[CR.]
	{
		\includegraphics[height=2.1in]{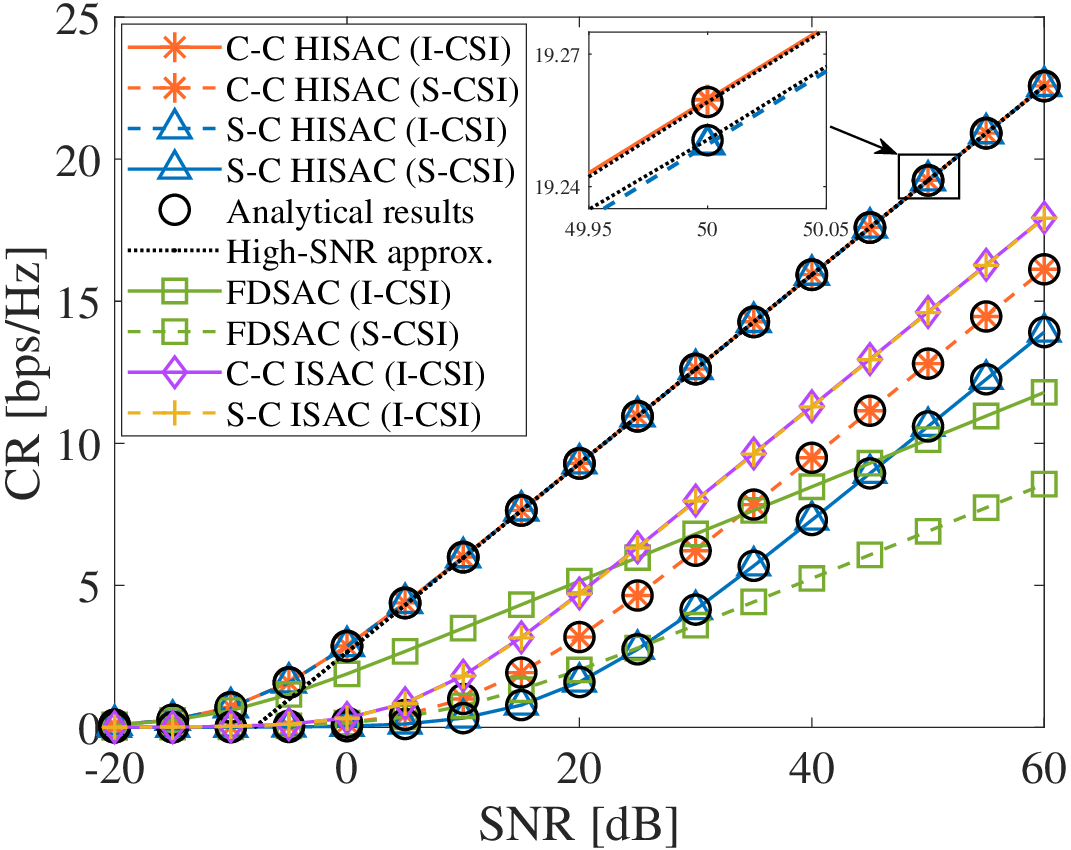}
		\label{up_CR}	
	}
	\subfigure[OP.]
	{
		\includegraphics[height=2.1in]{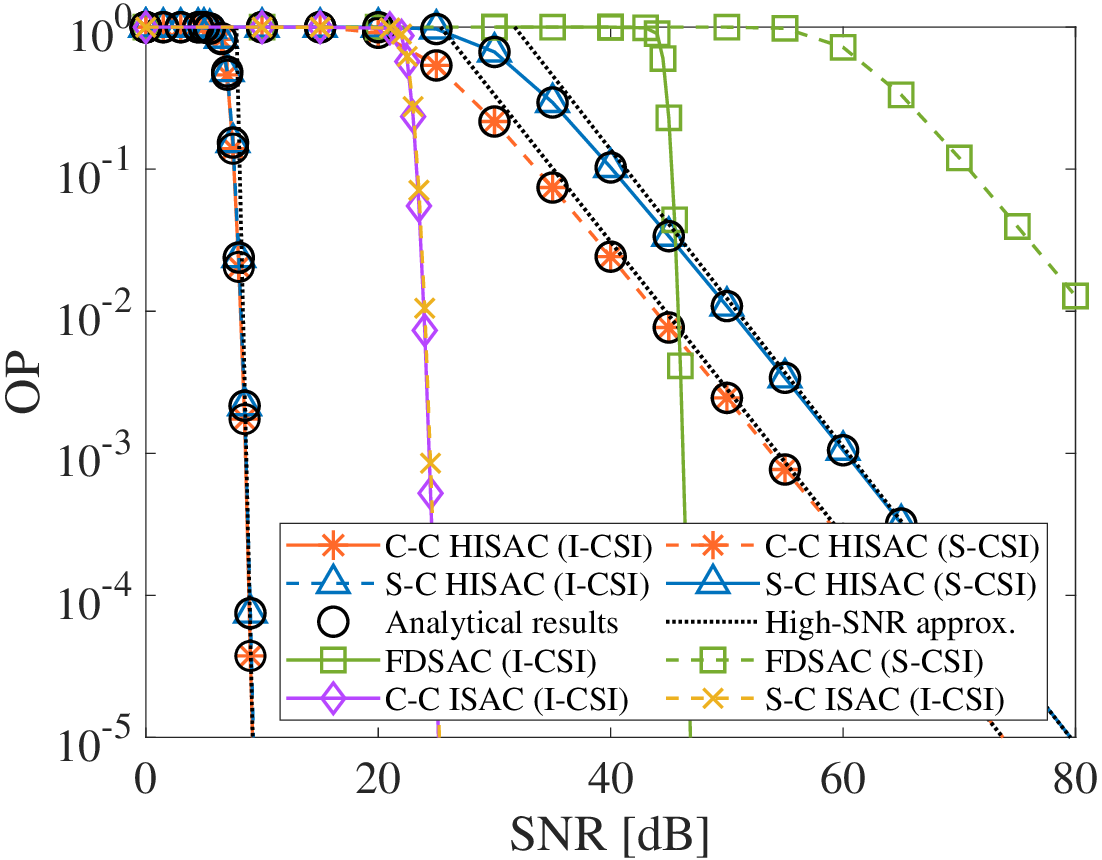}
		\label{up_OP}	
	}
	\caption{Uplink communication performance.}
	\label{up_commun}
	\vspace{-8pt}
\end{figure}

\subsection{Uplink Performance}
\begin{figure} [!t]
	\centering
	\includegraphics[height=2.1in]{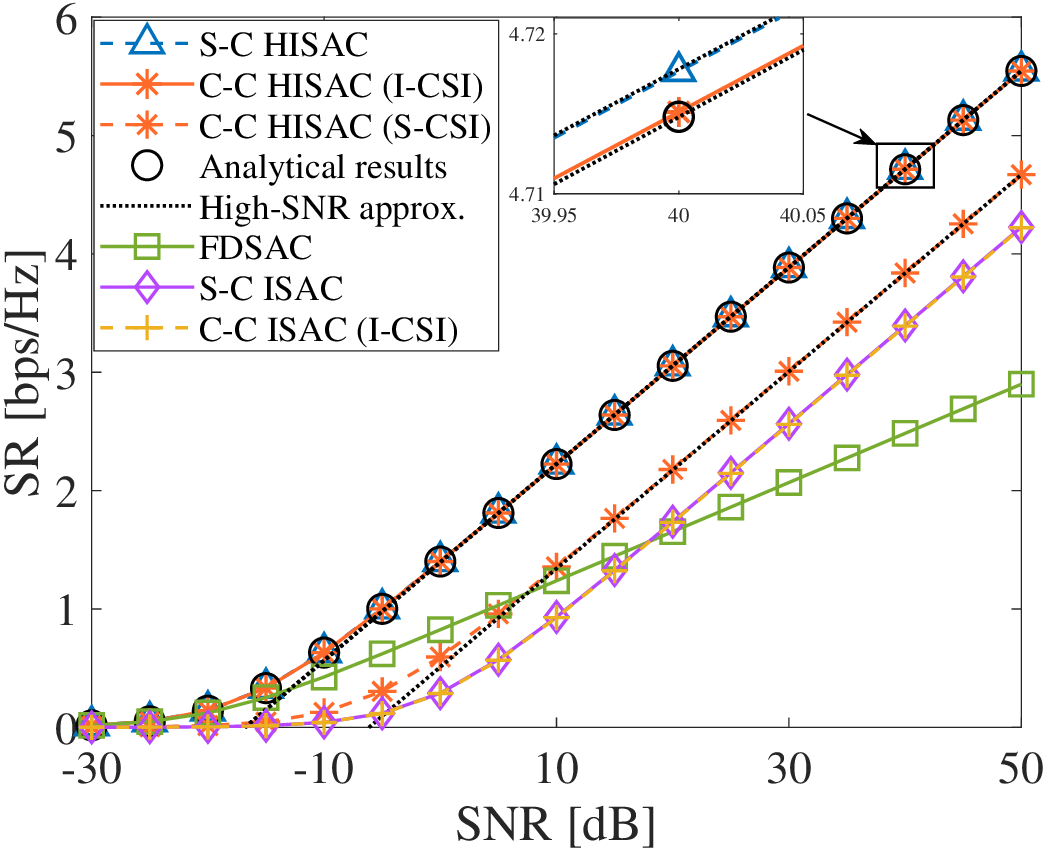}
	\caption{Uplink SR.}
	\vspace{-7pt}
	\label{up_SR}
\end{figure}

Turning our attention to the uplink outcomes, we examine Fig~\ref{up_CR} and Fig~\ref{up_OP}, which depict the uplink CR and OP with respect to the SNR $p_{\rm{c}}/\sigma_{\rm{u}}^2$, respectively. The analytical results closely match the simulation results, and the high-SNR approximations accurately capture the behavior in the high-SNR region. Notably, despite C-C HISAC with I-CSI exhibits the highest CR, the performance gap between uplink C-C HISAC and S-C HISAC under the I-CSI case is remarkably small. This occurs because the correlation between the sensing channel and communication channel is minimal, i.e., the S\&C channels are approximately orthogonal, which significantly reduces the IFI. Further, we observe that the CR of HISAC yields an identical high-SNR slope with both I-CSI and S-CSI, which surpasses that achieved by FDSAC. On the other hand, the diversity orders achieved by HISAC with I-CSI are higher then those achieved under the S-CSI case, validating the discussions presented in \textbf{Remark~\ref{up_ins_compare}} and \textbf{\ref{up_st_compare}}. Moreover, the diversity orders achieved by both HISAC and FDSAC are identical within the same CSI case, but the OPs of FDSAC are significantly lower than those of HISAC. In Fig.~\ref{up_SR}, we present the uplink SR as function of $p_{\rm{s}}/\sigma_{\rm{u}}^2$, which verifies the analytical and approximated results. It is worth noting that regardless of the SIC order, HISAC is capable of achieving higher high-SNR slopes than FDSAC, which is consistent with the statements in \textbf{Remark~\ref{up_fdsac}}. Importantly, compare to the conventional MIMO based ISAC, HISAC exhibits enhanced performance in S\&C for uplink scenario. This underscores the advantages that HMIMO brings to ISAC systems.

\begin{figure} [!t]
\centering
\setlength{\abovecaptionskip}{3pt}
\includegraphics[height=2.1in]{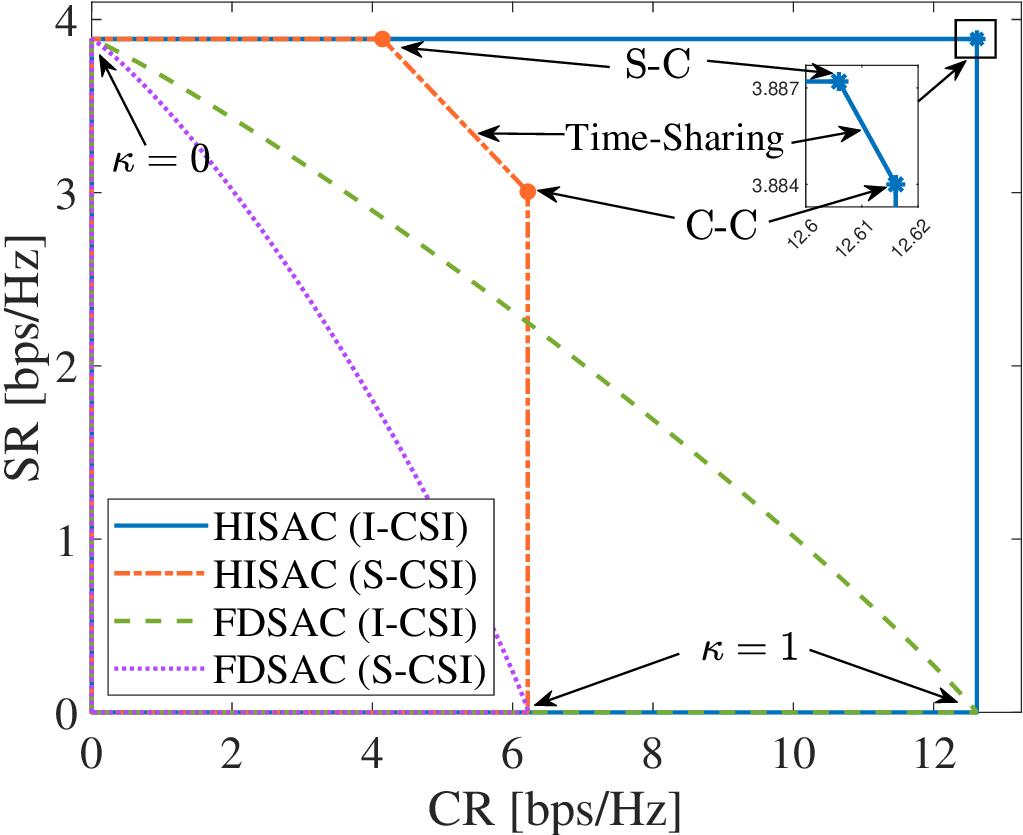}
 \caption{Uplink rate regions.}
 \vspace{-7pt}
 \label{up_region}
\end{figure}

\begin{figure} [!t]
	\centering
\setlength{\abovecaptionskip}{3pt}
	\includegraphics[height=2.05in]{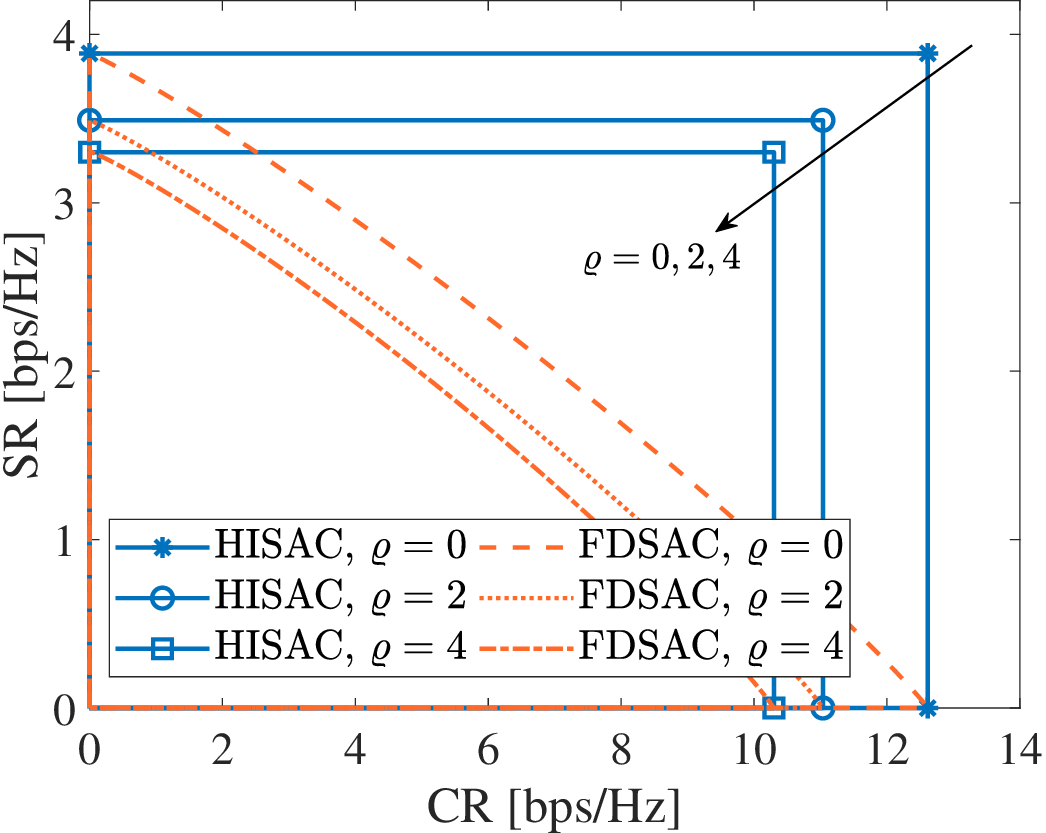}
	\caption{Uplink rate regions with I-CSI for different power levels of interference.}
	\vspace{-7pt}
	\label{interference}
\end{figure}

Moving to Fig~\ref{up_region}, we present the SR-CR regions attained by the uplink HISAC and FDSAC systems with both I-CSI and S-CSI. For the region of HISAC under either I-CSI or S-CSI case, the two points on the plot correspond to the rate pairs achieved by the S-C and C-C designs, respectively, while the line segment connecting these points represents the rates attainable through the time-sharing strategy between the two designs. A pivotal observation from the plot is that the rate region of uplink FDSAC is entirely encompassed within the rate region of uplink HISAC. This unequivocally demonstrates the superior performance of HISAC over FDSAC in the uplink scenario. Additionally, as mentioned before, in the I-CSI case where the IFI is negligible, the S\&C performance achieved by the S-C and C-C designs are nearly identical. This results in the corner points of the rate region being very close, giving the region an approximately rectangular shape.

Furthermore, to explore the impact of interference from other cells, which is very likely to occur in real-world scenario, we present the uplink attainable rate regions under different power levels of interference. Specifically, assuming that the interference follows $\mathcal{C} \mathcal{N} \left( \mathbf{0},\varrho\sigma_{\mathrm{u}}^2\mathbf{I}_N \right)$, Fig.~\ref{interference} plots the rate regions for both HISAC and FDSAC with different values of $\varrho$. It is clear that increased interference power results in reduced rate regions. Moreover, we note that even in the presence of interference, HISAC may still achieve a wide range of SR-CR pairs that surpass those of FDSAC without interference, underscoring the effectiveness of HISAC.

\section{Conclusion}\label{conclusion}
This paper has provided a comprehensive performance analysis of a HISAC system for both downlink and uplink scenarios. To account for the closed antenna spacing of the HMIMO, we considered the communication channels in a scattering environment as spatially correlated Rayleigh fading. Leveraging both I-CSI and S-CSI, we derived closed-form expressions and high-SNR approximations for the SRs, CRs, and OPs under the S-C and C-C designs. Furthermore, we identified the Pareto boundary of the downlink SR-CR region and characterized the uplink rate region using the time-sharing strategy. The high-SNR slopes and diversity orders of the proposed HISAC system and the baseline FDSAC system were obtained, providing further insights into system performance. The results have demonstrated that HISAC outperforms the conventional MIMO based ISAC and is able to achieve more DoFs and broader rate regions than the FDSAC in both downlink and uplink cases, underscoring the significant performance advantages offered by the HISAC system.

\begin{appendix}
\renewcommand{\theequation}{A.\arabic{equation}}
\setcounter{equation}{0}
\subsection{The Relationship Between Sensing Mutual Information and Estimation Error}\label{Appendix:0}
Given ${\mathbf{w}}$ and $\mathbf{s}$, the echo signal matrix is shown in \eqref{echo}, and its vectorization is expressed as follows:
\setlength\abovedisplayskip{4pt}
\setlength\belowdisplayskip{4pt}
\begin{align}\label{Original Signal}
\mathsf{vec}( \mathbf{Y}_{\mathrm{s}} ) =\sqrt{p}\mathbf{h}_{\mathrm{s}}^{\mathsf{T}}\mathbf{w}\mathsf{vec}( \mathbf{h}_{\mathrm{s}}\mathbf{s}^{\mathsf{H}} ) \beta +\mathsf{vec}( \mathbf{N}_{\mathrm{s}} )\triangleq {\mathbf{y}}_{{\mathrm{s}},{\mathsf{v}}}.    
\end{align} 
The mean-square error for an estimate, $f_{\mathsf{est}}({\mathbf{y}}_{{\mathrm{s}},{\mathsf{v}}})$, of the reflection coefficient $\beta$, based on the observation ${\mathbf{y}}_{{\mathrm{s}},{\mathsf{v}}}$, can be written as follows:
\begin{align}
{\mathsf{MSE}}={\mathbbmss{E}}\{\lvert f_{\mathsf{est}}({\mathbf{y}}_{{\mathrm{s}},{\mathsf{v}}})-\beta\rvert^2\}.\label{MSE_Def}
\end{align}
It is well known that the minimum value of \eqref{MSE_Def} is achieved by the following conditional mean estimator:
\begin{align}
f_{\mathsf{est}}({\mathbf{y}}_{{\mathrm{s}},{\mathsf{v}}})={\mathbbmss{E}}\{\beta|{\mathbf{y}}_{{\mathrm{s}},{\mathsf{v}}}\}.
\end{align}
To compute this conditional mean, we need to know the probability density function (PDF) of $\beta$ conditioned on ${\mathbf{y}}_{{\mathrm{s}},{\mathsf{v}}}$, which is denoted as $f_{\beta|{\mathbf{y}}_{{\mathrm{s}},{\mathsf{v}}}}(x|{\mathbf{y}})$. According to Bayes' theorem, then we have \cite{sengijpta1995fundamentals}
\begin{align}\label{PDF_MSE_Bayes}
f_{\beta|{\mathbf{y}}_{{\mathrm{s}},{\mathsf{v}}}}(x|{\mathbf{y}})=
\frac{f_{\beta}(x)f_{{\mathbf{y}}_{{\mathrm{s}},{\mathsf{v}}}|\beta}({\mathbf{y}}|x)}{f_{{\mathbf{y}}_{{\mathrm{s}},{\mathsf{v}}}}({\mathbf{y}})},
\end{align}
where $f_{\beta}(\cdot)$, $f_{{\mathbf{y}}_{{\mathrm{s}},{\mathsf{v}}}}(\cdot)$, and $f_{{\mathbf{y}}_{{\mathrm{s}},{\mathsf{v}}}|\beta}(\cdot|\cdot)$ represent the PDFs of $\beta$, ${\mathbf{y}}_{{\mathrm{s}},{\mathsf{v}}}$, and ${\mathbf{y}}_{{\mathrm{s}},{\mathsf{v}}}$ conditioned on $\beta$, respectively. Given that $\beta\sim{\mathcal{CN}}(0,\alpha_{\rm{s}})$ and $\mathsf{vec}( \mathbf{N}_{\mathrm{s}} )\sim{\mathcal{CN}}({\mathbf{0}},\sigma_{\rm{s}}^2{\mathbf{I}_{NL}})$, we have
\begin{subequations}\label{PDF_Used_MSE}
\begin{align}
f_{\beta}(x)&=\frac{1}{\pi \alpha_{\rm{s}}}{\rm{e}}^{-\frac{|x|^2}{\alpha_{\rm{s}}}},\\
f_{{\mathbf{y}}_{{\mathrm{s}},{\mathsf{v}}}|\beta}({\mathbf{y}}|x)&=\frac{1}{(\pi \sigma_{\rm{s}}^2)^{NL}}
{\rm{e}}^{-\frac{1}{\sigma_{\rm{s}}^2}({\mathbf{y}}-{\mathbf{h}}_{\mathsf{v}}x)^{\mathsf{H}}({\mathbf{y}}-{\mathbf{h}}_{\mathsf{v}}x)},\\
f_{{\mathbf{y}}_{{\mathrm{s}},{\mathsf{v}}}}({\mathbf{y}})&=
\frac{{\rm{e}}^{-{\mathbf{y}}^{\mathsf{H}}(\sigma_{\rm{s}}^2{\mathbf{I}_{NL}}
+\alpha_{\rm{s}}{\mathbf{h}}_{\mathsf{v}}{\mathbf{h}}_{\mathsf{v}}^{\mathsf{H}})^{-1}{\mathbf{y}}}}{\pi^{NL}\det(\sigma_{\rm{s}}^2{\mathbf{I}_{NL}}
+\alpha_{\rm{s}}{\mathbf{h}}_{\mathsf{v}}{\mathbf{h}}_{\mathsf{v}}^{\mathsf{H}})},
\end{align}
\end{subequations}
where ${\mathbf{h}}_{\mathsf{v}}\triangleq\sqrt{p}\mathbf{h}_{\mathrm{s}}^{\mathsf{T}}\mathbf{w}\mathsf{vec}( \mathbf{h}_{\mathrm{s}}\mathbf{s}^{\mathsf{H}} )\in{\mathbbmss{C}}^{NL\times1}$. Substituting \eqref{PDF_Used_MSE} into \eqref{PDF_MSE_Bayes} and using the facts that $(\sigma_{\rm{s}}^2{\mathbf{I}_{NL}}
+\alpha_{\rm{s}}{\mathbf{h}}_{\mathsf{v}}{\mathbf{h}}_{\mathsf{v}}^{\mathsf{H}})^{-1}=\frac{1}{\sigma_{\rm{s}}^2}({\mathbf{I}_{NL}}-\frac{\alpha_{\rm{s}}}{\sigma_{\rm{s}}^2+\alpha_{\rm{s}}\lVert{\mathbf{h}}_{\mathsf{v}}\rVert^2}
{\mathbf{h}}_{\mathsf{v}}{\mathbf{h}}_{\mathsf{v}}^{\mathsf{H}})$ and $\det(\sigma_{\rm{s}}^2{\mathbf{I}_{NL}}
+\alpha_{\rm{s}}{\mathbf{h}}_{\mathsf{v}}{\mathbf{h}}_{\mathsf{v}}^{\mathsf{H}})=(\sigma_{\rm{s}}^2)^{NL}(1+\frac{\alpha_{\rm{s}}}{\sigma_{\rm{s}}^2}\lVert{\mathbf{h}}_{\mathsf{v}}\rVert^2)$, we obtain
\begin{align}
f_{\beta|{\mathbf{y}}_{{\mathrm{s}},{\mathsf{v}}}}(x|{\mathbf{y}})=
\frac{1}{\pi \alpha_{\rm{s}}\frac{1}{\omega}\sigma_{\rm{s}}^2}{\rm{e}}^{-\frac{\omega}{\alpha_{\rm{s}}\sigma_{\rm{s}}^2}\lvert x - \frac{\alpha_{\rm{s}}}{\omega}{{\mathbf{h}}_{\mathsf{v}}^{\mathsf{H}}{\mathbf{y}}}\rvert^2},
\end{align}
where $\omega={\sigma_{\rm{s}}^2}+{\alpha_{\rm{s}}}\lVert{\mathbf{h}}_{\mathsf{v}}\rVert^2$. Consequently, we obtain
\begin{align}
{\mathbbmss{E}}\{\beta|{\mathbf{y}}_{{\mathrm{s}},{\mathsf{v}}}\}=\int_{0}^{\infty}xf_{\beta|{\mathbf{y}}_{{\mathrm{s}},{\mathsf{v}}}}(x|{\mathbf{y}}){\rm{d}}x
=\frac{\alpha_{\rm{s}}}{\omega}{{\mathbf{h}}_{\mathsf{v}}^{\mathsf{H}}{\mathbf{y}}}.
\end{align}
The above arguments imply that the conditional mean estimate of the Gaussian input $\beta$ is merely a linear transformation of the output ${\mathbf{y}}_{{\mathrm{s}},{\mathsf{v}}}$, namely
\begin{align}\label{MMSE_Estimator}
f_{\mathsf{est}}({\mathbf{y}}_{{\mathrm{s}},{\mathsf{v}}})={\mathbbmss{E}}\{\beta|{\mathbf{y}}_{{\mathrm{s}},{\mathsf{v}}}\}=
\frac{\alpha_{\rm{s}}}{{\sigma_{\rm{s}}^2}+{\alpha_{\rm{s}}}\lVert{\mathbf{h}}_{\mathsf{v}}\rVert^2}{{\mathbf{h}}_{\mathsf{v}}^{\mathsf{H}}{\mathbf{y}}_{{\mathrm{s}},{\mathsf{v}}}}.
\end{align}
Substituting \eqref{Original Signal} and \eqref{MMSE_Estimator} into \eqref{MSE_Def} gives
\begin{align}
{\mathsf{MSE}}&={\mathbbmss{E}}\left\{
\left\lvert
\left(\frac{\alpha_{\rm{s}}\lVert{\mathbf{h}}_{\mathsf{v}}\rVert^2}{{\sigma_{\rm{s}}^2}+{\alpha_{\rm{s}}}\lVert{\mathbf{h}}_{\mathsf{v}}\rVert^2}-1\right)\beta
+\frac{\alpha_{\rm{s}}{{\mathbf{h}}_{\mathsf{v}}^{\mathsf{H}}\mathsf{vec}( \mathbf{N}_{\mathrm{s}} )}}{{\sigma_{\rm{s}}^2}+{\alpha_{\rm{s}}}\lVert{\mathbf{h}}_{\mathsf{v}}\rVert^2}
\right\rvert^2
\right\}\nonumber\\
&={\mathbbmss{E}}\left\{
\left\lvert
\frac{-{\sigma_{\rm{s}}^2}\beta}{{\sigma_{\rm{s}}^2}+{\alpha_{\rm{s}}}\lVert{\mathbf{h}}_{\mathsf{v}}\rVert^2}
+\frac{\alpha_{\rm{s}}{{\mathbf{h}}_{\mathsf{v}}^{\mathsf{H}}\mathsf{vec}( \mathbf{N}_{\mathrm{s}} )}}{{\sigma_{\rm{s}}^2}+{\alpha_{\rm{s}}}\lVert{\mathbf{h}}_{\mathsf{v}}\rVert^2}
\right\rvert^2
\right\}.\label{MSE_Expression_Step2}
\end{align}
Since $\beta$ and $\mathbf{N}_{\mathrm{s}}$ are mutually independent, we can further simplify \eqref{MSE_Expression_Step2} as follows:
\begin{align}
{\mathsf{MSE}}=\frac{\alpha_{\mathrm{s}}}{1+\alpha_{\rm{s}}\lVert{\mathbf{h}}_{\mathsf{v}}\rVert^2/\sigma_{\rm{s}}^2}
=\frac{\alpha_{\mathrm{s}}}{1+\frac{pL\alpha_{\mathrm{s}}}{\sigma_{\rm{s}}^{2}}\lVert \mathbf{h}_{\mathrm{s}} \rVert ^2\lvert \mathbf{h}_{\mathrm{s}}^{\mathsf{T}}\mathbf{w} \rvert^2}.\label{MSE_Formulation}
\end{align}
Comparing \eqref{MSE_Formulation} with \eqref{SR_Downlink_Basic}, we can conclude that the beamforming vector $\mathbf{w}$, which maximizes the SR, also minimizes the mean-square error in estimating the reflection coefficient $\beta$. In other words, maximizing the SR is equivalent to minimizing the estimation error for the reflection coefficient $\beta$.
\subsection{Proof of Lemma~\ref{SR_lemma}}\label{Appendix:A}
Vectorizing the echo signal $\mathbf{Y}_{\mathrm{s}}$ yields \eqref{Original Signal}. It is important to highlight that the conditional MI between $\mathrm{vec}( \mathbf{Y}_{\mathrm{s}} )={\mathbf{y}}_{{\mathrm{s}},{\mathsf{v}}}$ and $\beta$ is equivalent to the capacity of the following multiple-input single-output (MISO) Gaussian channel with a Gaussian distributed input $\beta \sim \mathcal{C} \mathcal{N} \left( 0,\alpha _s \right) $: 
\begin{align}
{\mathbf{y}}_{{\mathrm{s}},{\mathsf{v}}}={\mathbf{h}}_{\mathsf{v}}\beta +\dot{\mathbf{n}},
\end{align}
where ${\mathbf{h}}_{\mathsf{v}}=\sqrt{p}\mathbf{h}_{\mathrm{s}}^{\mathsf{T}}\mathbf{w}\mathsf{vec}( \mathbf{h}_{\mathrm{s}}\mathbf{s}^{\mathsf{H}} )$ denotes the channel vector, and $\dot{\mathbf{n}}=\mathsf{vec}( \mathbf{N}_{\mathrm{s}} )\sim \mathcal{CN}\left( \mathbf{0},\sigma_{\rm{s}} ^2\mathbf{I}_{NL} \right) $. Consequently, the sensing MI can be expressed as 
\begin{align}
I\left( \mathbf{Y}_{\mathrm{s}};\beta |\mathbf{X} \right) =\log _2\det ( \mathbf{I}_{NL}+\alpha_{\rm{s}}{\mathbf{h}}_{\mathsf{v}}{\mathbf{h}}_{\mathsf{v}}^{\mathsf{H}}/\sigma_{\rm{s}}^2 ).
\end{align}
With the aid of Sylvester's identity, we have
\begin{align} 
 I( \mathbf{Y}_{\mathrm{s}};\beta |\mathbf{X} ) &=\log _2( 1+\alpha_{\rm{s}}/\sigma_{\rm{s}}^2{\mathbf{h}}_{\mathsf{v}}^{\mathsf{H}}{\mathbf{h}}_{\mathsf{v}} )   \notag\\
 &=\log _2( 1+{p}/{\sigma_{\rm{s}}^{2}}L\alpha _{\mathrm{s}}\lVert \mathbf{h}_{\mathrm{s}} \rVert ^2\lvert \mathbf{h}_{\mathrm{s}}^{\mathsf{T}}\mathbf{w} \rvert^2 ).  \label{sensing_MI}
\end{align}
Substituting \eqref{sensing_MI} into \eqref{SR_define}, we obtain the results in Lemma~\ref{SR_lemma}.

\subsection{Proof of Theorems \ref{SC_CR_theorem} and \ref{SC_OP_theorem}}\label{Appendix:B}
Notably, the communication channel can be written as $\mathbf{h}_{\mathrm{c}}=\mathbf{R}^{\frac{1}{2}}\overline{\mathbf{h}}$ with $\overline{\mathbf{h}}\sim \mathcal{C} \mathcal{N} \left( \mathbf{0},\mathbf{I}_N \right) $. Since $\mathbf{h}_{\mathrm{s}}$ is irrelevant to $\mathbf{h}_{\mathrm{c}}$, we have $\mathbf{h}_{\mathrm{s}}^{\mathsf{H}}\mathbf{h}_{\mathrm{c}}=\mathbf{h}_{\mathrm{s}}^{\mathsf{H}}\mathbf{R}^{\frac{1}{2}}\overline{\mathbf{h}}\sim \mathcal{CN} ( 0,\mathbf{h}_{\mathrm{s}}^{\mathsf{H}}\mathbf{R}\mathbf{h}_{\mathrm{s}}) $. Hence, $\lvert \mathbf{h}_{\mathrm{s}}^{\mathsf{H}}\mathbf{h}_{\mathrm{c}} \rvert^2$ is exponentially distributed with its probability density function (PDF) given by $\frac{1}{\mathbf{h}_{\mathrm{s}}^{\mathsf{H}}\mathbf{Rh}_{\mathrm{s}}}\mathrm{e}^{-\frac{x}{\mathbf{h}_{\mathrm{s}}^{\mathsf{H}}\mathbf{Rh}_{\mathrm{s}}}}$. Consequently, based on \eqref{SC_CR_def}, the ECR can be calculated as follows:
\begin{align}
\mathcal{R} _{\mathrm{d},\mathrm{c}}^{\mathrm{s}}=\frac{1}{\mathbf{h}_{\mathrm{s}}^{\mathsf{H}}\mathbf{Rh}_{\mathrm{s}}\ln 2}\int_0^{\infty}\!\!\ln \left( 1\!+p/\sigma _{\mathrm{c}}^{2}\left\| \mathbf{h}_{\mathrm{s}} \right\| ^{-2}x \right) \mathrm{e}^{-\frac{x}{\mathbf{h}_{\mathrm{s}}^{\mathsf{H}}\mathbf{Rh}_{\mathrm{s}}}}dx.    
\end{align}
With the aid of \cite[(4.337.2)]{integral}, the results in \eqref{SC_CR} can be derived. When $p \rightarrow \infty$, by applying the fact of $\lim_{x\rightarrow\infty}\log_2(1+x)\approx\log_2{x}$, we have 
\begin{align}
\!\!\mathcal{R}_{\mathrm{d},\mathrm{c}}^{\mathrm{s}}\!\approx\!\frac{1}{\mathbf{h}_{\mathrm{s}}^{\mathsf{H}}\mathbf{Rh}_{\mathrm{s}}\ln 2}\int_0^{\infty}\!\!\ln \left( p/\sigma _{\mathrm{c}}^{2}\left\| \mathbf{h}_{\mathrm{s}} \right\| ^{-2}x \right) \mathrm{e}^{-\frac{x}{\mathbf{h}_{\mathrm{s}}^{\mathsf{H}}\mathbf{Rh}_{\mathrm{s}}}}dx.
\end{align}
With the aid of \cite[(4.331.1)]{integral}, we can obtain \eqref{SC_CR_p}.

By substituting \eqref{SC_CR_def} into \eqref{OP_def}, the OP can be obtained as 
{
\begin{align}
\mathcal{P} _{\mathrm{d}}^{\mathrm{s}}=F_{\left| \mathbf{h}_{\mathrm{s}}^{\mathsf{H}}\mathbf{h}_{\mathrm{c}} \right|^2}\left( \sigma _{\mathrm{c}}^{2}\left\| \mathbf{h}_{\mathrm{s}} \right\| ^2\left( 2^{\mathcal{R} _{0}}-1 \right) /p \right)  ,   
\end{align}
}where $F_{\left| \mathbf{h}_{\mathrm{s}}^{\mathsf{H}}\mathbf{h}_{\mathrm{c}} \right|^2}\left( x \right) =1-\mathrm{e}^{-\frac{x}{\mathbf{h}_{\mathrm{s}}^{\mathsf{H}}\mathbf{Rh}_{\mathrm{s}}}} $ represents the cumulative distribution function (CDF) of $\left| \mathbf{h}_{\mathrm{s}}^{\mathsf{H}}\mathbf{h}_{\mathrm{c}} \right|^2$ following an exponential distribution. The high-SNR approximation can be easily obtained by applying $\underset{x\rightarrow 0}{\lim}\exp \left( -x \right) =1-x$ \cite[Eq. (1.211.1)]{integral}.

\subsection{Proof of Theorems \ref{CC_OP_theorem} and \ref{CC_CR_theorem}}\label{Appendix:C}
Based on \eqref{commun_model}, by defining $\overline{\mathbf{h}}\triangleq \left[ h_1,\ldots,h_{\mathsf{n}} \right]^\mathsf{H} $, the communication channel gain $\lVert \mathbf{h}_{\mathrm{c}} \rVert ^2$ can be expressed as follows:
{
\begin{align}
\lVert \mathbf{h}_{\mathrm{c}} \rVert ^2=\overline{\mathbf{h}}^{\mathsf{H}}\bm{\Sigma} \overline{\mathbf{h}}=\sum_{n=1}^{\mathsf{n}}{\lambda _n\lvert h_n \rvert^2}.
\end{align}
} Note that $\left\{h_n\right\}_{n=1,\ldots,{\mathsf{n}}}$ contains ${\mathsf{n}}$ independent complex Gaussian variables with $h_n\sim \mathcal{C} \mathcal{N} ( 0,1 )$. Therefore, the PDF of $\sum_{n=1}^{\mathsf{n}}{\lambda _n\lvert h_n \rvert^2}$, which is the sum of $\mathsf{n}$ independent exponentially distributed variables, is given by \cite{sum_gamma}
{
\begin{align}
f_{\lVert \mathbf{h}_{\mathrm{c}} \rVert ^2}\left( x \right) =\frac{\lambda _{\mathsf{n}}^{\mathsf{n}}}{\prod_{n=1}^\mathsf{n}{\lambda _n}}\sum_{k=0}^{\infty}{\frac{\delta _kx^{\mathsf{n}+k-1}}{\lambda _{\mathsf{n}}^{\mathsf{n}+k}\left( \mathsf{n}+k \right)!}}{\rm{e}}^{-\frac{x}{\lambda _{\mathsf{n}}}}.
\end{align}
}With the aid of \cite[(3.351.1)]{integral}, we can calculated the CDF of $\lVert \mathbf{h}_{\mathrm{c}} \rVert ^2$ as follows:
{
\begin{align}
F_{\left\| \mathbf{h}_{\mathrm{c}} \right\| ^2}\left( x \right) 
=\frac{\lambda _{\mathsf{n}}^{\mathsf{n}}}{\prod_{n=1}^{\mathsf{n}}{\lambda _n}}\sum_{k=0}^{\infty}{\frac{\delta _k\Upsilon ( \mathsf{n}+k,\frac{x}{\lambda _{\mathsf{n}}} )}{\left( \mathsf{n}+k-1 \right) !}}.    
\end{align}
}Subsequently, the OP is given by 
{
\begin{align}
\mathcal{P} _{\mathrm{d}}^{\mathrm{c}}\!=\!\mathrm{Pr}\!\left(\! \left\| \mathbf{h}_{\mathrm{c}} \right\| ^2\!<\!\frac{ 2^{\mathcal{R} _{0}}\!-\!1  }{p/\sigma_{\rm{c}}^2} \right)\! =\!F_{\left\| \mathbf{h}_{\mathrm{c}} \right\| ^2}\!\!\left( \frac{ 2^{\mathcal{R} _0}\!-\!1  }{p/\sigma_{\rm{c}}^2} \right) .
\end{align}
}When $p\rightarrow \infty$, by utilizing the asymptotic property of the lower incomplete gamma function \cite[(8.354.1)]{integral}, i.e., $\lim_{x\rightarrow \infty}\Upsilon \left( s,x \right) \simeq \frac{x^s}{s}$, within \eqref{CC_OP}, we can obtain \eqref{CC_OP_p}. Moreover, the ECR can be calculated as follows:
{
\begin{align}
\mathcal{R} _{\mathrm{d},\mathrm{c}}^{\mathrm{c}}=\int_0^{\infty}{\log _2\left( 1+p/\sigma_{\rm{c}}^2x \right) f_{\left\| \mathbf{h}_{\mathrm{c}} \right\| ^2}\left( x \right) dx}.
\end{align}
}With the aid of \cite[(4.337.5)]{integral}, we can obtain \eqref{CC_CR}. When $p\rightarrow \infty$, by leveraging $\lim_{x\rightarrow\infty}\log_2(1+x)\simeq\log_2{x}$ and \cite[(4.352.1)]{integral}, the high-SNR approximation can be derived.
\subsection{Proof of Theorem \ref{cc_SR_theorem}}\label{Appendix:D}
By performing some manipulations based on \eqref{CC_SR_def}, the average SR can be written as 
\begin{align}
\!\!\mathcal{R} _{\mathrm{d},\mathrm{s}}^{\mathrm{c}}\!= \!\mathbbmss{E} \{\overline{\mathcal{R}} _{\mathrm{d},\mathrm{s}}^{\mathrm{c}}\}\!= \!\frac{1}{L}\Big(\underset{P_1}{\underbrace{\mathbbmss{E} \{ \log _2( \overline{\mathbf{h}}^{\mathsf{H}}\mathbf{\Delta}\overline{\mathbf{h}} ) \}}}\!-\!\underset{P_2}{\underbrace{\mathbbmss{E}\{\log _2( \lVert \mathbf{h}_{\mathrm{c}} \rVert^2 ) \}}}\Big) ,
\end{align}
Upon applying ED to $\mathbf{\Delta}$, i.e., $\bm{\Delta}=\mathbf{Q}\bm{\Lambda} \mathbf{Q}^{\mathsf{H}}$, where $\mathbf{Q}$ is unitary and $\bm{\Lambda} =\mathsf{diag}\left\{ \lambda _{\Delta,1},\ldots,\lambda _{\Delta,\mathsf{t}},0,\ldots,0 \right\} $ with $\mathsf{t}=\mathrm{rank}\left( \mathbf{\Delta} \right) $, and defining $\mathbf{Q}^{\mathsf{H}}\overline{\mathbf{h}}\triangleq \left[ z_1,\ldots,z_N \right] ^{\mathsf{H}}$, where $\left\{z_n\right\}_{n=1,\ldots,N}$ contains $N$ independent complex Gaussian variables with $z_n\sim \mathcal{C} \mathcal{N} ( 0,1 )$, we can rewrite $P_1$ as follows: 
\begin{align}\label{e2}
P_1\!=\!\mathbbmss{E} \{ \log _2( \overline{\mathbf{h}}^{\mathsf{H}}\mathbf{Q}\bm{\Lambda} \mathbf{Q}^{\mathsf{H}}\overline{\mathbf{h}} ) \}\!=\!\mathbb{E} \left\{ \log _2\!\left( \sum_{t=1}^{\mathsf{t}}{\lambda _{\Delta,t}\left| z_t \right|^2} \right) \right\} .
\end{align}
Consequently, both \eqref{e2} and $P_2$ can be calculated by following the steps similar to Appendix~\ref{Appendix:C}. When $p\rightarrow \infty$, the results of \eqref{CC_SR_p} can be obtained by following the similar steps of the high-SNR approximation derivation in Appendix~\ref{Appendix:B}.  
\subsection{Proof of Theorem \ref{pareto_theo}}\label{Appendix:E}
Based on the Karush-Kuhn-Tucker conditions, we have
{
\begin{numcases}{}
\nabla ( -\mathcal{R} ) +\lambda \nabla ( \lVert \mathbf{w} \rVert ^2-1 ) +\mu _1\nabla f_1+\mu _2\nabla f_2=\mathbf{0},\label{c1}\\
\mu _1f_1=0,\ \mu _2f_2=0,\ \mu _1\geqslant 0,\ \mu _2\geqslant 0,
\end{numcases}
}where $f_1=2^{( 1-\tau ) \mathcal{R}}-1-\lvert \mathbf{h}_1^{\mathsf{T}}\mathbf{w} \rvert^2$, $f_2=2^{ \tau L  \mathcal{R}}-1-\lvert \mathbf{h}_2^{\mathsf{T}}\mathbf{w} \rvert^2$, and $\lambda$, $\mu_1$, $\mu_2$ are real Lagrangian multipliers. Based on \eqref{c1}, we obtain
{
\begin{numcases}{}
\mu _1\mathbf{h}_1^{*}\mathbf{h}_{1}^{\mathsf{T}}\mathbf{w}+\mu _2\mathbf{h}_2^{*}\mathbf{h}_{2}^{\mathsf{T}}\mathbf{w}=\lambda \mathbf{w},\label{c3}\\
\mu _1\lvert \mathbf{h}_1^{\mathsf{T}}\mathbf{w} \rvert^2+\mu _2\lvert \mathbf{h}_2^{\mathsf{T}}\mathbf{w} \rvert^2=\lambda \mathbf{w}^{\mathsf{H}}\mathbf{w}=\lambda ,\label{c4}\\
\mu _1 2^{( 1-\tau )\mathcal{R}}( 1-\tau ) \ln 2+\mu _2 2^{\tau L\mathcal{R}}\tau L \ln 2=1.\label{c5}
\end{numcases}
}It can be concluded from \eqref{c5} that $\mu_1$ and $\mu_2$ cannot be $0$ at the same time. Therefore, we discuss three cases as follows. 
\subsubsection*{Case 1: $\mu _1>0$ and $\mu _2=0$}
In this case, we have
{
\begin{numcases}{}
\mu _1\mathbf{h}_1\mathbf{h}_{1}^{\mathsf{H}}\mathbf{w}=\lambda \mathbf{w},\label{c6}\\
\mu _1\lvert \mathbf{h}_1^{\mathsf{T}}\mathbf{w} \rvert^2=\lambda ,\label{c7}\\
\mu _1 2^{( 1-\tau )\mathcal{R}}( 1-\tau)\ln 2=1,\label{c8}\\
f_2=0\Rightarrow \lvert \mathbf{w}^{\mathsf{H}}\mathbf{h}_1 \rvert^2= 2^{( 1-\tau ) \mathcal{R}}-1.  \label{c9}  
\end{numcases}
}Based on the above conditions, the optimal value of $\mathbf{w}$ and  $\mathcal{R}$ are given by $\mathbf{w}_\tau={\lVert \mathbf{h}_{\mathrm{c}}\rVert^{-1}}{\mathbf{h}_{\mathrm{c}}^{*}}$ and $\mathcal{R}^{\star}=\frac{1}{1-\tau}\log( 1+\lVert \mathbf{h}_1 \rVert ^2) ={\overline{\mathcal{R}}_{\mathrm{d},\mathrm{c}}^{\mathrm{c}}}/({1-\tau})$, respectively. Accordingly, the SR satisfies $\overline{\mathcal{R}}_{\rm{d},\mathrm{s}}^{\rm{c}}\geqslant \tau\mathcal{R} ^{\star}$, which yields $\tau \in [ 0,{\overline{\mathcal{R}}_{\rm{d},\mathrm{s}}^{\rm{c}}}/({\overline{\mathcal{R}}_{\mathrm{d},\mathrm{c}}^{\mathrm{c}}+\overline{\mathcal{R}}_{\rm{d},\mathrm{s}}^{\rm{c}}})] $.
\subsubsection*{Case 2: $\mu _1=0$ and $\mu _2>0$}
Similar to Case 1, we can obtain $\mathcal{R} ^{\star}=\frac{\mathcal{R}_{\mathrm{d},\mathrm{s}}^{\mathrm{s}}}{\tau}$ and $\mathbf{w}_\tau={\lVert \mathbf{h}_{\mathrm{s}}\rVert^{-1}}{\mathbf{h}_{\mathrm{s}}^{*}}$ for $\tau \in [{\mathcal{R}_{\mathrm{d},\mathrm{s}}^{\mathrm{s}}}/(\overline{\mathcal{R}}_{\mathrm{d},\mathrm{c}}^{\mathrm{s}}+\mathcal{R}_{\mathrm{d},\mathrm{s}}^{\mathrm{s}}), 1]$.
\subsubsection*{Case 3: $\mu _1>0$ and $\mu _2>0$}
In this case, we have $f_1=0\Rightarrow \lvert \mathbf{h}_1^{\mathsf{T}}\mathbf{w}\rvert^2= 2^{( 1-\tau ) \mathcal{R}}-1$ and $f_2=0\Rightarrow \lvert \mathbf{h}_2^{\mathsf{T}}\mathbf{w} \rvert^2= 2^{\tau L \mathcal{R}}-1$. From \eqref{c3}, we have $\mathbf{w}=\frac{\mu _1\mathbf{h}_{1}^{\mathsf{T}}\mathbf{w}}{\lambda}\mathbf{h}_1^{*}+\frac{\mu _2\mathbf{h}_{2}^{\mathsf{T}}\mathbf{w}}{\lambda}\mathbf{h}_2^{*}\triangleq a\mathbf{h}_1^{*}+b\mathbf{h}_2^{*}$, which further yields
{
\begin{align}\label{c12}
\mu _1( \lVert \mathbf{h}_1 \rVert ^2+{b}/{a}\rho ) =\mu _2( \lVert \mathbf{h}_2 \rVert ^2+{a}/{b}\rho^* ) =\lambda,
\end{align}
}where $\frac{a}{b}=\frac{\mu _1\sqrt{2^{\left( 1-\tau \right) \mathcal{R}}-1}}{\mu _2\sqrt{2^{\tau L\mathcal{R}}-1}{\rm{e}}^{-{\rm{j}}\angle \rho}}$. By combining \eqref{c12} and \eqref{c5}, we can derive $\mu _1=\frac{\xi _2}{\chi }$, $\mu _2=\frac{\xi _1}{\chi }$, and $\lambda =\frac{\lVert \mathbf{h}_1 \rVert ^2\lVert \mathbf{h}_2 \rVert ^2-\rho ^2}{\chi}$. Additionally, upon substituting the above results of $\mu_1$, $\mu_2$ and $\lambda$ into \eqref{c4}, we obtain an equation for $\mathcal{R}$ as follows:
\begin{align}\label{c13}
\xi _1( 2^{\tau L\mathcal{R}}-1 ) +\xi _2( 2^{( 1-\tau ) \mathcal{R}}-1 )= \lVert \mathbf{h}_1 \rVert ^2\lVert \mathbf{h}_2 \rVert ^2-\lvert\rho\rvert^2.
\end{align}
Since $\xi _1( 2^{\tau L\mathcal{R}}-1 ) +\xi _2( 2^{( 1-\tau ) \mathcal{R}}-1 )$ is a monotonic function with respect to $\mathcal{R}$, ranging from $0$ to $\infty $, the optimal solution $\mathcal{R}=\mathcal{R}^\star$ can be obtained by solving \eqref{c13}, and $w_\tau$ follows immediately.
\subsection{Proof of Lemma \ref{Lemma_S_CSI_Downlink}}\label{Appendix:F}
As BS dose not know $\mathbf{h}_{\mathrm{c}}$, the C-C design of $\mathbf{w}$ is irrelevant to $\mathbf{h}_{\mathrm{c}}$, which yields $\mathbf{w}^{\mathsf{H}}\mathbf{h}_{\mathrm{c}}=\mathbf{w}^{\mathsf{H}}\mathbf{R}^{\frac{1}{2}}\overline{\mathbf{h}}\sim \mathcal{CN} ( 0,\mathbf{w}^{\mathsf{H}}\mathbf{Rw}) $. Hence, $\lvert \mathbf{w}^{\mathsf{H}}\mathbf{h}_{\mathrm{c}} \rvert^2$ is exponentially distributed with its PDF given by $\frac{1}{\mathbf{w}^{\mathsf{H}}\mathbf{Rw}}{\rm{e}}^{-\frac{x}{\mathbf{w}^{\mathsf{H}}\mathbf{Rw}}}$. Consequently, the ECR in \eqref{Average_CR_C_C_Beam} can be expressed as follows:
{
\begin{align}
\mathbbmss{E}\{\overline{\mathcal{R}}_{\mathrm{d},\mathrm{c}}\}&=
\int_{0}^{\infty}\log_2(1+p/\sigma_{\rm{c}}^2x)\frac{1}{\mathbf{w}^{\mathsf{H}}\mathbf{Rw}}{\rm{e}}^{-\frac{x}{\mathbf{w}^{\mathsf{H}}\mathbf{Rw}}}dx
\notag\\
&=\int_{0}^{\infty}\log_2(1+p/\sigma_{\rm{c}}^2{\mathbf{w}^{\mathsf{H}}\mathbf{Rw}}x){\rm{e}}^{-x}dx,
\end{align}
}which is monotone increasing with ${\mathbf{w}^{\mathsf{H}}\mathbf{Rw}}$. Consequently, the problem in \eqref{Average_CR_C_C_Beam} is equivalent to 
{
\begin{align}
\mathbf{w}=\argmax\nolimits_{\lVert\mathbf{w}\rVert^2=1}{\mathbf{w}^{\mathsf{H}}\mathbf{Rw}}={\mathbf{u}},
\end{align}
}where ${\mathbf{u}}\in{\mathbbmss{C}}^{N\times1}$ is used to denote the principal eigenvector of $\mathbf{R}$. Furthermore, based on the ED of $\mathbf{R}$ shown in \eqref{correlation_matrix}, it is trivial that ${\mathbf{u}}={\mathbf{a}}_{\star}$.   
\subsection{Proof of Lemma~\ref{up_CC_SR_lem}}\label{Appendix:G}
The superposed uplink S\&C signal reads $\mathbf{Y}=\sqrt{p_{\rm{s}}}\beta \mathbf{h}_{\mathrm{s}}\mathbf{h}_{\mathrm{s}}^{\mathsf{T}}{\mathbf{w}}{\mathbf{s}}_{\mathrm{s}}^\mathsf{H}+\mathbf{Z}_{\mathrm{c}}$. Vectorizing this signal yields
{
\begin{align}\label{f1}
\mathsf{vec}( \mathbf{Y} ) =\sqrt{p_{\rm{s}}}\mathbf{h}_{\mathrm{s}}^{\mathsf{T}}\mathbf{w}{\mathbf{h}}_{\rm{v}} \beta +\mathbf{z}_{\mathrm{c}},    
\end{align}
}where ${\mathbf{h}}_{\rm{v}}=\mathsf{vec}( \mathbf{h}_{\mathrm{s}}{\mathbf{s}}_{\mathrm{s}}^\mathsf{H} )$ and $\mathbf{z}_{\mathrm{c}}=\mathsf{vec}( \mathbf{Z}_{\mathrm{c}} ) $. By treating \eqref{f1} as a MISO channel model with Gaussian noise $\mathbf{z}_{\mathrm{c}}$, the SR can be written as follows:
{
\begin{align}\label{f2}
\overline{\mathcal{R}} _{\mathrm{u},\mathrm{s}}^{\mathrm{c}}=L^{-1}\log _2( 1+\alpha_{\mathrm{s}}p_{\mathrm{s}}\lvert\mathbf{h}_{\mathrm{s}}^{\mathsf{T}}\mathbf{w}\rvert^2
{\mathbf{h}}_{\rm{v}}^{\mathsf{H}}{\mathbbmss{E}}^{-1}\{\mathbf{z}_{\mathrm{c}}\mathbf{z}_{\mathrm{c}}^{\mathsf{H}}\}{\mathbf{h}}_{\rm{v}}).
\end{align}
}Recall that $\mathbf{Z}_{\mathrm{c}}=\sqrt{p_{\mathrm{c}}}\mathbf{h}_{\mathrm{c}} \mathbf{s}_{\mathrm{c}}^\mathsf{H}+\mathbf{N}_{\mathrm{u}}$, which together with the fact that ${\mathbbmss{E}}\{\mathbf{s}_{\mathrm{c}}\mathbf{s}_{\mathrm{c}}^{\mathsf{H}}\}={\mathbf{I}_L}$, yields ${\mathbbmss{E}}\{\mathbf{z}_{\mathrm{c}}\mathbf{z}_{\mathrm{c}}^{\mathsf{H}}\}={\mathbf{I}_L}\otimes(p_{\mathrm{c}}\mathbf{h}_{\mathrm{c}}\mathbf{h}_{\mathrm{c}}^{\mathsf{H}}+\sigma_{\rm{u}}^2\mathbf{I}_N)$. By further exploiting the fact that $L^{-1}\lVert \mathbf{s}_{\mathrm{s}} \rVert^2=1$, we obtain
{
\begin{align}\label{f3}
{\mathbf{h}}_{\rm{v}}^{\mathsf{H}}{\mathbbmss{E}}^{-1}\{\mathbf{z}_{\mathrm{c}}\mathbf{z}_{\mathrm{c}}^{\mathsf{H}}\}{\mathbf{h}}_{\rm{v}}=
{\mathbf{h}}_{\rm{s}}^{\mathsf{H}}(p_{\mathrm{c}}\mathbf{h}_{\mathrm{c}}\mathbf{h}_{\mathrm{c}}^{\mathsf{H}}+\sigma_{\rm{u}}^2\mathbf{I}_N)^{-1}{\mathbf{h}}_{\rm{s}}\lVert \mathbf{s}_{\mathrm{s}} \rVert^2.     
\end{align}
}By harnessing the Woodbury matrix identity, we have $(p_{\mathrm{c}}\mathbf{h}_{\mathrm{c}}\mathbf{h}_{\mathrm{c}}^{\mathsf{H}}+\sigma_{\rm{u}}^2\mathbf{I}_N)^{-1}=\frac{1}{\sigma _{\mathrm{u}}^{2}}( \mathbf{I}_N-\frac{p_{\mathrm{c}}}{p_{\mathrm{c}}\left\| \mathbf{h}_{\mathrm{c}} \right\| ^2+\sigma _{\mathrm{u}}^{2}}\mathbf{h}_{\mathrm{c}}\mathbf{h}_{\mathrm{c}}^{\mathsf{H}} ) $. The final results follow immediately. 

\subsection{Proof of Theorem \ref{up_CC_SR_the}}\label{Appendix:H}
By defining
\begin{align}
	\mathbf{\Xi }&\triangleq\mathbf{I}_N+\frac{p_{\mathrm{s}}}{\sigma _{\mathrm{u}}^{2}}L\alpha _{\mathrm{s}}\left\| \mathbf{h}_{\mathrm{s}} \right\| ^2( \left\| \mathbf{h}_{\mathrm{s}} \right\| ^2\mathbf{I}_N-\mathbf{h}_{\mathrm{s}}\mathbf{h}_{\mathrm{s}}^{\mathsf{H}} )\notag\\
	&=\Big( 1+\frac{p_{\mathrm{s}}}{\sigma _{\mathrm{u}}^{2}}L\alpha _{\mathrm{s}}\left\| \mathbf{h}_{\mathrm{s}} \right\| ^4 \Big) \mathbf{I}_N-\frac{p_{\mathrm{s}}}{\sigma _{\mathrm{u}}^{2}}L\alpha _{\mathrm{s}}\left\| \mathbf{h}_{\mathrm{s}} \right\| ^2\mathbf{h}_{\mathrm{s}}\mathbf{h}_{\mathrm{s}}^{\mathsf{H}}, 
\end{align}
the matrix $\mathbf{\Theta}$ can be expressed as $\mathbf{\Theta}=\mathbf{R}^{\frac{1}{2}}\mathbf{\Xi R}^{\frac{1}{2}}$. Furthermore, based on the ED $\mathbf{h}_{\mathrm{s}}\mathbf{h}_{\mathrm{s}}^{\mathsf{H}}=\mathbf{P}\mathrm{diag}\{ \left\| \mathbf{h}_{\mathrm{s}} \right\| ^2,0,\ldots,0 \} \mathbf{P}^{\mathsf{H}}$ with $\mathbf{P}\in{\mathbbmss{C}}^{N\times N}$ being a unitary matrix, $\mathbf{\Xi }$ can be rewritten as follows:
\begin{align}
	\mathbf{\Xi }&=\nu \mathbf{PP}^{\mathsf{H}}-\mathbf{P}\mathrm{diag}\left\{ \frac{p_{\mathrm{s}}}{\sigma _{\mathrm{u}}^{2}}L\alpha _{\mathrm{s}}\left\| \mathbf{h}_{\mathrm{s}} \right\| ^4,0,\ldots,0 \right\} \mathbf{P}^{\mathsf{H}}\notag\\
	&=\mathbf{P}\mathrm{diag}\left\{ 1,\nu,\ldots,\nu \right\} \mathbf{P}^{\mathsf{H}}\succ {\mathbf{0}},
\end{align}
where $\nu =1+\frac{p_{\mathrm{s}}}{\sigma _{\mathrm{u}}^{2}}L\alpha _{\mathrm{s}}\left\| \mathbf{h}_{\mathrm{s}} \right\| ^4>0$. Therefore, it follows that $\mathbf{\Theta}\succeq {\mathbf{0}}$. Similarly, we can prove that $\tilde{\bm{\Theta}}\succeq{\mathbf{0}}$. Subsequently, the results of \textbf{Theorem \ref{up_CC_SR_the}} can be derived by following the steps similar to those in Appendix~\ref{Appendix:D}.
\vspace{-5pt}
\subsection{Proof of Theorem \ref{up_SC_CR_the}}\label{Appendix:I}
To achieve the optimal SR, we have $\mathbf{w}={\mathbf{h}}_{\rm{s}}\lVert{\mathbf{h}}_{\rm{s}}\rVert^{-1}$, and thus the received signal of the BS at the $\ell$th time slot reads
{
\begin{align}
\mathbf{y}_\ell=\sqrt{p_{\mathrm{c}}}\mathbf{h}_{\mathrm{c}}\mathrm{s}_{\mathrm{c},\ell}+\underset{\mathbf{z}_{\mathrm{s},\ell}}{\underbrace{\sqrt{p}_{\rm{s}}\beta \mathbf{h}_{\mathrm{s}}\lVert \mathbf{h}_{\mathrm{s}} \rVert \mathrm{s}_{\mathrm{s},\ell}+\mathbf{n}_{\mathrm{u},\ell}}}    
\end{align}
}for $\ell=1,\ldots,L$, where $\mathbf{n}_{\mathrm{u},\ell}\sim{\mathcal{CN}}(\mathbf{0},\sigma_{\rm{u}}^2{\mathbf{I}_N})$ is the $\ell$th column of $\mathbf{N}_{\mathrm{u}}$. By treating $\mathbf{z}_{\mathrm{s},\ell}$ as a zero-mean Gaussian random variable, we calculate the CR as follows:
{
\begin{align}\label{g2}
\overline{\mathcal{R}}_{\mathrm{u},\mathrm{c}}^{\mathrm{s}}= \log _2( 1+p_{\mathrm{c}}\mathbf{h}_{\mathrm{c}}^{\mathsf{H}}{\mathbbmss{E}}\{ \mathbf{z}_{\mathrm{s},\ell}\mathbf{z}_{\mathrm{s},\ell}^{\mathsf{H}}\} ^{-1}\mathbf{h}_{\mathrm{c}} ) 
\end{align}
}with ${\mathbbmss{E}}\{ \mathbf{z}_{\mathrm{s},\ell}\mathbf{z}_{\mathrm{s},\ell}^{\mathsf{H}}\}=p_{\mathrm{s}}\alpha_{\mathrm{s}}\lVert{\mathbf{h}}_{\rm{s}}\rVert^{2}{\mathbf{h}}_{\rm{s}}
{\mathbf{h}}_{\rm{s}}^{\mathsf{H}}+\sigma_{\rm{u}}^2{\mathbf{I}_N}$, which is independent with the time slot. Hence, the ECR is given by ${\mathcal{R}}_{\mathrm{u},\mathrm{c}}^{\mathrm{s}}={\mathbbmss{E}}\{\overline{\mathcal{R}}_{\mathrm{u},\mathrm{c}}^{\mathrm{s}}\}$, which can be analyzed by following the similar steps presented in Appendix~\ref{Appendix:C}.
\vspace{-5pt}
\subsection{Proof of Lemma \ref{Lemma_S_CSI_Uplink}}\label{Appendix:J}
As $\mathbf{h}_{\mathrm{c}}$ is unknown to the BS, the design of $\mathbf{v}$ is irrelevant to $\mathbf{h}_{\mathrm{c}}$, which yields $\mathbf{v}^{\mathsf{H}}\mathbf{h}_{\mathrm{c}}=\mathbf{v}^{\mathsf{H}}\mathbf{R}^{\frac{1}{2}}\overline{\mathbf{h}}\sim \mathcal{CN}( 0,\mathbf{v}^{\mathsf{H}}\mathbf{Rv})$. Hence, $\frac{{p}_{\mathrm{c}}\lvert \mathbf{v}^{\mathsf{H}}\mathbf{h}_{\mathrm{c}} \rvert^2}{p_{\rm{s}}\alpha _{\mathrm{s}}\lVert \mathbf{h}_{\mathrm{s}} \rVert ^2\lvert \mathbf{v}^{\mathsf{H}}\mathbf{h}_{\mathrm{s}} \rvert^2+\sigma_{\rm{u}}^2}$ is exponentially distributed. Based on the property of exponential distribution, maximizing the ECR $\mathbbmss{E}\{\overline{\mathcal{R}}_{\mathrm{u},\mathrm{c}}^{\mathrm{s}}\}$ is equivalent to maximizing ${\mathbbmss{E}}\left\{\frac{{p}_{\mathrm{c}}\lvert \mathbf{v}^{\mathsf{H}}\mathbf{h}_{\mathrm{c}} \rvert^2}{p_{\rm{s}}\alpha _{\mathrm{s}}\lVert \mathbf{h}_{\mathrm{s}} \rVert ^2\lvert \mathbf{v}^{\mathsf{H}}\mathbf{h}_{\mathrm{s}} \rvert^2+\sigma_{\rm{u}}^2}\right\}$, which yields
\begin{align}
\mathbf{v}_{\star}&=\argmax_{\lVert\mathbf{v}\rVert^2=1}\frac{{\mathbbmss{E}}\{\lvert \mathbf{v}^{\mathsf{H}}\mathbf{h}_{\mathrm{c}} \rvert^2\}}{p_{\rm{s}}/\sigma_{\rm{u}}^2\alpha _{\mathrm{s}}\lVert \mathbf{h}_{\mathrm{s}} \rVert ^2\lvert \mathbf{v}^{\mathsf{H}}\mathbf{h}_{\mathrm{s}} \rvert^2+1}\notag\\
&=\argmax_{\lVert\mathbf{v}\rVert^2=1}\frac{\mathbf{v}^{\mathsf{H}}\mathbf{Rv}}{\mathbf{v}^{\mathsf{H}}( p_{\rm{s}}/\sigma_{\rm{u}}^2\alpha _{\mathrm{s}}\left\| \mathbf{h}_{\mathrm{s}} \right\| ^2\mathbf{h}_{\mathrm{s}}\mathbf{h}_{\mathrm{s}}^{\mathsf{H}}+\mathbf{I}_N )\mathbf{v}}.
\end{align}
By applying the Rayleigh quotient theorem, we can obtain the expression of $\mathbf{v}_{\star}$. Besides, it can be concluded that $\frac{\mathbf{v}_{\star}^{\mathsf{H}}\mathbf{R}\mathbf{v}_{\star}}{\mathbf{v}_{\star}^{\mathsf{H}}( p_{\rm{s}}/\sigma_{\rm{u}}^2\alpha _{\mathrm{s}}\lVert \mathbf{h}_{\mathrm{s}} \rVert ^2\mathbf{h}_{\mathrm{s}}\mathbf{h}_{\mathrm{s}}^{\mathsf{H}}+\mathbf{I}_N )\mathbf{v}_{\star}}=\varkappa$, where $\varkappa$ denotes the principal eigenvalue of $( \frac{p_{\rm{s}}}{\sigma_{\rm{u}}^2}\alpha _{\mathrm{s}}\left\| \mathbf{h}_{\mathrm{s}} \right\| ^2\mathbf{h}_{\mathrm{s}}\mathbf{h}_{\mathrm{s}}^{\mathsf{H}}+\mathbf{I}_N)^{-\frac{1}{2}}{\mathbf{R}}(\frac{p_{\rm{s}}}{\sigma_{\rm{u}}^2}\alpha _{\mathrm{s}}\left\| \mathbf{h}_{\mathrm{s}} \right\| ^2\mathbf{h}_{\mathrm{s}}\mathbf{h}_{\mathrm{s}}^{\mathsf{H}}+\mathbf{I}_N)^{-\frac{1}{2}}$.
\end{appendix}
\vspace{-5pt}
\bibliographystyle{IEEEtran}
\bibliography{IEEEabrv}
\end{document}